\documentclass[a4paper,11pt]{article}
\usepackage{graphicx,amsmath,amsfonts,amssymb,dcolumn,amsthm,euscript}
\def\bea{\begin{eqnarray}}
\def\eea{\end{eqnarray}}
\def\be{\begin{equation}}
\def\ee{\end{equation}}
\usepackage[usenames,dvipsnames]{xcolor}
\usepackage{float}   

\usepackage{mathtools}
\DeclarePairedDelimiter\bra{\langle}{\rvert}
\DeclarePairedDelimiter\ket{\lvert}{\rangle}
\DeclarePairedDelimiterX\braket[2]{\langle}{\rangle}{#1\,\delimsize\vert\,\mathopen{}#2}

\newcommand\nn{\nonumber} 
\newcommand{\bq}{\begin{equation}}
\newcommand\eq{\end{equation}}

\newcommand\pa{\partial}
\def\hb{\hbar}

\usepackage{changepage}

\usepackage{graphicx}

\usepackage{slashed}
\usepackage{verbatim}
\usepackage{blindtext}
\usepackage{amsmath}
\usepackage{geometry}
\usepackage{pdflscape}   
\usepackage[nottoc, notlof, notlot]{tocbibind} 
\graphicspath{{./images/}}
\usepackage{mathtools}
\usepackage{jheppub}
\usepackage{cancel} 

\DeclareMathOperator{\Tr}{Tr}

\usepackage{booktabs}
\usepackage{caption}
\usepackage{float}
\usepackage{titlesec}
\usepackage{capt-of}

\usepackage{tcolorbox}
\usepackage{tabularx}
\usepackage{colortbl}
\tcbuselibrary{skins}

\newcolumntype{Y}{>{\raggedleft\arraybackslash}X}

\tcbset{tab/.style={enhanced, fonttitle=\bfseries,fontupper=\normalsize\sffamily,
colback=white,colframe=black,colbacktitle=white,
coltitle=black,center title}}

\usepackage{array}
\usepackage{arydshln}
\usepackage{mathtools}

\usepackage{hyperref}

\numberwithin{equation}{section}

\def\demi{\frac{1}{2}}

\def\Lie {\mathcal{L}}
\def\mpm {\mu^+_-}
\def\={&=&}

\title{BRST Covariant Phase Space and \\ Holographic Ward Identities}

\author{Laurent Baulieu}
\author{and Tom Wetzstein}
 \affiliation{LPTHE, Sorbonne Universit\'e, CNRS,\\4 Place Jussieu, 75005 Paris, France}

\emailAdd{baulieu@lpthe.jussieu.fr}
\emailAdd{twetzstein@lpthe.jussieu.fr }

\abstract{This paper develops an enlarged BRST framework to treat the large gauge transformations of a given quantum field theory.  It determines the associated infinitely many Noether charges stemming from a gauge fixed and BRST invariant Lagrangian, a  result   that cannot be obtained     from    Noether's  second theorem.    The geometrical significance of this result is highlighted by the construction of a trigraded BRST covariant phase space, allowing a BRST invariant gauge fixing procedure. This provides an appropriate framework for determining the conserved BRST Noether current of the global BRST symmetry and the associated global Noether charges.  The latter are found to be equivalent with the usual classical corner charges of large gauge transformations.  It allows one to prove the gauge independence of their physical effects at the perturbative quantum level. In particular,  the underlying BRST fundamental canonical relation provides the same graded symplectic brackets as in the classical covariant phase space.  A unified Lagrangian Ward identity for small and large gauge transformations is built. It consistently decouples into a bulk part for small gauge transformations, which is the standard BRST--BV quantum master equation,   and a boundary part for large gauge transformations. The boundary part provides a perturbation theory origin for the  invariance of the Hamiltonian physical $\mathcal{S}$-matrix under asymptotic symmetries.  Holographic anomalies for the boundary Ward identity are studied and found to be solutions of a codimension one Wess--Zumino consistency condition.  Such solutions are studied in the context of extended BMS symmetry.  Their existence clarifies the status of the $1$-loop correction to the subleading soft graviton theorem. }

\begin{document} 
\maketitle


\def\blue{  \color{blue}}
\def\red{  \color{red}}
\def\red{  \color{black}}
\def\blue{  \color{black}}

\def\bea{\begin{eqnarray}}
\def\eea{\end{eqnarray}}
\def\t{\tau}
\def\hb{\hbar}
\def\demi{\frac{1}{2}}
\def\blue{  \color{blue}}
\def\red{  \color{red}}
\def\black{ \color{black}}
\renewcommand\t{\tau}
\def\Diff{\mathrm{Diff}}
\def \m {\mu^z_{\bar z}}
\def \mb {\mu_z^{\bar z}} 
\def \mo {\mu^z_{0}}
\def \mbo {\mu_0^{\bar z}}
\def \mob{\mu_0^{\bar{z}}}
\def \mt {\mu^z_\t}
\def \mbt {\mu^\bz_\t}
\def \mut {\mu^z_\t}
\def \mubt {\mu^\bz_\t}
\def \mot {\mu^0_\t}
\def\w{\wedge}
\def\p{\pa_z}
\def\p0{\pa_0}
\def\bp{\pa_{\bar z}}
\def\pt{\pa_\t}
\def\bz{{\bar z}  }
\def\bZ{\bar{Z}}
\def\o {\omega}
\def\O {\Omega}
\def\vp {\varphi}
\def\vpb {\bar{ \vp}}
\def\Eo{ {\cal{ E}}^z}
\def\al{\alpha}
\def\Ce{C_\eta}
\def\Eob{  {\cal{ E}}^\bz}
\def\E0 { \mathcal{E}^0  }
\def\Eot {  \mathcal{E}^\t }
\def\Et{\mathcal{E} }
\def\Do{ {\cal {D}}_0  }
\def\Dz{ {\cal {D}}_z  }
\def\Dbz{ {\cal {D}}_\bz  }
\def\Lie {\mathcal{L}}
\def\Da{ {\mathcal{D}_\alpha} }
\def\mmb{1-\m\mb }
\def\Dt{ {\cal {D}}_\tau  }
\def \pbM  {   \begin{pmatrix}     }
\def \peM  {   \end{pmatrix}     }
\def \bM  {   \begin{matrix}     }
\def \eM  {   \end{matrix}     }
\def \Mo {\pbM  \mo & \mt \\ \mbo & \mbt  \peM}
\def \idd {\pbM 0 & 1 \\ -1 & 0  \peM}
\def \mmbo {\mo \mbt - \mbo \mt }
\def \mmo {\mbo \mt - \mo \mbt	}
\def \mpm {\mu^+_-}
\def\DD{\mathbb{D}}
\def\C{\mathsf{C}}
\def \V   { {V_{\rm BRST}  } }
\def \J {\star J_{\rm BRST}}
 \def \invMu { \begin{pmatrix}
1
&
-\m
 \\ 
-\mb
  &
1 
\end{pmatrix}  }
  \def \Mut
{ \begin{pmatrix}
1
&
\mb
 \cr 
\m
  &
1 
\end{pmatrix}  
}

  \def \Mu
{ \begin{pmatrix}
1
&
\m
 \cr 
\mb
  &
1 
\end{pmatrix}  
}

\date{
$ $
\today
}

\section{Introduction}

This paper  is a contribution   to  fill   part of a   yet unfinished chapter  in the BRST construction of Lorentzian space  quantum field theories, which concerns boundary effects.   Indeed, the standard Lagrangian quantization method for a gauge fixed BRST invariant action generally assumes trivial boundary conditions for the fields, namely that they vanish at infinity.  However, it is known that relaxing these conditions often leads to non trivial physics related to the infrared sector of the theory. This additional information results in an infinite dimensional enhancement of the symmetry group of the theory \cite{bondi,sachs,Strominger_2014,Strominger_BMS_scattering,BMS/CFT,Kapec,Campiglia}, which is related to soft theorems \cite{PhysRev.140.B516,Cachazo} and memory effects \cite{Pasterski:2015zua,BMS_memory,Pasterski}, forming the so-called \textit{infrared triangle} \cite{Strominger_lectures,BMS_soft_graviton,BMS_memory,campiglia2015asymptotic}. These new symmetries,  called asymptotic symmetries, determine field transformations that are called large gauge transformations and represent a physical sub-sector of the full gauge group.  They are   defined as the \textit{allowed} gauge transformations, which  leave the gauge fixing and boundary conditions of the fields invariant, modulo the \textit{trivial} (or \textit{small}) gauge transformations that vanish on the boundaries.  A heuristic and suggestive definition is hence given by  
\begin{equation}
\label{asg}
\textit{Asymptotic\ symmetry\ group} \equiv \frac{\text{Allowed\ gauge\ transformations}}{\text{Trivial\ gauge\ transformations}}.
\end{equation}
These asymptotic symmetries have to be accounted for in a BRST language.  This paper pursues this goal.

The study of the infinite number of global Noether charges associated with asymptotic symmetries is unavoidable in this quest.  For classical gauge theories, they arise through Noether's second theorem \cite{Noether_1971} that shows that they are
supported on corners of the spacetime manifold, i.e. codimension two manifolds.  Computing these charges and exploring their physical effects is a subtle task. In particular, finding integrable, conserved and finite charges is  very tricky  for open systems such as gravity in asymptotically flat spacetimes (see \cite{Compere:2018aar,Ciambelli_lectures} for reviews). Great progresses in this direction have been made in the recent years using the covariant phase space formalism \cite{Crnkovic:1986ex,Lee:1990nz,Wald:1999wa,Barnich:2001jy,Harlow:2019yfa},  which was designed to study the Hamiltonian dynamics of Lagrangian field theories without breaking Poincar\'e covariance.  This formalism provides a systematic definition of symplectic brackets, which lead to an  algebra of Noether charges that projectively represent the asymptotic symmetry algebra \cite{Brown:1986nw,Barnich_Charge_algebra,Campiglia:2020qvc, Compere:2020lrt,Freidel1,Freidel,Freidel:2020xyx,Ciambelli_2022_Embeddings,Donnay:2021wrk,Rignon-Bret:2024gcx}.  Such brackets also generate the action of the asymptotic symmetry group on the phase space of the theory.  The canonical quantization method then formally defines a quantum gauge theory by promoting the classical symplectic brackets  to quantum commutators.  However,  this traditional investigation does not possess the tools to properly fix the gauge, which is a crucial step in the definition of asymptotic symmetries, as seen  in \eqref{asg}.  Therefore,  it leaves open the question of the gauge independence of the underlying results computed in a given choice of gauge.
This paper aims to provide an answer to this question.

Given that   unphysical fields  such as ghosts, antighosts and Lagrange multipliers  are needed   to fix the gauge and quantize  gauge    theories in a  relativistic invariant way,  and since Noether's second theorem is not valid anymore after adding a  BRST invariant gauge fixing term to the gauge invariant classical Lagrangian, an extension of the classical approach was waiting to be done. 
As it is well known, the use of such extra unphysical fields   implies replacing 
  the concept of   infinitesimal local  field transformations   with that  of an  associated nilpotent  global   BRST transformation.  Then,  the principle of 
postulating the use of gauge fixed and BRST invariant Lagrangians systematically ensures explicit and consistent proofs  
of  gauge independence and unitarity.  Physical observables are also well defined from the cohomology of the BRST operator. 
This ``principle of BRST symmetry" \cite{Baulieu:1981sb,Alvarez-Gaume:1981klj} may  be  considered    as a  fundamental tool for quantizing any given gauge theory.

Thus, this  paper addresses  the question of 
   conserved charges for a gauge symmetry at the quantum and gauge fixed level within the BRST framework.    Their  definitions and   properties  are based  on  the  Noether current   $\star J_{\rm BRST}$ associated with the global BRST symmetry.  To mathematically define this BRST Noether current  for a gauge fixed Lagrangian and to prove that the classical corner Noether charges are recovered from this current, one must enlarge the classical covariant phase space (denoted as CPS from now on) and generalize its bigrading to a trigrading  in order to account for the ghost number and the existence of the BRST symmetry.   The classical fields $\{ \vp_{\rm cl} \}$ are to    be combined with    geometrical ghosts and doublets  made of trivial pairs   of antighosts and Lagrange multiplier fields, 
providing the enlarged  BRST multiplet    $\{ \vp_{\rm cl} \} \to \{ \vp_{\rm brst} \} \equiv \{ \vp_{\rm cl}, c , \bar{c} , b  \}$. 
This  paper  thus generalizes  the    classical   CPS   into    its   BRST version,   which is 
    based on  a trigraded    manifold  with grading coming from the spacetime form degree,  the field space form degree and the ghost number.  
    
 
In this new framework, one can study the impact of the $s$-exact gauge fixing term of the Lagrangian in determining  the associated BRST Noether current.  Given that Noether's second theorem does not hold for this current, and having understood the importance of the corner Noether charges for asymptotic symmetries, one might think that the gauge fixed theory has lost some precious information about the asymptotic symmetry group.  This paper solves this problem by showing that, at least in the relevant examples of Yang--Mills theory in covariant and axial gauges and of gravity in the Bondi--Beltrami gauge,\footnote{The cases of the Chern--Simons theory,  the topological Yang--Mills theory as well as the generalization of these results for generic BV systems involving the necessity of a BV gauge fixing of all antifields, in particular in the case of supergravity, will be published in a different work.} the BRST Noether current $\J$ reduces on-shell to an $s$-exact term reminiscent of the gauge fixing plus a corner term which is exactly the classical  Noether charge.  Even if a renormalization of the phase space is necessary to get integrable and finite charges, it is shown that it can still be done in the gauge fixed case precisely because of this $s$-exact term. In fact, the fully gauge fixed fundamental canonical relation is worked out and shown to lead to the same integrable charges and symplectic bracket structure as in the classical case.


The on-shell conservation of the Noether current $\J$ implies off-shell Ward identities for scattering amplitudes.
  In fact,   it will be shown that the Ward identity for the BRST symmetry of any given gauge theory with non trivial boundary conditions can be decomposed into the standard BRST master equation in the  bulk, which defines, order by order in perturbation theory, the BRST transformations of the  fields  as well as  the expression of the BRST invariant action,  and into an additional boundary Ward identity, which ensures the conservation of  the Noether current $\J$.  This boundary Lagrangian Ward identity justifies the Hamiltonian  commutation relation $[Q,\mathcal{S}]=0$ between the physical $\mathcal{S}$-matrix  and the global Noether charge $Q$ operator, which expresses the invariance of any scattering process under asymptotic symmetries \cite{Strominger_BMS_scattering}. It also makes the link between asymptotic symmetries and soft theorems \cite{Strominger_lectures}.  Most importantly, this new derivation explicitly verifies the gauge independence  of the Noether charges $Q$ entering this identity.\footnote{One may note that the  presented  BRST analysis confirms the   \cite{Rejzner:2020} expectation  that  a boundary BV identity may directly imply Weinberg's soft theorems. More work is needed to compare the methods used in \cite{Rejzner:2020} and our standard BRST methodology.}


 An other major advantage of this construction is that one can now use the BRST technology to study (possibly harmless) anomalies of the boundary Ward identity. 
   These are anomalies for the large gauge symmetry.  They are obtained as  solutions of a codimension one Wess--Zumino consistency condition. 
   For this reason, this paper calls them holographic anomalies. 
   One can suggestively compare and possibly match them with anomalies for the dual boundary theory,  in the same way as central charges are matched in AdS3/CFT2 \cite{Ryu:2006bv}.  Relating these potentially anomalous boundary Ward identities with soft theorems will also 
shed  new light on the all-loop behavior of soft theorems.  This  will be studied in details for the leading and subleading soft graviton theorem \cite{PhysRev.140.B516,Cachazo}.

The  paper is organized as follows.   Section [\ref{Section_trigraded}] extends the bigraded structure of the covariant phase space  to a trigraded one to incorporate the BRST symmetry.   A conjectured BRST Noether 1.5th theorem for  BRST invariant gauge fixed theories is exposed.   Moreover,  a fundamental canonical relation for the gauge fixed and non degenerate symplectic two form is derived.  A second conjecture is made, to be verified case  by case, in order for 
this canonical relation  to be equivalent with the classical one on-shell.
\\
\indent Section [\ref{Section_WI}] derives the Ward identities for small and large gauge transformations.  It also shows that, if Noether's 1.5th  is valid, the Ward identity for large gauge transformations implies the formal Hamiltonian relation $[Q, \mathcal{S}]=0$. Consistent anomalies for this identity are studied. 
 \\
\indent As a first application,   Section [\ref{Section_Yang_Mills}] applies the BRST CPS framework to Yang--Mills theory  in various gauges.  The two conjectures are explicitly checked in these cases. 
\\
\indent Section [\ref{Section_gravity}] considers  the case of first order gravity in the Bondi--Beltrami gauge.  
Both conjectures are again verified  in this technically  different example.  The boundary Ward identities for large gauge transformations, which in this case are extended BMS transformations,  are related to the soft graviton theorem.  The 1-loop correction of the subleading soft graviton theorem is then studied in this BRST context. 
\\
\indent Appendix [\ref{Annexe_A}] contains a compact derivation of Noether's second theorem.  It points out which assumption fails to hold for the residual global BRST symmetry of a gauge fixed action. 
\\
\indent Appendix [\ref{Annex_anomaly}]  comments on the extension of the classical CPS anomaly operator in the case of the BRST CPS.  It shows that this extension cannot be used to generically derive the off-shell fundamental canonical relation of a gauge fixed theory.

\section{The Trigraded BRST Covariant Phase Space}
\label{Section_trigraded}

This section is to clarify  the properties of the   enlarged  BRST  CPS that   
  generalizes  the  geometry of the classical bigraded CPS  with   a  trigraded structure  involving the BRST symmetry.  The purpose is to unify the various   operations that one must perform in   BRST invariant quantum field theories such as doing arbitrary field transformations, BRST transformations,  contractions of field space forms along ghost graded field space vector fields, etc.  The rather formal  mathematical discussion of this section  
    appears to be  useful for a   deeper geometrical visualization of the   BRST symmetry  and  the definition of the global charges of classical gauge theories at the perturbative quantum level.

\subsection{Classical bigraded covariant phase space formalism}

Let us start with a  review of the geometry of the classical covariant phase space and its relevance to gauge theories in view of identifying its limitations and eventually extending it in the next subsection. Here the term ``classical"  means both that no quantization procedure has yet been performed  and that the field space contains only classical fields $\vp_{\rm cl}$,  and no ghosts, antighosts and Lagrange multipliers.  

Part of the logic of the CPS is that a field theory arbitrarily given variation of  $\vp_{\rm cl}$, denoted as $\delta \vp_{\rm cl}$, is to be interpreted as a field space 1-form.  The quantity $\delta \vp_{\rm cl}$ is thus an    infinite dimensional generalization  of the definition of $dx^\mu $ as the anticommuting differential of
 $x^\mu$,  with     $dx^\mu\equiv d(x^\mu)$, so that  $d^2=   (dx^\mu\pa_\mu)^2=0$. One needs a better description of physicist fields $\vp^A = \vp_{\rm cl}^A$ to rigorously implement this idea.
To do so, one often 
 considers the  vector bundle $E$ with local coordinates $(x^\mu,\vp^A)$.  For instance in Yang--Mills,  the classical field is just the gauge field $\vp^A = A$  with all its gauge algebra and spacetime indices.  One  can  actually make more precise the physicist  interpretation and  define the fields as elements of $\mathcal{F} \equiv \Gamma(E,M)$, which are sections of the  jet space $E$, with projection $\pi : E \rightarrow M$ over the spacetime manifold $M$. 
  Locally, the fibers are defined as $F = \pi^{-1} (M)$ and one has the isomorphism $E \cong M \times F$. This allows one  to specify the  local coordinates on $E$. Fields $\{\Phi : M \rightarrow E \} \in \mathcal{F}$ can thus be seen as a map $\Phi : x \mapsto (x,\vp^A(x))$.  
  
  To be able to treat fields and their derivatives as independent variables in variational calculus, one needs to work in
  the  jet bundle $JE$ that generalizes $E$ by extending each  point $\vp^A$ into a  ``jet" made of  $\vp^A$ and all its spacetime derivatives.
The extension of $E$ into  the jet bundle $JE \cong M \times F^\infty$ is therefore such that the local coordinates on $F^\infty$ are now given by $\vp^A, \vp^A_{(\mu_1)}, \vp^A_{(\mu_1 \mu_2)}$, etc..., and such that a section $\Phi \in \mathcal{F}$ induces a section of $JE$ through 
\begin{equation}
\left. \vp^A_{(\mu_1 ... \mu_n)} \right\vert_\Phi = \frac{\pa^n \vp^A(x)}{\pa x^{\mu_1}...\pa x^{\mu_n}} .
\end{equation}
To evaluate fields on a jet, one then defines 
\begin{align}
j^\infty : \mathcal{F} \times M &\rightarrow JE
\nn \\
 (\Phi, x ) &\mapsto j^\infty \Phi(x) = (x,\vp^A(x), \left. \vp^A_{(\mu_1)} \right\vert_\Phi, \left. \vp^A_{(\mu_1 \mu_2)} \right\vert_\Phi,...) .
\end{align}
This leads  to the dual map $(j^\infty)^* : \O(JE) \rightarrow \O(\mathcal{F} \times M)$ and to the definition of forms on $JE$, with a bigrading inherited from the splitting between horizontal and vertical forms of $JE$. The corresponding bicomplex is $(\O^{\bullet, \bullet}(JE),d_V,d_H)$.

Following Anderson \cite{Anderson1992IntroductionTT} and  e.g.  \cite{Mnev:2019ejh} one can define \textit{local forms} that are spacetime and field space differential forms on $\mathcal{F} \times M$ as elements of $\O^{\bullet,\bullet}_{\rm loc}(\mathcal{F} \times M)$ endowed with the differentials $\delta$ and $d$ by the bicomplex 
\begin{equation}
(\O^{\bullet,\bullet}_{\rm loc}(\mathcal{F} \times M),\delta,d) = (j^\infty)^* (\O^{\bullet, \bullet}(JE),d_V,d_H) 
\end{equation}
with 
\begin{equation}
\delta (j^\infty)^* \alpha = (j^\infty)^* d_V \alpha, \quad d (j^\infty)^* \alpha = (j^\infty)^* d_H \alpha
\end{equation}
for $\alpha \in \O^{\bullet, \bullet}(JE) $.  With these definitions, $d$ raises the spacetime $M$ form degree by one unit and $\delta$ the field space $\mathcal{F}$ form degree by one unit. The statistics of $\delta \vp^A$ is thus opposed to that of $\vp^A$. One also has the graded property 
\begin{equation}
(d + \delta)^2=0 \quad \iff \quad d^2=0, \quad d\delta+\delta d=0,      \quad \delta^2=0  .
\end{equation}
Field space vector fields $U$ can be  defined as elements of the tangent bundle $\mathfrak{X}(\mathcal{F}) \equiv {\O_{\rm loc}^{1,0}}^*$, which is  dual to the bundle made of field space one forms  ${\O_{\rm loc}^{1,0}}$. To be more specific,  one  uses the dual basis to that of  the field space $1$-forms 
$\delta \vp^A$  so that 
any given field space vector field $U$  can be written  as 
\begin{equation}\label{fgfv}
U[\vp] = \int_M d x \ U^A( \vp(x)) \frac{\delta}{\delta \vp^A(x)}  \  .
\end{equation} 
The definition \eqref{fgfv} is    the      infinite dimensional    generalization of the spacetime vector field formula
$v(x) =   v^\mu(x) \pa_\mu$. 
 The  graded interior product in field space $I_U$ then  satisfies  
\begin{align}
\label{field_space_contraction}
I_U \vp^A &= 0 ,
\nn \\
I_U \delta \vp^A &= U(\vp^A) .
\end{align}
Eq.\eqref{field_space_contraction} generalizes the definition of the finite dimensional contraction operators $i_v$ along spacetime vector fields $v$, that is   
$i_v f(x) = 0$ and $i_v dx^\mu = v^\mu (x)$.  


All this strongly suggests  to define the graded  Lie derivative in field space as
\begin{equation} 
L_U \equiv [I_U, \delta].
\end{equation}
$L_U$  consistently applies to any local form in  $\O^{\bullet,\bullet}_{\rm loc}(\mathcal{F} \times M)$.

 In fact,  once graded  field space forms and field space vectors have been precisely defined,  one  can  define  a Cartan calculus on $\mathcal{F}\times M$. Table [\ref{table_1}]   summarizes its graded rules. 
All  commutators $[\bullet,\bullet ]$ appearing in this table are   graded ones.   The    bigrading of any entity    is  the   sum of the  field  space and spacetime form degrees. The grading   effect is defined modulo two.  

Notice that,  in Table [\ref{table_1}],   there are no different vector fields
 $U_1,U_2$ or $v_1,v_2$ for formulas where two of them appear because they will eventually become objects with an arbitrary grading coming from ghost number.\footnote{\label{footnote1}For commuting $U$'s, one has automatically $\{U,U \}=0$ so this trivial  identity carries no information about the algebraic structure of the transformations possibly represented by the vector $U$. 
To capture that information, one must rather consider two different vector fields $U_1,U_2$ such that $\{U_1,U_2 \}\neq0$.  However, it is not necessary if they have ghost number one since in this case,  $\{U,U\}   \neq 0$.}  For the moment,  $I_U$ and $i_v$ are contraction operators that lower the degrees of field space forms  and 
      spacetime forms by one unit respectively.   
\begin{table}[ht!] 
\centering
\begin{small}
\begin{tabular}{@{}lrrrr@{}}\toprule 
 &  \textbf{Field space} $\mathcal{F}$&  \textbf{Spacetime} $M$   \\ \midrule
 \textbf{Definitions}  &  &  \\ \midrule
\textit{Exterior derivative} &  $\delta$  & $d$   \\ \hdashline
\textit{Vector fields} & $U \in \mathfrak{X}(\mathcal{F}) \equiv {\O_{\rm loc}^{1,0}}^*$  & $v \in \mathfrak{X}(M) \equiv {\O_{\rm loc}^{0,1}}^*$  \\ \hdashline
\textit{Interior product} &  $I_U$  & $i_v$    \\ \hdashline
\textit{Lie derivative} &  $L_U = [I_U,\delta]$ & $\Lie_v = [i_v,d]$  \\ \hdashline
\textit{Vector field bracket} &  $ \{ U,U \} \equiv L_U U$ & $ \{v, v \} \equiv \Lie_v v $     \\ \midrule
\textbf{Properties}  &  &  \\ \midrule
\textit{ } &  $\delta^2=0$  & $d^2=0$                  \\ \hdashline
\textit{ } &  $[\delta,d]=0$  &   $[I_U,i_v]=0$     \\ \hdashline
\textit{ } &  $[\delta,L_U]=0$  &   $[d, \Lie_v]=0$     \\ \hdashline
\textit{ } &  $[ I_U, I_U ] = 0$  & $[i_v, i_v] = 0$                  \\ \hdashline
\textit{ } &  $[ L_U, L_U ] = L_{\{ U,U \}} $  & $[\Lie_v, \Lie_v] = \Lie_{\{ v, v \}}$                  \\ \hdashline
\textit{ } &  $[ L_U, I_U ] = I_{\{ U,U \}} $  & $[\Lie_v, i_v] = i_{\{ v, v \}}$                  \\ \hdashline
\textit{ } &  $[ L_U, \Lie_v ] = \Lie_{ L_U v  } $  & $[L_U, i_v] = i_{L_U v }$                  \\ \hdashline
\textit{ } &  $[\delta, \Lie_v ] = \Lie_{ \delta v  } $  & $[\delta, i_v] = i_{\delta v }$    \\ \hdashline
\textit{ } &  $[d,L_U]=0$  &   $[d,I_U]=0$    
 \\  \bottomrule
\end{tabular}
\end{small}
\caption{\textit{Cartan calculus on $\mathcal{F}\times M$ } \label{table_1} }
\end{table}

The basic constituents of any given classical field theory on spacetime $M$ of dimension $d$ are then defined as the pair $(\mathcal{F}, S)$, namely  the field space of the theory and  its action $S[\vp]$, defined from a local Lagrangian density according to
\begin{equation}
\label{lagrangian}
S \equiv \int_M \left.L[\vp^A]\right\vert_{\Phi}.
\end{equation}
The Lagrangian density $\left.L[\vp^A]\right\vert_{\Phi} \in \O_{\rm loc}^{0,d}$ is  a spacetime top form.  For the sake of simplicity, one may
  drop the  symbol $\Big|_{\Phi}$ for ``evaluation at a section $\Phi$", which simply means  that 
  $\int_{M} \left.L[\vp^A]\right\vert_{\Phi}  \equiv  \int _M  
  L(\vp^A(x))$.
The equations of motion associated with the action \eqref{lagrangian} are defined by the following key identity   of the covariant phase space formalism, that is
\begin{equation}
\label{L_variation}
\delta L (\vp) = E^A_\vp ( \vp ) \delta \vp^A + d \theta (\vp , \delta \vp) .
\end{equation}
Here the ${E^A_\vp}$'s are the Euler--Lagrange equations of the field $\vp^A$ and $\theta \in \O^{1,d-1}_{\rm loc}$ is the local presymplectic potential.  There is an implicit summation over all the fields $\vp^A$ in $E \equiv E^A_\vp   \delta \vp^A \in \O^{1,d}_{\rm loc}$. Very importantly,   $\theta$ is uniquely determined by the choice of Lagrangian \cite{Freidel:2020xyx}. In fact, the apparent ambiguity that $\theta$ could be defined up to a $d$-exact term $\theta \sim \theta + d\nu$ is resolved by writing \eqref{L_variation} for two Lagrangians $L,L'$ that differ by a boundary term: $L'-L=dl$.  As these two Lagrangians have the same equations of motion, their respective local presymplectic potentials differ by $\theta'- \theta = -  \delta l + d\nu$. So $\nu$ can simply be seen as the local presymplectic potential for the boundary Lagrangian $l$.


Out of the local presymplectic potential, one can  construct the local presymplectic two form 
\begin{equation}
\o \equiv \delta \theta \in \O^{2,d-1}_{\rm loc} .
\end{equation}
Global versions of these objects can also be considered  by integration  over a $(d-1)$-dimensional hypersurface $\Sigma$.  It provides the presymplectic potential 
\begin{equation}
\Theta = \int_\Sigma \theta \in \O^{1,0}_{\rm loc}
\end{equation}
and the presymplectic two form
\begin{equation}
\label{symp_two_form}
\O = \delta \Theta = \int_\Sigma \o  \in \O^{2,0}_{\rm loc} . 
\end{equation}
It is interesting to mention that these definitions are not invariant under the choice of codimension one hypersurfaces $\Sigma$ \cite{Harlow:2019yfa}.  When going from one Cauchy hypersurface to another, the evolution of $\O$ is basically measured by the symplectic flux $\mathcal{F}^\theta_\xi \equiv \int_S i_\xi \theta$ supported on the boundaries of the Cauchy hypersurfaces \cite{Freidel}.  The $\Sigma$ in \eqref{symp_two_form} can be spacelike, timelike or null and the full boundary of any compact sub-region of spacetime $M$ is actually a union of such hypersurfaces.   

The field space 2-form $\O$  is called presymplectic  rather than   symplectic   because   it can  be degenerate.
Indeed, in some cases one finds that  a  field space vector field   $V \neq 0$ may exist such that $I_V \O = 0$.  Such $V$'s can typically be the one associated with the invariance of a theory under a local symmetry (internal gauge symmetry or spacetime symmetries).  The representation of this invariance under infinitesimal transformations    $\vp_{\rm cl}(x) 
\to \vp_{\rm cl}(x) + \delta_\lambda \vp_{\rm cl}(x)$  
with the   infinitesimal local parameter   $\lambda(x)$ by a field space  vector field $V_\lambda$ is given by 
\begin{equation}
\label{V_infinitesimal}
V_\lambda = \int_M dx \ (\delta _\lambda \vp^A )(x) \frac{\delta}{\delta \vp^A(x)}.
\end{equation}
This definition indeed provides the expected transformation law as 
\begin{equation}
\label{infinitesimal_transformation}
\delta_\lambda \vp^B = \frac{\delta \vp^B}{\delta \vp^A} \delta_\lambda \vp^A  = V_\lambda (\vp^B) = I_{V_\lambda} \delta \vp^B
= L_{V_\lambda}  \vp^B ,
\end{equation}
where the properties \eqref{field_space_contraction} were used.

The relevance of $\theta$ and $\O$ is understood by constructing the Noether current associated with the invariance of the Lagrangian under \eqref{infinitesimal_transformation}. 
In order for the equations of motion \eqref{L_variation} to be invariant under this symmetry transformation, the Lagrangian \eqref{lagrangian} must at least be semi-invariant under this symmetry, namely 
\begin{equation}
\label{semi_sym}
L_{V_\lambda} L = I_{V_\lambda} \delta L = d B_\lambda ,
\end{equation}
where $B_\lambda \in \O^{0,d-1}_{\rm loc}$.  One can then build an on-shell conserved current by contracting the identity \eqref{L_variation} with $I_{V_\lambda}$, leading to 
\begin{equation}
\label{construction_noether}
I_{V_\lambda} \delta L = I_{V_\lambda} E + d I_{V_\lambda} \theta \ \hat{=} \ d I_{V_\lambda} \theta  \quad \Longrightarrow \quad d ( I_{V_\lambda} \theta - B_\lambda ) \ \hat{=} \ 0 . 
\end{equation}
Here and elsewhere, the notation $\hat{=}$ means that the equality is only valid on-shell, namely when $E=0$.
Finally,   noticing that 
\begin{equation}
\label{def_Noether_constraints1}
I_{V_\lambda} E = I_{V_\lambda} (E^A_\vp \delta \vp^A) = d ( B_\lambda - I_{V_\lambda} \theta ) \equiv d C_\lambda \quad \Longrightarrow \quad I_{V_\lambda} \theta - B_\lambda \equiv - C_\lambda + d q_\lambda
\end{equation}
by using  \eqref{semi_sym}, \eqref{construction_noether} and the algebraic Poincar\'e lemma,  one gets\footnote{The $\star$ operation is the Hodge dual in spacetime so that in four spacetime dimensions, $\star J_\lambda$ is a spacetime $3$-form and $q_\lambda$ a $2$-form.} 
\begin{equation}
\label{classical_Noether_current}
\star J_\lambda = I_{V_\lambda} \theta - B_\lambda = - C_\lambda + d q_\lambda \quad \text{with}  \quad d ( \star J_\lambda ) \ \hat{=} \ 0 .
\end{equation}
This is Noether's first theorem, where $J_\lambda$ is the conserved Noether current\footnote{For a Lagrangian that depends on the fields and their first order derivatives only,  the local presymplectic potential is given by $\theta^\mu  =    \frac { \delta L   }{  \delta ( \pa_\mu \vp^A ) } \delta \vp^A$. Therefore,   the Noether current \eqref{classical_Noether_current} is equivalent to the well known form  $J_{\lambda}^\mu \equiv \frac { \delta L  }{  \delta (\pa_\mu \vp^A)}  \delta_\lambda \vp^A - B^\mu_\lambda$.} associated with the local symmetry parametrized by $\lambda(x)$ and the $q_\lambda$'s are the associated Noether charges.  The $C_\lambda$'s are called the Noether constraints.  Noether's second theorem is then given by 
\begin{equation}
\label{Noether_2}
C_\lambda \ \hat{=} \ 0 \quad \Longrightarrow \quad \star J_\lambda \ \hat{=} \   d q_\lambda ,
\end{equation}
which means that the Noether current associated with a local symmetry reduces to a pure corner (codimension 2) term on-shell.  For a proof of this, see Noether's original paper \cite{Noether_1971}, more recent discussions \cite{Karatas:1989mt,Avery_Schwab,Miller:2021hty}, or Appendix [\ref{Annexe_A}].  

The classical global Noether charges 
\begin{equation}
\label{global_noether_charges}
Q_\lambda \equiv \int_{\pa \Sigma} q_\lambda
\end{equation}
then serve as a probe for asymptotic symmetries.  Indeed, it was shown in  \cite{Lee:1990nz} that $Q_\lambda$ vanishes for any $\lambda$ associated with trivial gauge transformations.\footnote{For this statement to be non trivial, one needs to consider hypersurfaces with corners $S \equiv \pa \Sigma \neq \emptyset$ and trivial gauge transformations that do not necessarily vanish at infinity.}  Large gauge transformations or asymptotic symmetry transformations are then indexed by the $\lambda$'s such that $Q_\lambda \neq 0$.  This is understood by studying  the fundamental canonical relation associated with the presymplectic two form $\O$.  As a matter of fact,   \textit{integrable} (or \textit{Hamiltonian}) global Noether charges $Q_\lambda$ are such that
the fundamental canonical relation  takes the form 
 \begin{equation}
 \label{hamiltonian_charges}
 - I_{V_\lambda} \O \ \hat{=} \ \delta Q_\lambda .
 \end{equation}
For such charges, the Poisson bracket  is constructed from  
 \begin{equation}
 \label{poisson_brackets}
 \{ Q_{\lambda_1} , Q_{\lambda_2}  \} \equiv L_{V_{\lambda_2}} Q_{\lambda_1} = I_{V_{\lambda_2}} \delta Q_{\lambda_1} \ \hat{=} \ - I_{V_{\lambda_2}} I_{V_{\lambda_1}} \O = I_{V_{\lambda_1}} I_{V_{\lambda_2}} \O ,
 \end{equation}
 and then the action of the large gauge symmetry  on phase space is  canonically generated by
 \begin{equation}
  \{ Q_{\lambda} ,  \vp_{\rm cl}  \} = \delta_\lambda \vp_{\rm cl} . 
 \end{equation}
 Therefore, any  charges $Q_\lambda \neq 0$  have a physical non trivial  action on the whole phase space.

 However,   when one deals with open systems with dissipation at the boundary and/or field dependent symmetry parameters $\lambda$ (such that $\delta \lambda \neq 0$),  the relation \eqref{hamiltonian_charges} is not valid anymore.  Instead, it takes the more general form    
 \begin{equation}
 \label{fund_with_fluxes}
 - I_{V_\lambda} \O \ \hat{=} \ \delta Q_\lambda - \mathcal{F}_\lambda . 
 \end{equation}
 The explicit expression of the flux $\mathcal{F}_\lambda$ can be worked out (it is done in Appendix [\ref{Annex_anomaly}]) and turns out to be supported on corners as well \cite{Freidel}.  This means that  one cannot use the simple brackets \eqref{poisson_brackets} in this case. There are essentially two ways out \cite{Ciambelli_lectures}.  One consists in defining new brackets that take care of the split \eqref{fund_with_fluxes} and faithfully\footnote{Up to some field dependent central extensions. } represent the symmetry algebra, see for example \cite{Barnich_Charge_algebra,Freidel}. The other one is to enlarge the field space with additional new fields, called edge modes \cite{Freidel:2020xyx,Donnelly:2016auv,Freidel:2020svx,Freidel:2020ayo,Speranza:2017gxd,Ball:2024hqe},  such that the relation \eqref{hamiltonian_charges} is valid in this enlarged space and the Poisson brackets \eqref{poisson_brackets} can be used \cite{Ciambelli_2022_Embeddings,freidel2021canonical,Ciambelli:2021vnn}. Both solutions crucially rely on the explicit form of $\mathcal{F}_\lambda$ and the fact that it is a corner quantity.  In both cases as well, quantizing the phase space is formally done by promoting the defined brackets to quantum commutators.  It is therefore relevant to go ahead and check   whether or not \eqref{fund_with_fluxes} holds in the BRST invariant gauge fixed theory, namely when $\O$ stands for the symplectic 2-form of  a gauge fixed and    BRST invariant   action.

In the case of \eqref{hamiltonian_charges},  the degeneracy of the presymplectic two form comes from the  trivial transformations $V_\lambda$'s such that $I_{V_\lambda} \O = 0$. The latter  are true redundancies of the theory in the sense that their corresponding Hamiltonian charge is zero and therefore their action on phase space is trivial.  This point was noted in the Hamiltonian formalism where the gauge symmetry is expressed by the existence of constrains, meaning that one of the basic consequences of a Lagrangian gauge symmetry is that some classical field components have no canonically conjugate momenta.
It turns out that  in this case, as first noticed by Dirac, and much before the delivering introduction of Faddeev--Popov ghosts,  using  appropriate definitions of the Poisson brackets for defining the quantization is an hard task, 
in particular for determining conserved charges and consistently checking their commutation relations.
This task actually consists in transforming the 
presymplectic 2-form $\Omega$ into an invertible symplectic one  by restricting the field space $\mathcal{F}$ to the quotient 
\begin{equation}
\label{quotient}
\hat{\mathcal{F}} \equiv \mathcal{F} \big/ \mathcal{G} .
\end{equation}
Here $\mathcal{G}$ is the group of trivial transformations.  The non degenerate symplectic 2-form $\hat{\O}$ is then such that $\O$ is the pullback of $\hat{\O}$ from $\hat{\mathcal{F}}$ to $\mathcal{F}$.  All remaining $V_\lambda$'s of the form \eqref{V_infinitesimal} that ``survive" the quotient procedure are now vector field on $\hat{\mathcal{F}}$ generating the transformations under the asymptotic symmetry group as defined in \eqref{asg}.\footnote{This story is not impacted by the potentially non vanishing flux of \eqref{fund_with_fluxes}.}

In fact,  all these complications disappear when one uses the BRST quantization method for computing the  global charges of asymptotic symmetries.  Indeed,   the 
BRST CPS  is a non trivial extension  of the classical CPS  by ghost degrees of freedom that allow  one to compensate for the contributions of the unphysical modes of a gauge theory.  One therefore obtains a non degenerate BRST symplectic  2-form in the enlarged field space $\tilde{\mathcal{F}}$, which contains the classical fields, ghosts, antighosts and Lagrange multipliers,  by construction.  Information of the type \eqref{quotient} is then given in the space $\tilde{\mathcal{F}}$ by the cohomology at ghost number zero of the BRST operator $H^0(s) \equiv \text{Ker}(s) \big/ \text{Im}(s)$.

\subsection{BRST symmetry: a third grading}
\label{Section_adding_ghosts}

When   perturbatively quantizing  gauge   theories one must extend  the classical field
     $\vp_{\rm cl}$     into its BRST completion 
 $ \vp_{\rm cl} \to \vp^A \equiv( \vp_{\rm cl}, c, \bar c, b)$,  where $c$ stands for all geometrical ghost fields  associated with the   classical
  parameters of  the gauge symmetry;  and the antighost   $\bar c $ and Lagrange multiplier fields $b$    stand for trivial BRST doublets,  allowing one to  provide gauge fixing term with a  trivial BRST invariance. The pair   $(\bar c, b)$   is called  a trivial one because  it undergoes  the action of the BRST operator 
  $s$  as $s\bar c= b$ and $sb=0$.  
The doublet $(\vp_{\rm cl},c)$ is non trivial. 
Indeed, the  $s$  transformation  of   $\vp_{\rm cl}$ 
  is    a 
  gauge transformation  where one replaces the  infinitesimal parameter $\lambda(x)$ by 
 the ghost $c(x) $ and the action of $s$ on $c$ is such that one has 
 $s^2=0$  both on $\vp_{\rm cl}$ and $c$. 
 The property $s^2=0$ is independent  of the choice of   gauge function. The geometrical significance of the gauge symmetry is thus encoded in the non trivial 
 part of the BRST symmetry algebra only.

This section thus  completes the information carried by Table [\ref{table_1}] when one generalizes the classical bigraded CPS to the trigraded BRST CPS, where the third grading is given by the conserved ghost number and where the field space of the classical fields $\mathcal{F}$ is enlarged into that of all BRST fields $\tilde{\mathcal{F}}$.

One also replaces  the  vector field $V_\lambda$ by the  BRST vector field $\V$, which generates the action of 
the    global BRST symmetry that is relevant when  one considers   a  gauge fixed Lagrangian $L_{\rm GF}$,   for any given consistent choice of  gauge fixing term, with   its complete ghost, antighost and Lagrange multiplier field dependence. 
The  BRST Cartan calculus rules are to provide  all needed 
technicalities to compute the BRST Noether current associated with the Lagrangian $L_{\rm GF}$.


Consider therefore a field theory $(\tilde{\mathcal{F}}, S)$ with a nilpotent BRST operator $s$.  This differential operator acts on all fields $\vp^A = (\vp_{\rm cl}, c,\bar{c},b)$ with the defining properties 
\begin{align}
\label{defining_s}
s^2 = 0 ,
\quad \quad 
[s,d] = sd + ds = 0 ,
\quad  \quad 
s x^\mu = 0, \quad  \quad  [s,\pa_\mu] = 0.
\end{align}
For instance in Yang--Mills theory,  the fields  are   $\vp^A = ( A,c,\bar{c},b  )$ with all their gauge algebra and spacetime indices.  One can add as many trivial pairs $(\bar{c},b)$ as one wishes,  depending on the chosen gauge.

The action of the BRST operator $s$ on any local forms in $\O^{\bullet,\bullet}_{\rm loc}(\tilde{\mathcal{F}} \times M)$ is  not entirely defined by the properties \eqref{defining_s}. In fact, \eqref{defining_s} only defines its action on  $\O^{0,\bullet}_{\rm loc}(\tilde{\mathcal{F}} \times M)$.  However, 
 the  consistent CPS framework developed in the previous subsection  implies adopting  the same  grading convention  for $\delta$ as for $s$.  Indeed, the $s$ transformation is nothing 
but one among all possible $\delta$  transformations.  
So to extend the definition of the action of $s$ on arbitrary local forms,  one postulates that  
$s$  must square to zero and anticommute with  $d$ and $\delta$, namely 
\begin{equation}
\label{nil3}
(\delta + d +s)^2 = 0
\end{equation}
so that the grading coming from  ghost number is consistently incorporated in $\O_{\rm loc}^{q,p,g}(\tilde{\mathcal{F}} \times~M)$. Therefore, one can define 
an enlarged Cartan 
structure with three gradings;  the 
field space form degree $q$,  the spacetime form degree $p$  and the ghost number $g$.  

Any local form $X[\vp]$  in $\O_{\rm loc}^{q,p,g}(\tilde{\mathcal{F}} \times M)$ 
has a  total grading
\begin{equation}
\label{grading_1}
\mathfrak{g}(X) = q + p + g 
\end{equation}   
and can thus be  denoted as 
$X_p^{g,q}$.  The statistics of such a local form is given by $(q + p + g)$ $\mod$~2.  The same occurs for graded operators that act on local forms.  Indeed, 
a transformation  by  either $d$, $\delta$  or  $s$ of $X_p^{g,q}$ provides a local form with total degree 
$q+p+g+1$ where  either $p$, $q$  or  $g$
 has been increased by one unit, respectively.  
For any $X,Y$ in $\O_{\rm loc}^{q,p,g}(\tilde{\mathcal{F}} \times M)$,  the graded commutator between such local forms is defined as\footnote{The product $XY$ has to be thought of as an exterior product in $\O_{\rm loc}^{q,p,g}(\tilde{\mathcal{F}} \times M)$.} 
\begin{equation}
\label{graded_commutator1}
[ X ,  Y ] = X Y  - (-1)^{\mathfrak{g}(X) \mathfrak{g}(Y)} Y X 
\end{equation}
and the graded Leibniz rule for any graded derivation $a$ that acts on $\O_{\rm loc}^{q,p,g}(\tilde{\mathcal{F}} \times M)$ as
\begin{equation}
a ( X Y  ) = a(X)  Y + (-1)^{\mathfrak{g}(a) \mathfrak{g}(X)} X  a(Y).
\end{equation}
 
To make the postulate \eqref{nil3} concrete,  one has to identify a differential on $\O^{\bullet,\bullet}_{\rm loc}(\tilde{\mathcal{F}} \times M)$ which is compatible with $d$ and $\delta$.  To fix this issue,  one may define a field space vector field  $V_{\rm BRST}$ with ghost number one as 
\begin{equation}
\label{V_BRST}
 V \equiv V_{\rm BRST} =  \int_M dx \ s \vp^A(x) \frac{\delta}{\delta \vp^A(x)} ,
\end{equation}
where the $\vp^A(x)$'s are now elements of $\tilde{\mathcal{F}}$ and satisfy
\eqref{defining_s}.  From now on, field space vector fields $V$ without any subscript refer to anticommuting field space vector fields with ghost number one defined by \eqref{V_BRST}. This justifies the notations of Table [\ref{table_1}] and the footnote [\ref{footnote1}].  It is interesting to notice that the apparent simplicity of \eqref{V_BRST} hides the whole non trivial determination of the action of $s$ on all fields $\vp^A$.  

One has therefore   the analogous of \eqref{infinitesimal_transformation} for BRST transformations, namely
\begin{equation}
\label{s_0-forms1}
s \vp^B = I_V \delta \vp^B.
\end{equation}
The nilpotency of the BRST operator is equivalent to the vanishing of the field space vector field bracket  
\begin{equation}
\label{VV_commutators}
\{V,V\} \equiv L_V V = I_V \delta V = \int_M dx \ s^2 \vp^A(x) \frac{\delta}{\delta \vp^A(x)} = 0 .
\end{equation}

%

The relation \eqref{VV_commutators} allows one to find a realization of \eqref{nil3} in the space of local forms.  Indeed,  Table [\ref{table_1}] indicates that for $\{ V,V \}=0$, the only candidate for a differential that raises the ghost number by one unit, squares to zero and is compatible with $d$ and $\delta$ is $L_V$.  This means that \eqref{nil3} is consistently satisfied for 
\begin{equation}
\label{definition_s}
s \equiv L_V = [I_V , \delta] = I_V \delta - \delta I_V
\end{equation} 
with $V \equiv V_{\rm BRST}$  given by \eqref{V_BRST}. Note that \eqref{definition_s} is nothing but \eqref{s_0-forms1} when acting on field space zero forms.  It is also interesting to notice that this $V_{\rm BRST}$ coincides with the degree one evolutionary and cohomological field space vector field $Q$ of \cite{Mnev:2019ejh} that defines a \textit{lax BV-BFV theory}.

Introducing the vector field $V_{\rm BRST}$ is thus the key for extending the bicomplex of local forms on $\tilde{\mathcal{F}} \times M$ to 
\begin{equation}
\label{tricomplex}
(\O_{\rm loc}^{\bullet,\bullet,\bullet}(\tilde{\mathcal{F}} \times M),\delta,d,s) 
\end{equation}
and consistently take into account the ghost number.

As $\delta$, $d$ and $s$ are graded differential   operators     with $(d+\delta+s)^2=0$, each one with a total grading equal to 1,  one can safely consider arbitrary exterior products of local forms 
\begin{equation}
\label{field_exterior_product}
\delta \vp^{A_1} \curlywedge .... \curlywedge \delta \vp^{A_n} 
\end{equation}
as elements of $\O_{\rm loc}^{\bullet,\bullet,\bullet}(\tilde{\mathcal{F}} \times M)$,
whose grading is equal to 
\begin{equation}
\label{grading}
n + \sum_{k=1}^n (g_{\vp^{A_k}} +  p_{\vp^{A_k}}) . 
\end{equation}

The definition of a field theory on spacetime $M$ is extended to the pair $(\tilde{F},S)$ where $S$ is still given by \eqref{lagrangian} but with $L$ in $\O_{\rm loc}^{0,d,0}(\tilde{\mathcal{F}} \times M)$.  Moreover, one can now add an $s$-exact gauge fixing term to $L$ and analyze the consequences of the key relation \eqref{L_variation}, which is  still valid for a gauge fixed Lagrangian in this enlarged space.


The cases of primary interest in this paper are those where the BRST symmetry stands for a local internal (say gauge) and/or spacetime (say reparametrization) symmetries. It is therefore important to generalize Table [\ref{table_1}] for \eqref{tricomplex} with the relevant objects appearing in  explicit computations in theories with such invariances. 
Typically, for theories invariant under reparametrization, the action of $s$ can be compared with that of the Lie derivative in spacetime along a specific vector field.  In general relativity for example, where the BRST field space contains the metric $g_{\mu \nu} (x)$ and the reparametrization vector ghost $\xi(x)$,  the nilpotent BRST operation is given by 
\begin{align}
\label{s_RG}
s g_{\mu \nu} &= L_V g_{\mu \nu} = \Lie_\xi g_{\mu \nu} ,
\nn \\
s \xi^\mu &= L_V \xi^\mu = \xi^\nu \pa_\nu \xi^\mu  = \demi \Lie_\xi \xi^\mu = \demi \{\xi,\xi\}^\mu  ,
\end{align}
where the factor $\demi$ is primordial to ensure the nilpotency of $s$.\footnote{The $g_{\mu\nu}$ and $\xi^\mu$  $s$ transformations 
\eqref{s_RG} are  often    obtained   by changing the  infinitesimal reparametrization parameter $\epsilon ^\mu \to \xi^\mu$ 
  into a  $\Diff_0$ transformation 
  $\delta  _\epsilon g_{\mu \nu}  =\Lie_\epsilon g_{\mu \nu}  \to  s g_{\mu \nu} =  \Lie_\xi g_{\mu \nu}  $ and then one computes $ s \xi^\mu  = \xi^\nu \pa_\nu \xi^\mu$ so that   $ s^2=0$. However
\cite{Baulieu:1985gy,Baulieu:1985md}        geometrically     define  the anticommuting   vector    ghost   $\xi^\mu(x)$   through   first order  gravity   horizontality equations for the vielbein and its  torsion,  $\tilde{e}^a = \exp (i_\xi)  e^a \equiv e^a + c^a$ and 
$\tilde{T}^a = \exp (i_\xi)  T^a$. 
 One can then  naturally  introduce  $c^a =e^a_\mu \xi^\mu \in \O_{\rm loc}^{0,0,1}(\tilde{\mathcal{F}} \times M)$. 
  Choosing~either~$\xi^\mu$~or~$c^a$~as the fundamental   ghost field  for  the reparametrization symmetry   is a matter of convenience in the
    quantum field theory  path integral formalism. Whatever the ghost  choice is, one must define  appropriate antighosts  that are compatible with 
the   gauge choice, so that  the ghost antighost contributions in closed loops allows for the gauge invariance and unitarity of the theory.
}  
This motivates the extension of Table [\ref{table_1}] by adding the graded commutation rules of $L_V$,  $\Lie_\xi$ and their building blocks for anticommuting $V$'s and $\xi$'s.  It is still worth keeping the commutation rules for a commuting vector field in spacetime $\phi$. Indeed, in topological gravity, one  generalizes the   definition of the reparametrization ghost as  
\begin{equation}
s\xi  =    \demi   \Lie_\xi \xi  \longrightarrow   s\xi  =      \demi \Lie_\xi \xi +\phi ,
\end{equation}
 where $\phi$ is a  commuting ``ghost of ghost" vector  field  with  ghost number 2 satisfying $\{ \phi , \phi \} = 0$.  The property $s^2=0$  holds    true provided  one defines 
\begin{equation}
s \phi = \Lie_\xi \phi \equiv \{ \xi,\phi\} = - \{ \phi, \xi \}.
\end{equation}
Then, $\phi$ becomes a relevant element of the BRST covariant phase space.   

One may find interesting to compare the action of $s \equiv L_{V_\xi}$ and that of $\Lie_\xi$.  To do so,  one can introduce the contraction $I_{V_{\delta \xi}}$ and define the so-called anomaly operator $\Delta_\xi \equiv   s - \Lie_{\xi} + I_{V_{\delta \xi}}$ \cite{Freidel,Hopfmuller:2018fni,Chandrasekaran:2020wwn,Odak:2022ndm}.  Such an operator   provides a relevant commuting field space vector field $U=V_{\delta \xi}$ valued in the field space 1-forms.  
Defining the action of $I_{V_{\delta \xi}}$ on a field space 1-form with ghost number zero  is an easy task by doing a BRST transformation of this field and replacing $\xi$ by $\delta \xi$ (as in the classical case).     It is however not clear how this generalizes for transformations of  the non classical fields   (ghosts,  antighosts and Lagrange multiplier fields)  since the form of the  BRST transformations of the latter are of course completely  different to that of a gauge transformation.
This issue is addressed in Appendix [\ref{Annex_anomaly}], where it is proven that the classical formulas in which the anomaly operator and $V_{\delta \xi}$ are useful  are no longer valid when ghost and antighost fields are involved.  Therefore, one may skip introducing  the graded commutation rules of $I_{V_{\delta \xi}}$ in  Table  [\ref{table_2}].\footnote{Quantities depending on $\delta \xi$ will  naturally occur in the practical computations by making use of \eqref{nil3}.}

The relevant generalization  of Table [\ref{table_1}]  for the gauge fixed and BRST invariant    situation    is therefore as follows,  
where  \eqref{graded_commutator1} and the knowledge of 
 the gradings of all entries in  Table [\ref{table_2}]  determines  case by case 
the  graded   nature  of  $[X,Y]$ (commutator or anticommutator):

\begin{table}[h] 
\centering
\begin{tcolorbox}[tab,tabularx={X||Y|Y|Y|Y|Y|Y|Y|Y},boxrule=0.9pt] 
$[X,Y]$ & $d$     & $\delta$     & $s = L_V$   & $I_V$     &   $\Lie_\xi$    & $\Lie_\phi$   & $i_\xi$     & $i_\phi$   
 \\\hline\hline
$d$   & 0     & 0     & 0   & 0     & 0      & 0   & $-\Lie_\xi$     & $-\Lie_\phi$
 \\\hline
$\delta$ & 0     & 0     & 0   & $- L_V$     & $\Lie_{\delta \xi}$      & $\Lie_{\delta \phi}$   & $i_{\delta \xi}$     & $i_{\delta \phi}$ 
\\\hline
$s = L_V$  & 0     & 0     & 0   & 0     & $\Lie_{s\xi}$      & $\Lie_{\{\xi,\phi\}}$   & $i_{s \xi}$     & $i_{\{\xi,\phi\}}$
\\\hline
$I_V$  & 0     & $L_V$     & 0   & 0     & 0      & 0   & 0     & 0 
\\ \hline
$\Lie_\xi$  & 0     & $\Lie_{\delta \xi}$    & $\Lie_{s\xi}$   & 0     & $\Lie_{\{ \xi, \xi \}}$     & $\Lie_{\{ \xi, \phi \}}$   & $i_{\{\xi,\xi\}}$     & $i_{\{\xi,\phi\}}$
\\\hline
$\Lie_\phi$  & 0     & $-\Lie_{\delta \phi}$     & $-\Lie_{\{\xi,\phi\}}$   & 0     & $\Lie_{\{\xi,\phi \}}$      & $\Lie_{\{\phi,\phi \}}$   & $-i_{\{\xi,\phi\}}$     & 0
\\\hline
$i_\xi$  & $\Lie_\xi$     & $-i_{\delta \xi}$     & $-i_{s\xi}$   & 0     & $- i_{\{\xi,\xi\}}$      & $i_{\{\xi,\phi\}}$   & 0     & 0
\\\hline
$i_\phi$ & $\Lie_\phi$     & $i_{\delta\phi}$     & $i_{\{\xi,\phi\}}$   & 0     &  $i_{\{\xi,\phi\}}$      & 0   & 0     & 0
\end{tcolorbox}
\caption{\textit{Graded commutators in the BRST Covariant Phase Space} \label{table_2} }
\end{table}

It must be noted  that one might find relevant to split $\delta$ in two pieces to define  
a general field transformation modulo a BRST transformation.   To tackle this question, it sounds natural to define a new hatted differential operator 
\begin{equation}
\label{hated_delta}
\hat{\delta} \equiv \delta -  s = \delta - L_{V},
\end{equation}
which satisfies $\hat{\delta}^2 = \hat{\delta} s + s \hat{\delta} = 0$ because of \eqref{nil3}.  Such a decomposition is  analogous to that occurring for decomposing \textit{internal symmetries}$\times$\textit{reparametrization symmetries} 
into internal symmetries  modulo reparametrization symmetries,  according to $\hat{s} = s - \Lie_\xi$.

In the BRST CPS  case,  $\hat{\delta}$ can be seen as a ``gauge covariant field space variation", in the language of \cite{Gomes:2016mwl}.
Subtracting  the intrinsic gauge symmetry of field space to the general variation $\delta$ can become relevant to consistently  and separately analyze  the ``vertical" and ``horizontal" parts of the symplectic form  as in \cite{Gomes:2016mwl,Riello:2019tad}.

\subsection{Gauge fixing and BRST Noether 1.5th theorem}
\label{subsection_BRST_Noether}

The framework developed in the previous subsection allows one to quantitatively analyze the consequences on the symplectic potential \eqref{L_variation} of adding a gauge fixing term to the Lagrangian \eqref{lagrangian}.

The generic BRST gauge fixing procedure begins with a classical Lagrangian $L_{\rm cl} \in~\O_{\rm loc}^{0,d,0}$ invariant under the following infinitesimal gauge symmetry transformation parametrized by $\lambda(x)$
\begin{equation}
\label{classical_gauge_transf}
\vp_{\rm cl}(x) 
\to \vp_{\rm cl}(x) + \delta_\lambda \vp_{\rm cl}(x). 
\end{equation}
The global BRST symmetry that captures this gauge invariance is obtained by introducing anticommuting ghost fields $c(x)$ for every infinitesimal local parameters $\lambda(x)$ and  a global anticommuting parameter $\eta$.  A BRST transformation $\delta_{\rm BRST} \equiv \eta s$ of the classical fields is then defined as a gauge transformation of the form \eqref{classical_gauge_transf} modulo the replacement $\lambda \to \eta c$. Therefore,  as $\eta$ is constant, one has
\begin{equation}
s \vp_{\rm cl} \equiv \delta_\lambda \vp_{\rm cl} \big\vert_{\lambda = c}
\end{equation}
and $L_{\rm cl}$ is clearly invariant under such a transformation, namely $sL_{\rm cl} = dB$.  The BRST transformation of the ghost fields $sc$ is adequately defined to ensure the nilpotency of $s$ on $\vp_{\rm cl}$ and $c$.  This means that one can consistently add any terms of the form 
$
L_{\rm cl} \longrightarrow L_{\rm cl} + s \Psi$ 
to the Lagrangian without spoiling its BRST invariance.  The denomination $\Psi$ is not an accident. It must be an anticommuting quantity with ghost number minus one in order for the Lagrangian to have ghost number zero.  Introducing trivial BRST pairs $(\bar{c},b)$ with ghost number $(-1,0)$ respectively and transforming as $s\bar{c}=b$, $sb =0$ is thus unavoidable to build such a $\Psi$.\footnote{Here we consider a BV system of rank 1, the generalization for higher rank will be published elsewhere.}  The BRST gauge fixing $\mathcal{F}_{\rm gauge}(\vp^A)=0$ of the classical Lagrangian is often reached by considering 
\begin{align}
\label{gauge_fixing}
L_{\rm cl} (\vp_{\text{cl}}) \longrightarrow L_{\rm GF} (\vp^A) &=  L_{\rm cl} (\vp_{\text{cl}}) + s \Big( \bar{c} \ \mathcal{F}_{\rm gauge}(\vp^A) \Big) 
\nn \\
&= L_{\rm cl} (\vp_{\text{cl}}) + b \  \mathcal{F}_{\rm gauge}(\vp^A) - \bar{c} \ s \mathcal{F}_{\rm gauge}(\vp^A)  . 
\end{align}
The $b$ dependence of the Lagrangian localizes the gauge field configurations around the gauge choice 
and the global BRST symmetry is   a residual symmetry of the gauge fixed Lagrangian $L_{\rm GF}$.  The term proportional to the $s$-variation of the gauge fixing functional provides the ghost interactions and the consistency of the gauge choice.

The BRST CPS is perfectly suited to analyze the consequences of the equations of motion of the classical fields, the ghosts, antighosts and Lagrange multipliers stemming from the gauge fixed Lagrangian \eqref{gauge_fixing}, keeping track of all different gradings. In what follows, any $L$ without subscript refers to $L_{\rm GF}$ as defined in \eqref{gauge_fixing}.  The key relation \eqref{L_variation} is still valid for this Lagrangian, that is 
\begin{equation}
\label{L_GF_variation}
\delta L  = E^A_\vp   \delta \vp^A + d \theta   .
\end{equation}
The differences with the classical case reside in both terms $E^A_\vp$ and $\theta$. Indeed, from \eqref{gauge_fixing}, it is obvious that one can split the local symplectic potential as 
\begin{equation}
\label{split_theta_gauge}
\theta (\vp) = \theta_{\rm cl} (\vp_{\rm cl},c) + \theta_{\rm gauge} (\vp^A).
\end{equation}
From now on, any subscripts ``cl" or ``gauge" refer to quantities that depend only on $(\vp_{\rm cl},c)$ or $\vp^A= (\vp_{\rm cl},c, \bar{c},b)$, respectively.  By construction, any ``gauge" quantity vanishes when one sets $\bar{c}=b=0$. 

One may now derive the BRST Noether current of the gauge fixed Lagrangian \eqref{gauge_fixing}. 
One starts with  
\begin{equation}
sL = s L_{\rm cl} = d B_{\rm cl} .
\end{equation}
Since the Lagrangian $L$ is a spacetime top-form, the boundary term $B=B_{\rm cl}$ is  a spacetime $(d-1)$-form with ghost number one.\footnote{
It is interesting to notice that a series  of cocycle equations  follows from  the 
definition of the BRST invariance of  $S_{\rm BRST}$, 
$s  L^0_d  \equiv  d B^1_{d-1}$ as 
\begin{equation}
\label{B_cocycle}
sB^i_{d-i}  = dB^{i+1} _{d-i-1}  
\end{equation}
for $i>0$.  These equations are direct consequences of the algebraic Poincar\'e lemma 
\begin{equation}
sX=dY   \Longrightarrow  0=  s^2 X    = sd Y=-dsY\Longrightarrow    sY=dZ
\end{equation}
that was used in \eqref{def_Noether_constraints1} and will be repeatedly used in the rest of the paper. It is only valid for off-shell equalities.  }
The classical expression of the Noether current~\eqref{classical_Noether_current}\footnote{The change of sign convention in the definition of $q_{\rm cl}$ compared to the original one in \eqref{classical_Noether_current} is arbitrary and made to match the value of the classical charge modulo $c \to + \lambda$ in the Yang--Mills case.}
\begin{equation}
\label{J_BRS_cl}
\star J_{\rm cl} = I_V \theta_{\rm cl} - B_{\rm cl} = - C_{\rm cl} - dq_{\rm cl} 
\end{equation}  
is obviously no longer  conserved on-shell since the constraints $C_{\rm cl}$ do not vanish on-shell because the equations of motion for the classical fields $E^A_{\vp_{\rm cl}}$ are modified and now depends on the unphysical fields. To get a conserved current, one must therefore extend the definition of the classical Noether current to its full BRST expression  
\begin{equation}
\label{NT1}
\star J_{\rm BRST} = I_V \theta - B =  I_V \theta_{\rm cl} +  I_V \theta_{\rm gauge} - B_{\rm cl} \quad \text{with} \quad d(\star J_{\rm BRST}) \ \hat{=} \ 0   ,
\end{equation}
obtained by applying the first Noether theorem for the global BRST invariance of the gauge fixed BRST invariant Lagrangian \eqref{gauge_fixing}.
The series of equations \eqref{def_Noether_constraints1}  is still valid for the full EoM's \eqref{L_GF_variation}, leading to
\begin{equation}
\label{def_Noether_constraints}
I_V ( E^A_\vp   \delta \vp^A ) = d (B_{\rm cl} - I_V \theta ) \equiv d C  .
\end{equation}
Here, one may call $C = C_{\rm cl} + C_{\rm gauge}$   the \textit{BRST Noether constraints}.
Combining \eqref{def_Noether_constraints} with \eqref{J_BRS_cl} provides
\begin{equation}
\label{J_BRST_gauge_dep?}
\star J_{\rm BRST} = I_V \theta_{\rm cl} +  I_V \theta_{\rm gauge} - B_{\rm cl}  = -C_{\rm cl} - C_{\rm gauge} - d( q_{\rm cl} + q_{\rm gauge} ) ,
\end{equation}
where $I_V \theta_{\rm gauge} = - C_{\rm gauge} - dq_{\rm gauge}$. 
This expression for the BRST Noether current implies that the  corner Noether charges that are defined off-shell by the $d$-exact term in
\eqref{J_BRST_gauge_dep?} are possibly gauge dependent.  If it happens,  it would spoil the interpretation of these charges as physical quantities defining the asymptotic symmetry group no matter how the gauge is fixed.

However,  it turns out that for all explicit computations presented in this paper and performed in different theories with different gauge fixing choices, one gets
\begin{equation}
\label{vanishing_q_gauge}
I_V \theta_{\rm gauge} = - C_{\rm gauge} \quad \Longrightarrow \quad  q_{\rm gauge} = 0    .
\end{equation}
This is a most satisfying result.
Indeed,  given a local  gauge symmetry parametrized by $\lambda(x)$,  one has  $C_\lambda \ \hat{=} \ 0$ and therefore $\star J_\lambda \ \hat{=} \ d q_\lambda$ defines  non vanishing corner charges
$q_{\rm cl} = q_\lambda \big\vert_{\lambda=c}$. Once integrated over a corner, the non vanishing pieces of these charges have been understood as defining  the physical sub-sector of this local gauge symmetry.
 The BRST CPS therefore allows one to investigate this aspect of gauge theories and use  their   BRST invariant  gauge fixed formulation to study the interplay between these global corner charges and the bulk physics (cf. Section [\ref{Section_WI}]).

The BRST symmetry  generalizes the gauge symmetry at the quantum  level. It must at least  provide  the same information about the physical relevance of the charges $q_{\rm cl}$. 
    A~first~evidence of this fact is provided by \eqref{vanishing_q_gauge}. In fact,         one must also  check that the charges $q_{\rm cl}$~canonically generate the asymptotic symmetry. 
For this    property to be true, 
it~will be explained  in the next subsection that the full constraints $C = C_{\rm cl} + C_{\rm gauge}$ must be $s$-exact on-shell.  It turns out that this is the case in the examples  studied  in this paper. 
A main result of this paper is thus that one has in these examples:  \begin{equation}
\label{Noether_1.5}
\star J_{\rm BRST}  =   -C_{\rm cl} - C_{\rm gauge} - d q_{\rm cl}  \ \hat{=} \    (s G)_{\rm gauge} - d q_{\rm cl} \  .
\end{equation}
 Here  $G= G[\vp^A] $ is a ghost number zero local functional of all  BRST fields that is gauge dependent.  It indicates that the  whole gauge choice  dependence   
 concentrates in  $sG$,  a property that we hope to prove in more generality in a separate publication. 
 The property that $\star J_{\rm BRST}$ is a sum of a $d$-exact and a $s$-exact term on-shell is conjectured to be generically true, and we  
call it BRST Noether $1.5$th theorem.\footnote{  Notice that the 1.5th terminology  in this article  has nothing to do with that of \cite{Miller:2021hty}.  Here,  it is used to underline that the quantum definition of the Noether charges $q_{\rm cl}$, which were discovered at the classical level from   Noether's second theorem,  is actually derived from  Noether's first theorem when one uses as one should in the gauge fixed situation the BRST symmetry methodology. }     
 
 As a matter of fact,  \eqref{NT1} and   \eqref{Noether_1.5}     imply the following on-shell identity for $G \equiv G^0_{d-1}$
\begin{equation}
 \label{on-shell_cocycle}
ds   G^0_{d-1}  \  \hat{=} \  0  \quad \Longrightarrow \quad dG^0_{d-1} \ \hat{=} \ s G^{-1}_d ,
\end{equation}
where $s G^{-1}_d$ is possibly zero. This term has the same grading as the $s$-exact term \eqref{gauge_fixing} that one adds to the classical Lagrangian to fix the gauge.  In fact, modulo integration by parts,  one finds 
\begin{equation}
\label{sG_s-exact}
s G^{-1}_d \ \hat{=} \ s \big( \bar{c} \ \mathcal{F}_{\rm gauge}(\vp^A) \big),
\end{equation}
which gives a strong intuition for the splitting \eqref{Noether_1.5} at the level of the BRST Noether current.

It must be noted that the    spacetime  $(d-2)$-form $q_{\rm cl}$  coincides with $q_{\lambda}$ in  \eqref{Noether_2} when one    replaces the ghost field  $c$  by  the  associated infinitesimal   gauge symmetry parameter~${\lambda}$.
 Since  $q_{\rm cl}$ is ghost dependent,    questioning    its gauge covariance is  an  irrelevant   notion;  
 rather, the  understanding of the   BRST covariance  of $q_{\rm cl}$  is an important question.
 For instance,    it will be proven   in  the Yang--Mills case   that  $q_{\rm cl}   = c \star F$  and~$s q_{\rm cl} = \demi  [c,c] \star F$ while  $\delta_{\rm gauge} ( c \star F)$ is of course  meaningless due to the non existence of ghost gauge transformations.

 In fact, the next subsection  will  consider the integrated  charges $Q =\int _{\pa \Sigma}  q_{\rm cl} \neq 0$ (in the sense of \eqref{hamiltonian_charges}) and show  that  $s Q$ can be understood in terms of  a Barnich--Troessaert bracket \cite{Barnich_Charge_algebra},  due to $sQ =  I_V I_V \O    \equiv   \{ Q ,  Q \} $.
 All this  allows one to separate a complete   
 BRST transformation into a trivial one and 
 a large one, in the sense of \eqref{asg}. In fact,  motivated by  the definition of  large gauge transformations parametrized by $\lambda(x)$ as those leading to $\int_{\pa \Sigma} q_{\lambda} \neq 0$, one defines the large  BRST operator $  s_{\rm large}$  that  generates the asymptotic symmetry as 
\begin{equation}
\label{def_s-large}
s_{\rm large} \equiv s \big\vert_{c = c_{\rm large}} \quad   \text{with} \quad \int_{\pa \Sigma} q_{\rm cl}(\vp_{\rm cl}, c_{\rm large}) \neq 0.
\end{equation}
The trivial   BRST operator   $s_0  $ is such that $s_0 \equiv s \big\vert_{c = c_0} $ with  $\int_{\pa \Sigma} q_{\rm cl}(\vp_{\rm cl}, c_0) = 0$.
The large BRST transformations  \eqref{def_s-large} do change the physics because their action on phase space is non trivial.  
This will be seen explicitly in the next subsection.  Therefore, the physical states are defined as the elements 
of  $H^0(s_0)$  and not  of $H^0(s)$.
The action of  $s_{\rm large} $  and of  $s_{ 0} $ on  the BRST trivial doublet $(\bar{c},b)$  are 
 identical.  However they are not identical when acting on  the geometrically non trivial    doublet $(\vp_{\rm cl}, c)$ and one has  
\begin{align}
s_0^2 \vp_{\rm cl} &= 0 ,   &    s_{\rm large}^2  \vp_{\rm cl} &=0 , 
\nn \\
s_0^2 c_0 &= 0 ,  & s_{\rm large}^2 c_{\rm large} &=0 . 
\end{align}
The decomposition $s= s_{ 0} + s_{\rm large} $  is  conceptually important but not relevant for  explicit computations. 
One   will thus derive the  Ward identities for the complete BRST operator $s$ and avoid as much as possible the use  of this decomposition.

\subsection{Gauge fixed fundamental canonical relation}
\label{subsection_gauge_fixed_fund}

In order to recover   the   same information   from  the ghost dependent equation \eqref{Noether_1.5}   for the Noether charge $q\equiv q_{\rm cl}$
as  from the classical equation  \eqref{fund_with_fluxes},  one needs to study   the fundamental canonical relation $I_V \O$ for a symplectic two form 
\begin{equation}
\label{BRST_2-form}
\O \equiv \delta \left( \int_{\Sigma} \theta \right) = \delta \left( \int_{\Sigma} \theta_{\rm cl}  + \theta_{\rm gauge}  \right)
\end{equation}
 stemming from the symplectic potential $\theta$ of a gauge fixed and    BRST invariant   action~\eqref{split_theta_gauge}.  
To do so,  one starts with the definition of the BRST Noether current \eqref{NT1}   as 
\begin{equation}
\star J_{\rm BRST} = I_V \theta - B = - C - dq .
\end{equation}
Then,  one   uses the definition $s \equiv L_V$ of the BRST CPS and $\o \equiv \delta \theta$ to obtain
\begin{align}
\label{s_theta_BRS1}
s \theta &= I_V \o - \delta I_V \theta = I_V \o - \delta ( B  - C - dq ) .
\end{align}
Since $\delta L = E +d\theta$ and $sL=dB$,  one   also defines a new  boundary local form $Z \in \O_{\rm loc}^{1,d-1,1}(\tilde{\mathcal{F}} \times M)$ from 
\begin{equation}
\label{def_Z}
d \delta B = s \delta L = sE - ds\theta \quad \Longrightarrow \quad sE \equiv dZ ,
\end{equation}
which in turn defines $Y \in \O_{\rm loc}^{1,d-2,1}(\tilde{\mathcal{F}} \times M)$ as 
\begin{equation}
\label{s_theta_BRS2}
d ( s \theta + \delta B  - Z) = 0 \quad \Longrightarrow   \quad s \theta + \delta B - Z \equiv d Y .
\end{equation}
Comparing \eqref{s_theta_BRS1} with \eqref{s_theta_BRS2} and integrating  over $\Sigma$   leads to
\begin{equation}
\label{off-shell_BRST_fund}
 I_V \O =  - \delta \Big(  \underbrace{  \int_\Sigma C + Q }_{ \text{\tiny Integrable\ part} } \Big)  +  \underbrace{ \int_\Sigma ( Z + dY ) }_{\text{\tiny Non\ integrable\ part}}  ,
\end{equation}
where $Q \equiv  \int_\Sigma d q$. 
This is the off-shell fundamental canonical relation for the ``gauge fixed" symplectic two form \eqref{BRST_2-form} in the BRST CPS.
The terminology  ``integrable" and ``non integrable" is inherited from the classical definition of an integrable charge \eqref{hamiltonian_charges}.  Indeed,  this  relation must determine whether or not the BRST Noether charges $Q$ are integrable, namely, whether or not    they  canonically generate the large gauge symmetry~\eqref{def_s-large}.
One~cannot expect    \eqref{off-shell_BRST_fund} to   reduce on-shell to the simplest Hamiltonian definition of charges \eqref{hamiltonian_charges} since the  status of   ghost fields  in the   BRST CPS   is  completely analogous to that of classical  fields and  therefore $\delta c$ never vanishes.  
 So one must rather compare the on-shell projection of \eqref{off-shell_BRST_fund} with the classical fundamental relation  \eqref{fund_with_fluxes}, which has a non vanishing non integrable part, and check if it leads to the same bracket structure. This classical relation, which is only valid for an ungauge fixed theory,  is derived off-shell in Appendix [\ref{Annex_anomaly}] and takes the form\footnote{Although it is derived for a spacetime symmetry parametrized by $\xi(x)$ and involves an additional $i_\xi \theta$ term in the non integrable part (see \eqref{fundamental_canon}),  the derivation equally applies to an internal symmetry parametrized by $\lambda(x)$ if one forgets this additional term. In any case, the following arguments and conjectures do not depend on the nature of the symmetry and also hold for a spacetime symmetry. } 
 \begin{equation}
\label{fundamental_canon_2}
  I_{V_\lambda} \O =  - \delta \Big(  \int_\Sigma C_\lambda + Q_\lambda \Big)  - \int_\Sigma \Big( C_{\delta \lambda} + d ( q_{\delta \lambda}  - A_\lambda ) \Big) .
\end{equation}
A few comments are in order.  Noether's second theorem \eqref{Noether_2} implies $C_\lambda \ \hat{=} \ 0$ and $C_{\delta \lambda} \ \hat{=} \ 0$. The on-shell non integrable part is then $\mathcal{F}_\lambda \equiv - \int_S (  q_{\delta \lambda}  - A_\lambda)$, which is the quantity appearing in \eqref{fund_with_fluxes} and is sometimes called the flux. 
The term $A_\lambda$ in this flux comes from the use of the anomaly operator \cite{Freidel}.  Its explicit definition is given in Appendix [\ref{Annex_anomaly}] and is not needed for the following reasoning.

Let us now point out the similarities and differences between  \eqref{off-shell_BRST_fund} and \eqref{fundamental_canon_2}:
\vspace{0.2cm}

\noindent \textit{ (i) The integrable part}

\noindent - Ungauge fixed theory: \ $I_{V_\lambda} \o \big\vert_{\text{\tiny int}} = - \delta (  C_\lambda + dq_\lambda ) \underset{\text{\tiny Noether 2}}{\hat{=}}  d \delta q_\lambda$.  The asymptotic symmetry algebra is governed by the $\lambda$'s that are such that $Q_\lambda = \int_\Sigma dq_\lambda \neq 0$.  The Poisson brackets providing a representation of this symmetry algebra are then given by $I_{V_\lambda} I_{V_\lambda} \o \big\vert_{\text{\tiny int}} \ \hat{=} \ d s q_\lambda$.  This simple Poisson bracket construction works only if there is no non integrable part in $I_{V_\lambda} \o$.  \\ 
- Gauge fixed theory: \ $I_V \o \big\vert_{\text{\tiny int}}  = - \delta ( C + dq )  \underset{\text{\tiny Noether 1.5}}{\hat{=}}   \delta s G + d \delta q$ with $q=q_\lambda\big\vert_{\lambda=c}$.  
The symplectic two form is non degenerate since even for trivial gauge transformations where $Q_{\lambda=c} = 0$, one still has $I_V \O \ \hat{=} \  \delta \int_\Sigma sG  \neq 0$.
The Poisson brackets are given by 
\begin{equation}
\label{delta_s-exact}
I_V I_V \o \big\vert_{\text{\tiny int}} \ \hat{=} \  s^2 G + d sq = d s q.  
\end{equation}

It shows that the integrable parts of the gauge fixed and ungauged fixed equations    \eqref{off-shell_BRST_fund} and \eqref{fundamental_canon_2} provide  the   same information about  the charges and their bracket structure. 

\vspace{0.2cm}

\noindent \textit{  (ii) The non integrable part}

\noindent - Ungauge fixed theory: \ $I_{V_\lambda} \o \big\vert_{\text{\tiny non-int}} = - C_{\delta \lambda} - d ( q_{\delta \lambda}  - A_\lambda ) \underset{\text{\tiny Noether 2}}{\hat{=}}  - d ( q_{\delta \lambda}   - A_\lambda )$.  The fact that it reduces to a $d$-exact term on-shell that has this specific expression is crucial in order to define a new bracket that provides a representation of the asymptotic symmetry algebra in the presence of fluxes \cite{Barnich_Charge_algebra,Freidel}; or in order to add codimension two fields, called edge modes,  to the phase space that cancel the flux part of $I_{V_\lambda} \o$ such that the Poisson brackets can be used \cite{Ciambelli_2022_Embeddings,freidel2021canonical,Ciambelli:2021vnn}.
\\ 
- Gauge fixed theory: \ $I_V \o \big\vert_{\text{\tiny non-int}}  = Z + dY  \ \hat{=} \   d Y$? One understands that on-shell, $Z$ has to vanish or at the very least to reduce to a $d$-exact term otherwise the above construction of a bracket for systems with fluxes cannot be used and therefore the phase space cannot be quantized.  However,  the algebraic Poincar\'e lemma is not valid on-shell so $sE = dZ \ \hat{=} \ 0$ neither implies $Z\ \hat{=} \ 0$ nor $Z \ \hat{=} \ dW$.  It is again an explicit computation of $Z$ for different gauge fixed theories that supports the important conjecture 
\begin{equation}
\label{claim_Z}
Z \ \hat{=} \ 0.
\end{equation}
Notice that this postulate is in the exact same spirit as the one for Noether's 1.5th \eqref{Noether_1.5}, which is $C \ \hat{=}  -sG$. The explicit decomposition of $Y$ into $\mathcal{F}_{\rm cl} = ( q_{\delta \lambda}   - A_\lambda )\big\vert_{\lambda=c}$   also turns out to be true in the various examples, unfortunately without having a generic origin for~it. 
One has thus  also recovered the  same information for the non integrable part of the fundamental canonical relation in the gauge fixed case \eqref{off-shell_BRST_fund} as in the ungauge fixed case~\eqref{fundamental_canon_2}.  

\vspace{0.2cm}

The quantization of the phase space which consists in promoting the classical brackets to quantum commutators can therefore be safely performed in the gauge fixed case with a consistent introduction of the ghosts, antighosts and Lagrange multiplier fields.\footnote{It is worth mentioning  that the BRST CPS might give access to another quantization prescription for the phase space,  which could  be equivalent to the canonical one. It  amounts 
to do a path integral quantization of the geometric action $S = \int \theta$ for the gauge fixed symplectic potential $\theta$ \eqref{split_theta_gauge},  see e.g. \cite{Alekseev:1988vx,Alekseev:1988ce}.
  Such an action has been constructed in a ghost independent way  for four dimensional   gravity  in asymptotically flat non radiative spacetimes \cite{Barnich:2022bni}, but is not yet known for the obviously more realistic case of radiative spacetimes.  We plan to investigate this issue with our new BRST tools in the future. }  
  At the present stage,  one still has a lack of generality in the    gauge fixed case because the  needed relations $C \ \hat{=} - sG$,  $Z \ \hat{=} \ 0$ and $Y = \mathcal{F}_{\rm cl}$   are physically mandatory postulates that have only  been checked in a few non trivial examples.
One hopes that a generic proof of these statements will be found in the near future.

Another important feature of the gauge fixed fundamental canonical relation \eqref{off-shell_BRST_fund} is that it is invariant under boundary shifts of the Lagrangian.  To show this, one has to write \eqref{off-shell_BRST_fund} for the new pair $(L',\theta')$ \begin{align}
L' &= L +dl  ,
\nn \\
\theta' &= \theta - \delta l + d \nu  .
\end{align}
Since the equations of motion are unaffected by a boundary shift of the Lagrangian, one has $C'=C$ and $Z'=Z$.  Then,  one has 
\begin{align}
\o' &= \o - d \delta \nu ,  &  B' &= B - sl ,
\nn \\
q' &= q - I_V \nu,   &  d Y' &= dY  - d s \nu.
\end{align}
Hence, the  relation \eqref{off-shell_BRST_fund} for the pair $    (L',\theta')$ writes
\begin{align}
\label{boundary_shift_fund}
&I_V \o' = - \delta \Big(  C' + dq'  \Big)  + Z' + dY'
\nn \\
\iff \ \  &  I_V \o - d I_V \delta \nu =  - \delta \Big(  C + dq - d I_V \nu  \Big)  + Z + dY - d I_V \delta \nu + d \delta I_V \nu 
\nn \\
\iff \ \ & I_V \o = - \delta \Big(  C + dq  \Big)  + Z + dY . 
\end{align} 
This equivalence is important in our construction since some gauge fixing terms differ from one another by a boundary term while imposing the same gauge fixing condition, as will be seen for instance  in \eqref{boundary_term_diff}.  This implies  that choosing a $s$-exact term or a $(s+d)$-exact term to fix the gauge has no impact on the split between charges and fluxes, and consequently has no impact on the construction of the bracket made from this split.  This also implies that one can renormalize the Noether charges $Q$ that have no reasons to be finite by adding boundary terms to the Lagrangian. This will become important in the next section as they will appear in the boundary Ward Identity $[Q,\mathcal{S}]=0$, which can be made finite thanks to~\eqref{boundary_shift_fund}.

The  next section will  introduce  sources  
for inserting  Noether currents  $\star J_{\rm BRST}$ satisfying \eqref{Noether_1.5}
in  correlation functions.   Their introduction is compulsory in order 
  to    derive both bulk and boundary  Ward identities  of the full BRST symmetry. 
  Eventually, these  identities will 
    make explicit  the physical consequences of the existence of $\star J_{\rm BRST}$ and of its on-shell decomposition \eqref{Noether_1.5}.

\section{Holographic BRST Ward Identities}
\label{Section_WI}

 To use  functional methods for  
determining  and     exploring     the  Ward  identities for the correlation  functions of the fields  with  all possible insertions of the BRST Noether current,  one may  extend the  method exposed in
  \cite{Baulieu:1984iw,Baulieu:1986hv} for a simpler example of a genuine  current~algebra.
  By doing so,  one is to find boundary Ward identities for the current $\star J_{\rm BRST}$ that reduce to the well known Ward identities  for the physical global charges $\bra{\rm out}  [Q,\mathcal{S} ]  \ket{\rm in}   = 0 $ associated with the invariance of the $\mathcal{S}$-matrix under large gauge transformations.  
  This is done directly in the Lagrangian formulation, while taking  care of the consequences of the invariance of the bulk physics under small gauge transformations as well.  One also gets 
   a transparent description of the   possible anomalies for the boundary Ward identities,  which are purely holographic quantities and are eventually related to   
  loop corrections to soft theorems.

\subsection{Sourcing the global BRST Noether current}

The  global BRST symmetry    possesses a constant anticommuting parameter $\eta$.  It implies the existence of the  local  operator $ \J$ as shown in the previous section.
       One  may therefore  introduce an abelian  commuting 1-form       $\al \equiv \alpha^{-1} _\mu dx^\mu$  with ghost number $-1$ as a source for   the operator
  $I_V \theta^\mu = \frac {\delta  L(\vp, \pa_\mu \vp)}{\delta  \pa_\mu \vp} s\vp \subset J_{\rm BRST}^\mu$ .  One also introduces its
   spectator 
      commuting  0-form ghost   $\Ce$ with ghost number 0.  This provides  a classical gauging of the global BRST symmetry $ \vp \to \vp + \eta(x)  s \vp $, with  a   ghostification  $ \eta \to \eta(x) \to  \Ce(x)$.  
        $\eta$ being an anticommuting parameter,   the ghost  $\Ce$ is  commuting as  well as the $1$-form $\al$.  Obviously,  $\alpha$ and $\Ce$ are not propagating, they are just sources.

%
          
One may then define the nilpotent `` BRST${}^2$ "        graded    differential operator $r$ \cite{Baulieu:1986hv}
     \begin{align}
     \label{rsym}
      r \al (x)   &=- d  \Ce   (x) ,
     \nn\\
        r\Ce   (x)     &= 0 ,
         \nn\\
        r\vp    (x)     &=  \Ce   (x)     s\vp (\vp, {\Da}_\mu \vp) ,
     \end{align}
     with the notation $X(\vp, \Da_\mu \vp)\equiv 
      X(\vp, \pa_\mu \vp)  \big|  _{\pa_\mu \vp = {\Da}_\mu   \vp }$ for any functional $X$.
Introducing   $\al$ basically gauges the BRST symmetry and  defines the 
   ``BRST covariant derivative" $\Da$,   according to  
     \begin{equation}
     \label{minimal_coupling}
     \pa_\mu \to \Da_\mu \equiv \pa_\mu -\al_\mu s \quad  \iff \quad d \to d +  \al s . 
     \end{equation}
    Changing $d$ into $d \to d +  \al s$  realizes a minimal coupling.    
     One has consistently 
\begin{equation}
 r^2= 0, \quad rd + dr =0  , \quad d^2=0 
    \end{equation} 
  on  all fields $\vp$, $\alpha$ and $\Ce$, as a consequence of \eqref{rsym} and $s^2 = sd +ds = 0$.
  
  In fact, given any field functional $F(\vp, \pa_\mu \vp)$, one has 
  \begin{equation}
  \eta sF(\vp, \pa_\mu \vp )   = \eta \left(  \frac{\delta  F(\vp, \pa_\mu \vp ) }{\delta  \vp} s\vp +  \frac{\delta  F(\vp, \pa_\mu \vp ) }{\delta \pa _\nu  \vp} \pa_\nu s\vp \right)
 \end{equation}
  where $\eta$ is  constant, so       the ghostified   minimal coupling \eqref{minimal_coupling}  with $r\al(x)=-d\Ce(x)$     imply
\begin{equation}
r F(\vp, (\pa_\mu  -  \al_\mu   s)  \vp )   =  \Ce(x)    s F (\vp, \pa_\mu \vp) \big|  _{ \pa_\mu \vp = (\pa_\mu   -  \al_\mu   s)  \vp } .
 \end{equation} 
 This computation uses an other useful identity in order to understand why $\al_\mu(x)$ can be interpreted as a source for  part of the Noether current, that is
 \begin{equation}
 \label{alpha_identity}
    \frac {\delta  F  (\vp   ,    \pa_\mu \vp         )  }
    {\delta \pa_\nu \phi} 
    s\phi
    = - \left. \frac {\delta  F( \vp  ,   \Da_\mu \vp        )  }
    {\delta \al_\nu}   \right\vert_{\al=0} .
    \end{equation}
   \def\eps{\epsilon}
   It gives the $I_V \theta$ piece of the Noether current when applied to $F=L$ indeed.


The   generating functional  $W_c$ of  connected correlation functions of the fields $\vp$ with insertions of their BRST transformation  operators  $s\vp$ and of the  BRST  Noether current $\J$ is 
  \begin{align}
  \label{exp_W}
    \exp W_c[ j_\vp, v_\vp,\al,    j_B,  v_B    ]
   \equiv& \int_{\vp_i}^{\vp_f}  [ \delta  \vp] \exp\int_M
    \Big ( L   (\vp   ,   \Da \vp     )     
    + j_\vp \vp + v_\vp   s \vp (\vp,\Da\vp)   \nn\\
    & + j_{  B }   B ( \vp, \Da\vp     ) 
 +  v_{  B }   s B (\vp, \Da  \vp     )       
   \Big).
   \end{align}
   The  BRST   invariant path integral measure   $ [ \delta  \vp] $   stands for 
 $[ \delta  \vp]  \equiv \Pi_{all\  fields\  \vp^A} [ \delta  \vp^A]$ and $\vp_i$, $\vp_f$ are the boundary conditions of the fields.  
   The additional sources $j_B$ and $v_B$ are for the local operator $B(\vp,\Da_\mu \vp)$ and its BRST variation. 
   These additional couplings are necessary to insert the missing piece $B$ of the Noether current into correlation functions.  

\subsection{Unified Ward identities for small and large gauge transformations}

The complete set of Ward identities is to be determined from the invariance of the path
 integral \eqref{exp_W} under  
the following change of   field variables
\begin{equation}
  \label{change_variable}
  \vp (x) \to \vp'(x) =  \vp(x) +\eta(x)   s\vp (\vp, \Da\vp)(x) .
 \end{equation}
 The $\vp$ dependence of \eqref{exp_W} decomposes in a series of building blocks that transform  as follows
under \eqref{change_variable}
  \begin{align}
  \label{transforms}
    [ \delta  \vp'] =& \    [ \delta  \vp] ,
    \nn\\
    L   (\vp',  \Da\vp' )
   =& \ 
     L    (\vp,  \Da\vp) 
     - \eta d   B    (\vp, \Da   \vp)
      +d   \eta  \frac {\delta   }{\delta \al}     L       (\vp, \Da  \vp   )   ,
     \nn\\
    j_\vp \vp' + v_\vp     s \vp  (\vp', \Da\vp')   =& \ 
    j_\vp \vp +   v_\vp   s \vp  (\vp, \Da \vp)  -  \eta j_\vp    s\vp   (\vp, \Da \vp)  
    \nn \\
    &+ d \eta     
     \frac{\delta }{\delta \alpha  }   \Big( v_\vp  s\vp   (\vp, \Da\vp) \Big) ,
 \nn\\
  j_{  B }   B ( \vp', \Da\vp'      )
 +  v_{  B }   s B (\vp', \Da  \vp'     ) 
=& \ j_{  B }   B ( \vp, \Da\vp       )
 +  v_{  B }   s B (\vp, \Da  \vp     ) +\eta  j_{  B }   sB ( \vp, \Da\vp      )
   \nn \\
 &+d\eta
      \frac{\delta }{\delta \alpha  } \Big(
      v_{  B }   s B (\vp, \Da  \vp     ) +
      j_{  B }   B ( \vp, \Da\vp      )
      \Big) .
    \end{align}
    The path integral \eqref{exp_W}      that  defines     the generating  functional  $W_c$ 
   can be equivalently 
   written with  field integration variables $\vp$ or $\vp'$. 
  So by  using the change of variable \eqref{change_variable} in \eqref{exp_W},  as well as    \eqref{transforms},
        one gets  
      the Ward identity     satisfied by $\exp W_c[ j_\vp, v_\vp,\al,    j_B,  v_B    ]$ as~follows  
       \begin{align}
       \label{ouf}
    0 &=  \int_M   
          \bigg[ \eta(x) \Big (
       -  j_\vp(x)  s\vp (\vp, \Da\vp) +
         j _{  B }(x) s B  (\vp, \Da\vp) 
      -   d     \frac{\delta W_c}{\delta j_B(x)}  \Big )
     +  d\eta(x)
      \frac  {\delta}{\delta \al(x)}
      \bigg]
   \exp W_c
   \nn \\
   &=   \int_M \bigg[  \eta(x)
          \Big (
        - j_\vp(x)
        \frac {\delta}{ \delta v_ {\vp}(x)}
           +
         j _B (x)
         \frac {\delta}{ \delta v_B (x)}
      -    d     \frac{\delta W_c}{\delta j_{B}(x)}  \Big ) 
     +  d\eta(x)
      \frac  {\delta}{\delta \al (x)}
       \bigg ]
   \exp W_c \   .
      \end{align}
      
      The    generating functional      
     $\Gamma [  \vp, v_\vp,\al,    j_B,  v_B    ]$ of  the   connected    1PI correlation functions   
   is obtained by Legendre transformation of $j _\vp \to \vp$.     Since 
     \begin{equation}
      \Gamma [  \vp, v_\vp,\al,    j_B,  v_B ]    = - \int   j_\vp (x) \vp(x)  + W_c[   j_\vp ,  v_\vp,\al,    j_B,  v_B ]
      \end{equation}
      with $j_\vp = - \frac {\delta \Gamma[   \vp, v_\vp,\al,    j_B,  v_B ]  }{ \delta \vp}$, 
     one gets  from 
      \eqref{ouf}
         the   following Ward identity   for all relevant connected 1PI Green functions  
      \begin{equation} \label{wi2}
      0 = \int_M    \eta(x) \Big (   
     \frac {\delta \Gamma}{ \delta \vp(x)}
        \frac {\delta \Gamma}{ \delta v_ {\vp}(x)} 
           +
         j _{  B}(x)
         \frac {\delta \Gamma}{ \delta v_ {B }(x)}
   -
     d     \frac{\delta \Gamma}{\delta j_{B  }(x)}  \Big)
     +  d\eta(x)
      \frac  {\delta \Gamma}{\delta \al(x)}
     .
      \end{equation}
     Integrating by parts  the last term provides
  \begin{equation}
  \label{WI_1}
  0 = \int_M    \eta(x) \Big (   
      \frac {\delta \Gamma}{ \delta \vp(x)}
        \frac {\delta \Gamma}{ \delta v_ {\vp}(x)} +  j _{  B}(x)
         \frac {\delta \Gamma}{ \delta v_ {B }(x)} \Big)
   +
     d \Big( \eta(x)   \frac{\delta \Gamma}{\delta \alpha(x)}  \Big)
     +  \eta(x) d
    \Big(   \frac  {\delta \Gamma}{\delta \al(x)}  -  \frac{\delta \Gamma}{\delta j_{B  }(x)}   \Big)   .
  \end{equation}

At this stage,  the $\eta$ dependence of \eqref{WI_1} is arbitrary.  However, it has to be restricted consistently with the requirement that  
the field boundary conditions $\vp_i$, $\vp_f$ of the path integral \eqref{exp_W} must remain invariant when performing  the change of  field variables \eqref{change_variable}.  In fact, as large gauge transformations  do not leave invariant the field boundary conditions defined on $\pa M$ and are incorporated in the BRST operator \eqref{change_variable}, one has to set $\eta(x)$ to zero on $\pa M$.  

\begin{figure}[ht!]
\includegraphics[width=0.35\textwidth]{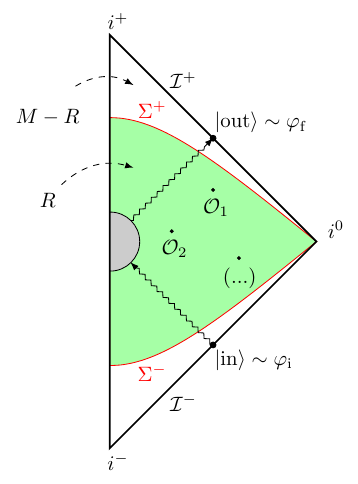}
\centering
\caption{\textit{Sub-region $R$ of the full Minkowski spacetime $M$.
 \label{figure_1}}}
\end{figure}

To deal with this issue,  and in the spirit of \cite{Avery_Schwab}, one can choose $\eta(x)$ to be an indicator function in the closed sub-region $R$ of the full spacetime $M$. This region is defined in Fig [\ref{figure_1}] such that $\pa R = \Sigma \cup i^0  \equiv \Sigma^+ \cup \Sigma^- \cup i^0$ where $\Sigma^\pm$ are spacelike hypersurfaces and $i^0$ is spacelike infinity.  Given that the main application of this formalism will be for asymptotically flat spacetimes,  Fig [\ref{figure_1}] is specified to be a Penrose diagram of Minkowski space in which the past and future boundary conditions of the path integral are defined on $\mathcal{I}^- \cup i^-$ and $\mathcal{I}^+ \cup i^+$ respectively.  The region $R$ is defined such that $R \cap \pa M = \pa R \cap \pa M = i^0$.  Notice that $\eta(x)$ does not necessarily need to vanish on $i^0$ as the path integral boundary conditions are not defined here. 

The indicator function is then defined as
\begin{align}
\label{eta_function}
\eta(x) &= \left\{ \begin{aligned} & \kappa, \quad x \in R  \\
 &0 , \quad x \in M-R 
 \end{aligned} \right.
\end{align}
where $\kappa$ is an anticommuting constant. 
In particular,  the indicator function is chosen such that it equates $\kappa$ on the boundary of $R$ and one has the following properties 
\begin{align}
\int_M  \eta(x) d  f(x)  &=  \kappa \int_{\pa R} f(x)   ,
\nn \\
\int_M  d \Big(  \eta(x)  f(x) \Big) &=  \kappa \int_{\pa M \cap R} f(x) =  \kappa \int_{i^0} f(x) . 
\end{align}

When $\eta$ is given by \eqref{eta_function},  one can safely perform  the change of field variables in \eqref{exp_W} and arrive at \eqref{WI_1}, which now rewrites 
  \begin{equation}
  \label{WI_2}
  0 =  \kappa \int_R    \Big (   
      \frac {\delta \Gamma}{ \delta \vp(x)}
        \frac {\delta \Gamma}{ \delta v_ {\vp}(x)} +  j _{  B}(x)
         \frac {\delta \Gamma}{ \delta v_ {B }(x)} \Big)
         - \kappa \int_{i^0} \frac{\delta \Gamma}{\delta j_{B  }(x)}
     +  \kappa
    \int_{\Sigma} \Big(   \frac  {\delta \Gamma}{\delta \al(x)}  -  \frac{\delta \Gamma}{\delta j_{B  }(x)}   \Big)   .
  \end{equation}
  The   Ward identity   \eqref{WI_2} generates all relevant  relations between  1PI correlation functions to be enforced  in  explicit computations.
One gets  their expressions by  writing all possible  differentiations  of \eqref{WI_2} with respect to the field sources    $\vp$,  the   field   operators   sources   
 $v_\vp, \al,    j_B,  v_{B}$ and  taking them equal to zero afterward.  Using \eqref{alpha_identity}, one has   
\begin{equation}
\label{NC_source}
\J =  \left. \left( \frac  {\delta \Gamma}{\delta \al(x)}  -  \frac{\delta \Gamma}{\delta j_{B  }(x)} \right) \right\vert_{\text{sources} = 0},
\end{equation}
which means that the last term proportional to $\kappa$ in the r.h.s.  of \eqref{WI_2} inserts a BRST Noether current integrated on the boundary $\Sigma$ into correlation functions.  This fact paves the way on how to properly decompose \eqref{WI_2} in two independent Ward identities that 
 complete one another. 
One identity is to    study the local properties in the bulk and adjust    the short  distance  renormalization while consistently  respecting  the BRST symmetry everywhere; 
  and  the other one is to provide constraints for the BRST Noether current in the boundaries. Those constraints  are 
    essential to guarantee  the possibility of a consistent  resolution of the IR problem for certain scattering process  involving inobservable clouds of  soft  massless particles  around observable hard ones, a la Faddeev--Kulish \cite{Kulish:1970ut}.
    
This split is justified by recalling that in order to renormalize the local ultraviolet divergencies of 
a gauge theory in the bulk,  one only needs to use the small gauge transformation Ward identities in the bulk.  The latter are derived by performing the change of field variable 
  \begin{equation}
  \label{change_variable_trivial}
  \vp (x) \to \vp'(x) =  \vp(x) + \kappa   s_0 \vp (x) 
 \end{equation}
 in the generating functional \eqref{exp_W} at $\alpha=0$.  In \eqref{change_variable_trivial},  $\kappa$ is an anticommuting constant that does not vanish on the boundaries and $s_0$ is the BRST operator generating small gauge transformations, namely leaving invariant the boundary conditions of the path integral by definition.  Repeating all the previous steps for this change of variable leads to 
   \begin{equation} 
   \label{wi_bulk_trivial}
      0 = \int_M   \Big (   
      \frac {\delta \Gamma}{ \delta \vp(x)}
        \frac {\delta \Gamma}{ \delta v_ {\vp}(x)} 
           +
         j _{  B}(x)
         \frac {\delta \Gamma}{ \delta v_ {B }(x)}
   -
     d     \frac{\delta \Gamma}{\delta j_{B  }(x)}  \Big)   .
      \end{equation}
This equation is  nothing but  the standard  BRST--BV  bulk master equation,   distorted by  the addition of  the sources $j_B$ and $v_B$, 
  which  originates from the  boundary refinement $s_0 L~=~d B_0 \neq 0$ that implies the introduction of a source for the relevant local operator $B_0$.   To understand why these new terms do not alter the renormalization procedure, it is interesting to invert \eqref{wi_bulk_trivial} at tree level,  providing
  \begin{equation}
  0 = \int_M \Big( s_r L - d B_0 + 2 j_B s_r B_0 + v_\vp s_r^2 \vp - v_B s_r^2 B_0     \Big) . 
\end{equation}   
This equation must be valid for all values of the sources and therefore \textit{defines} a nilpotent operator $s_r$ which is a symmetry of the tree level Lagrangian, namely
\begin{align}
\label{tree-level_renormalize}
s_r^2 \vp &= 0 ,
\nn \\
s_r L &= d ( B_0 + B' ) ,
\nn \\
s_r B_0 &= 0 .
\end{align}
It is interesting to notice that the last equation is consistent with the condition $s^2 L^0_4 = -d s B^1_3 = 0$, leading to the cocycle equations $sB^{i}_{d-i}=dB^{i+1} _{d-i-1}$ for $i>0$ given in \eqref{B_cocycle}. To properly take it into account, one needs to add sources for all the  $B^{i+1} _{d-i-1}$'s and their BRST variations.  Also, the just built operator $s_r$  reproduces the variation $s_0 L = dB_0$ only up to another boundary term $B'$.  This is not a problem since the local operator $B_0$ might not be multiplicatively renormalized.  Indeed, perturbatively imposing \eqref{wi_bulk_trivial} provides a renormalized nilpotent BRST operator $s_r$, which is a symmetry of the renormalized Lagrangian $L_r$  and  only differs from the tree level  operation $s$ by consistent $Z$ factors in its transformation laws.  Actually, the $j_B$ and $v_B$ dependence of  \eqref{wi_bulk_trivial} allows one to take care of the necessary renormalization of the ultraviolet divergences of $B_0$, providing $B_r$ such that $s_r L_r = d B_r$.  It is important to realize that there are no reasons for a multiplicative renormalization of the form $B_r = Z_0 B_0$ to hold, meaning that the $B'$ term in \eqref{tree-level_renormalize} is really not a concern. In any case, the ultraviolet renormalized BRST Noether current will be build out of $B_r$ and not $B_0$.  

All this actually means that \eqref{wi_bulk_trivial} implies the usual Ward identity 
\begin{equation}
0 = \int_M   \Big (   
      \frac {\delta \Gamma}{ \delta \vp(x)}
        \frac {\delta \Gamma}{ \delta v_ {\vp}(x)}  \Big) ,
\end{equation}
which completely defines the action of the renormalized BRST symmetry $s_r$ on the fields.

Taking \eqref{wi_bulk_trivial} as a guideline and remembering that whatever  the field boundary conditions are, 
$\Gamma(\vp, v_\vp,  j_B,v_B,\alpha)$ undergoes the same renormalization scheme as  when 
 the  BRST Noether current  was not introduced, \eqref{WI_2} must be split as 
 \begin{align}
 \label{Bulk_WI_final}
 \text{SGT} \leftrightarrow \text{Bulk\ Ward\ Identity:}&    \   \int_R    \Big (   
      \frac {\delta \Gamma}{ \delta \vp(x)}
        \frac {\delta \Gamma}{ \delta v_ {\vp}(x)} +  j _{  B}(x)
         \frac {\delta \Gamma}{ \delta v_ {B }(x)} \Big)
         -  \int_{i^0} \frac{\delta \Gamma}{\delta j_{B  }(x)} = 0 ,
  \\ 
  \label{Boundary_WI_final}
 \text{LGT} \leftrightarrow \text{Boundary\ Ward\ Identity:}&  \   
    \int_{\Sigma^+} \Big(   \frac  {\delta \Gamma}{\delta \al(x)}  -  \frac{\delta \Gamma}{\delta j_{B  }(x)}   \Big) -  \int_{\Sigma^-} \Big(   \frac  {\delta \Gamma}{\delta \al(x)}  -  \frac{\delta \Gamma}{\delta j_{B  }(x)}   \Big)  = 0   .
 \end{align}
 
 The bulk Ward identity \eqref{Bulk_WI_final} is readily understood  by pushing $\Sigma^\pm$ arbitrarily close to $\mathcal{I}^\pm$. In this case, the region $R$ tends to  the full spacetime $M$, so that \eqref{Bulk_WI_final} is simply \eqref{wi_bulk_trivial} in which the boundary term is not defined on the full boundary $\pa M$ but rather on $\pa M \cap R = i^0$. Therefore, by the previous arguments, \eqref{Bulk_WI_final} provides a renormalization of the theory in the bulk.  This procedure entirely comes from the change of variables \eqref{change_variable_trivial} that involves small gauge transformations (SGT) only, which is the reason why \eqref{Bulk_WI_final} is referred to as the Ward identity for SGT.

Once the bulk identity \eqref{Bulk_WI_final} is imposed,  the boundary Ward identity \eqref{Boundary_WI_final} can be enforced.  In fact, this bulk-boundary decomposition is automatically implemented in the full identity \eqref{WI_2}. Indeed, the boundary piece obviously expresses that the on-shell conserved  BRST Noether current at tree level is as in formula \eqref{NT1}; but by \eqref{Noether_1.5} and \eqref{def_s-large}, the only non vanishing part of this boundary integral is for ghosts corresponding to large gauge transformations (LGT).  Both Ward identities \eqref{Bulk_WI_final} and \eqref{Boundary_WI_final} are therefore mutually independent. 
At loop level,  the boundary Ward identity might be broken by consistent anomalies.  
Such potential breaking of the large gauge symmetry
 is harmless for the physical interpretation of the theory,  and would simply express the necessity of computing the  quantum  corrections to the Noether current,  order by order in perturbation theory.\footnote{Analogous harmless anomalies are actually welcome in the case e.g. of the  anomaly of the gauge current algebra of Yang--Mills symmetry,   as a  key reason for    the non zero mass of the pion.}
      
To link \eqref{Boundary_WI_final} more precisely with asymptotic symmetries, one has to be careful with the integration over $\Sigma^\pm$.  If one works in retarded Finkelstein--Eddington coordinates in four dimensions $(u = t - r, r,z,\bar{z})$,   $i^0$ and $i^+$ are just points in the Penrose diagram Fig [\ref{figure_1}]. This means that when pushing $\Sigma^+$ to $\mathcal{I}^+$,  which is topologically $S^2 \times \mathbb{R}$, the integral over $\Sigma^+$ writes 
\begin{equation}
\label{Sigma_integration}
\int_{\Sigma^+} d^3 x = \int_{-\infty}^{+ \infty} du \int_{S^2} d\O , 
\end{equation}
where $d\O$ is the measure on the two sphere.  The important point is that since $\Sigma^+ \cup \Sigma^-$ does not contain $i^0$ by definition,  the integration bound at $u=-\infty$ in \eqref{Sigma_integration} coincides with $\mathcal{I}^+_-$, which is precisely where the charges that generate the action of the asymptotic symmetry group on $\mathcal{I}^+$ are defined.  The integration bound at $u=+\infty$ is not as well defined as it differs from $\mathcal{I}^+_+$ and actually cuts $i^+$. This does, however,  allow massive  particles to be incorporated.  For a fully rigorous treatment of the fives boundaries $\pa M = i^+ \cup \mathcal{I}^+ \cup i^0 \cup \mathcal{I}^- \cup i^-$ of spacetime,  in which the region $R$ can be better defined, see \cite{Compere:2023qoa}.

Using \eqref{NC_source}, the functional boundary Ward identity \eqref{Boundary_WI_final} provides all possible constraints for  the correlation functions between  arbitrary number of physical observables $\mathcal{O}$'s  and   one insertion of   the BRST Noether current $\star J_{\rm BRST}$  at the boundary under the form\footnote{The minus sign comes from the different orientations of $\Sigma^+$ and $\Sigma^-$.}  
\begin{equation}
      \label{Noether_ward_identity}
      \left\langle \left( \int_{\Sigma^+} \J \right)    \mathcal{O}_1 ... \ \mathcal{O}_n    \right\rangle - \left\langle  \mathcal{O}_1 ... \ \mathcal{O}_n  \left( \int_{\Sigma^-} \J \right) \right\rangle = 0 . 
      \end{equation}
The physical observables are defined as elements of the cohomology of $s_0$ with ghost number zero.  The integrated Noether currents directly act on the vacua because the insertions of these currents  occur   before and after the insertions of the $\mathcal{O}$'s,  as advocated in Fig [\ref{figure_1}]. 
In order to enforce this constraint in a gauge invariant way and relate it to asymptotic symmetries,  getting that the part of  $\J$ that depends on the gauge choice is  an $s$-exact term is a necessary requirement.  In fact,  using the conjectured Noether 1.5th theorem \eqref{Noether_1.5}, the properties of correlation functions
\begin{equation}
\left\langle ( s G )    \mathcal{O}_1 ... \ \mathcal{O}_n    \right\rangle  = 0 ,  \quad  \quad \left\langle f(\text{EoM})   \mathcal{O}_1 ... \ \mathcal{O}_n   \right\rangle  = 0
\end{equation}
and applying LSZ reduction formula to  \eqref{Noether_ward_identity} leads to
      \begin{equation}
      \label{charge_ward_identity}
      \bra{\rm out}  [ Q,\mathcal{S} ]  \ket{\rm in} = 0 ,
      \end{equation}
with the definition $\left. \left\langle      \mathcal{O}_1 ... \ \mathcal{O}_n    \right\rangle \right\vert_{\rm LSZ} = \bra{\rm out}   \mathcal{S}  \ket{\rm in}$.  A few comments are in order.  The commutator in \eqref{charge_ward_identity} is defined as $[Q,\mathcal{S} ] \equiv Q^+ \mathcal{S} -  \mathcal{S} Q^-$ where $Q^\pm \equiv \int_{\Sigma^\pm} d q_{\rm cl}$.  
This is where the fact that the lower integration bound on $u$ in \eqref{Sigma_integration} corresponds to $\mathcal{I}^+_-$ is crucial and actually guarantees that the finite part of $Q^+$ is truly the charge generating asymptotic symmetries on $\mathcal{I}^+$.  Let us also mention that any ``holographic renormalization" procedure consisting of adding a boundary term to the Lagrangian in order to make the charge $Q^+$ finite is consistent with these Ward identities. One simply needs to find this boundary term $l$, add it to the Lagrangian $L \to L^{\rm HR} = L + d l$, write the new  local operator $B^{\rm HR} = B - s l$ such that $sL^{\rm HR} = d B^{\rm HR}$ and start again with the generating functional $W_c$ written with these new quantities.  
Noether's 1.5th must still be valid in this case and the new finite charges generate the same asymptotic symmetry because of \eqref{boundary_shift_fund}.

One must insist on the fact that   the Ward identity \eqref{charge_ward_identity} carries ghost number one. 
However,    $q_{\rm cl}$ is linear in the ghost fields $c(x)$ and  a ghost number zero  identity  can be computed by applying the operation $\int dy \lambda(y) \frac {\delta}{\delta c(y)}$ to \eqref{Noether_ward_identity}  where $\lambda(y)$ is the ghost number zero infinitesimal parameter of a gauge transformation associated with $c(y)$.  This is only at this step of the derivation that one needs to impose the antipodal matching condition of the ``large"  ghost field components $\left.c(z,\bar{z})\right\vert_{\mathcal{I}^+_-} = \left.c(z,\bar{z})\right\vert_{\mathcal{I}^-_+}$.
The standard antipodal matching of the large gauge symmetry parameter $\left.\lambda(z,\bar{z})\right\vert_{\mathcal{I}^+_-} = \left.\lambda(z,\bar{z})\right\vert_{\mathcal{I}^-_+}$ of \cite{Strominger_BMS_scattering} naturally follows. 
One~can afterward set the ghost field sources  to zero  by performing analogous manipulations 
as those used  when   deriving   standard 1PI correlation function Ward identities from the  ghost number one bulk  BRST Ward identity 
$\{ \Gamma , \Gamma \} = 0$. 

Another subtle step is needed to   obtain   non trivial constraints on all physical $\mathcal{S}$-matrix elements such as \eqref{charge_ward_identity}.
 Indeed,    when one uses    the LSZ reduction formula  the states $\ket{\rm in}$ and $\ket{\rm out}$ are not necessarily in the same asymptotic symmetry frame of the degenerate~vacuum.
  Another matching condition is needed  in order to equate these two frames and allows one to derive  soft theorems from  \eqref{charge_ward_identity} using standard perturbative quantum field theory techniques.  This subtlety is discussed in \cite{He:2020ifr} by considering  scattering amplitudes between $\ket{\rm in}$ and $\ket{\rm out}$ states that belong to different vacuum frames.

   This series of steps therefore provides a  perturbative  Lagrangian  proof of the gauge invariance of the Hamiltonian Ward identity \eqref{charge_ward_identity}, which expresses the invariance of the $\mathcal{S}$-matrix under asymptotic symmetries while being fully compatible with the quantum master equation in the bulk.  This proof  is one of the major results of this paper.  The~formalism   developed   above   also allows one to properly treat potential loop corrections to \eqref{charge_ward_identity}. This is the subject of the next subsection.

\subsection{Anomaly considerations}
\label{Holographic_anomalies}

 \def\e{{\epsilon_{uv}}}

As soon as one finds an identity of the form \eqref{charge_ward_identity} with non vanishing charges, there is an associated soft theorem \cite{Strominger_lectures}.  However, very little is known about the all loop validity of \eqref{charge_ward_identity}.  One actually infers the answer by looking at the possible loop corrections to the corresponding soft theorem.  The formalism developed in the previous subsection allows one to bypass the use of soft theorems and treat the all loop behavior of \eqref{charge_ward_identity} on its own by looking for consistent anomalies of the boundary Ward identity \eqref{Boundary_WI_final}. 

 The  whole Ward identity  \eqref {WI_2} is to  generate all relevant  relations to be enforced  in  explicit computations.  
As already mentioned,  one must perform ultraviolet and infrared regularizations and an appropriate ultraviolet renormalization.
The infrared regularization involves a regulator $\epsilon$ with a physical interpretation, that of a very low energy scale under which detectors are not able to distinguish between two different events.
 Here one supposes that    the   theory has no short distance anomaly,  otherwise it would be meaningless to even begin with and to try to define  a BRST Noether current.
This means that at any given order in perturbation theory,  one can add counterterms such that the generating functional $\Gamma^{\rm r}  (\vp, v_\vp ,     \epsilon)$ is finite and ultraviolet renormalized, and  
where  the infrared divergencies are regulated by the infrared regulator $\epsilon$.    For the sake of simplicity,  one  denotes
               $\Gamma^ {\rm r}(\vp, v_\vp ,     \epsilon)$ as~$\Gamma$.
     
More precisely,  having no ultraviolet anomaly at order $n-1$ of perturbation theory in the bulk means that  one has been able to  compute a renormalized $\Gamma^{n-1}$ satisfying the bulk Ward identity 
\begin{equation}
\label{trivial_BRST_WI}
\int_M   
       \sum_{p+q=n-1} \left( \frac {\delta \Gamma^p}{ \delta \vp(x)}
        \frac {\delta \Gamma^{q}}{ \delta v_ {\vp}(x)} \right)   = \mathcal{O}(\hbar^n) . 
     \end{equation}
The  only thing that may happen at order $n$ of perturbation theory and provide an irreducible gauge anomaly is when the further  renormalization at 
order $n$  is broken  by  a local $d$-form  with ghost number one  $\Delta^1_d (\vp)$, such that 
\begin{equation}
\label{wiwwwwano} 
\int_M   
       \sum_{p+q=n} \left( \frac {\delta \Gamma^p}{ \delta \vp(x)}
        \frac {\delta \Gamma^{q}} { \delta v_ {\vp}(x)} \right)
    = \mathfrak{a}^n \hbar ^n \int_{M}   \Delta^1_d (\vp) + \mathcal{O}(\hbar^{n+1}) . 
\end{equation}
The local form $\Delta^1_d (\vp)$ is further constrained by making use of the graded operation
 \begin{equation}
 \label{symplectic_bracket}
\mathcal{S}_\Gamma =  \{ \Gamma , \bullet \} \equiv \int_M  \frac {\delta \Gamma}{ \delta \vp(x)}
        \frac {\delta  }{ \delta v_ {\vp}(x)}  +  \frac {\delta \Gamma}{ \delta v_ {\vp}(x)}  \frac {\delta }{ \delta \vp (x)}   ,
 \end{equation}
 which satisfied $\mathcal{S}_\Gamma^2 \Gamma = \big\{ \Gamma ,   \{   \Gamma,  \Gamma  \}  \big\} = 0$ by the Jacobi identity of the  bracket.  Notice that the zero$^{\rm th}$ order
 action of $\mathcal{S}_\Gamma$ on fields reproduces that of the renormalized BRST operation.  Then,  the Ward identity \eqref{trivial_BRST_WI} simply writes
 \begin{equation}
 \label{bulk_WI_symplectic_bracket}
  \sum_{p+q=n-1} \mathcal{S}_{\Gamma^p} \Gamma^{q} = \{ \Gamma , \Gamma \}^{n-1} = \mathcal{O}(\hbar^n) . 
 \end{equation}
 Applying the operation  $\mathcal{S}_\Gamma = \sum_{k=0}^{n} \{  \Gamma^k, \bullet \}$ on both sides of \eqref{wiwwwwano} leads to the consistency condition for a possibly non vanishing bulk local anomaly\footnote{All the $\Gamma^k$'s for $k \leq n-1$ have already been made finite by the renormalization procedure at the lowest orders, so the only potentially divergent and non trivial piece at order $n$ is contained in $\mathcal{S}_{\Gamma^0}^2 \Gamma^{n}=0$.}
 \begin{equation}
 \label{bulk_anomaly}
 \mathcal{S}_{\Gamma^0} \int_M  \Delta^1_d (\vp) = 0 \quad \Longrightarrow \quad s \Delta^1_d = d \Delta^2_{d-1} ,
 \end{equation}
 if $\Delta^2_{d-1}$ vanishes on $\pa M$. A trivial solution for this equation is given by $\Delta^1_d =  s \Delta^0_d +  d \Delta^1_{d-1}$ if $\Delta^1_{d-1}$ vanishes on the boundary of $M$.  For such solutions,  $\Delta^1_d$ can be eliminated from \eqref{wiwwwwano} by adjusting the local counterterms that  one uses when going from  order $n-1$ to  order $n$ of perturbation theory.  Any other solutions of \eqref{wiwwwwano} are consistent and cannot be perturbatively eliminated. 
In this case,  $\mathfrak{a}$  is a model dependent  constant that can be computed in perturbation theory.  If $\mathfrak{a} \neq 0$ at any given order  of  perturbation theory,   one has a gauge anomaly beyond this level and the theory breaks down.

The significance of  having a theory with no bulk  ultraviolet anomaly is therefore that either the cocycle equation \eqref{bulk_anomaly} has no non trivial solution or, otherwise,   its perturbatively computable  coefficient $\mathfrak{a}$ vanishes  to all  orders.\footnote{ 
See   
 \cite{Barnich:2000zw} for 
a  derivation  of the 
general solution of \eqref{bulk_anomaly} in Yang--Mills gauge theory. }
It is important to realize that adding the sources $j_B$ and $v_B$ to $\Gamma$ does not impact the renormalization procedure in the bulk, as explained around   \eqref{tree-level_renormalize}, and therefore cannot impact the ultraviolet anomaly condition \eqref{bulk_anomaly}.  Thus, assuming that \eqref{trivial_BRST_WI} is valid at all loops leads to \eqref{Bulk_WI_final} being valid at all loops as well.

Now comes the question of determining the   consistency equation of a possible anomaly that could correct   the boundary Ward identity \eqref{Boundary_WI_final}  when there is no bulk anomaly.  In this situation, one would have
 \begin{equation}
 \label{potential_anomaly}
 \int_{\Sigma^+ - \Sigma^-}  \ \left(   
      \frac  {\delta }{\delta \al(x)} - \frac{\delta }{\delta j_{B  }(x)} \right) \Gamma^{n}
 = \mathfrak{a}^n_{\rm boundary} \hbar^{n} \int_{\Sigma^+ - \Sigma^-} \Delta^1_{d-1} (\vp)  +  \mathcal{O}(\hbar^{n+1}) . 
 \end{equation}
 One may call   $\Delta^1_{d-1}$ an ``holographic anomaly" because it is a codimension one object which is entirely determined by the symmetry group of the bulk theory and which satisfies the following anomaly consistency condition 
  \begin{equation}
 \label{holographic_anomaly}
 \mathcal{S}_{\Gamma^0} \int_{\Sigma^+ - \Sigma^-}  \Delta^1_{d-1} (\vp) = 0 \quad \Longrightarrow \quad s \Delta^1_{d-1} = d \Delta^2_{d-2} 
 \end{equation}
 for a vanishing $\Delta^2_{d-2}$ on $\pa \Sigma^\pm$.  In particular, it gives the correct interpretation for the cocycle of extended BMS symmetry found in \cite{Barnich_BRST,Baulieu_Tom_BMS}.  To derive \eqref{holographic_anomaly}, one computes 
 \begin{align}
  \mathcal{S}_\Gamma \int_{\Sigma^\pm} \frac  {\delta \Gamma }{\delta \al(x)} &= \int_M  \int_{\Sigma^\pm} \left( \frac {\delta \Gamma}{ \delta \vp(y)}
        \frac {\delta^2 \Gamma  }{ \delta v_ {\vp}(y) \delta \al(x) }  +  \frac {\delta \Gamma}{ \delta v_ {\vp}(y)}  \frac {\delta^2 \Gamma }{ \delta \vp (y) \delta \al(x)}  \right) 
        \nn \\
        &= \int_M  \int_{\Sigma^\pm} \left( \frac {\delta^2 \Gamma  }{  \delta \al(x) \delta v_ {\vp}(y) } \frac {\delta \Gamma}{ \delta \vp(y)}
          +  \frac {\delta \Gamma}{ \delta v_ {\vp}(y)}  \frac {\delta^2 \Gamma }{ \delta \al(x) \delta \vp (y) }  \right) 
          \nn \\
          &= - \demi \int_{\Sigma^\pm} \frac {\delta   }{  \delta \al(x) }  \{ \Gamma , \Gamma \} = 0 ,
 \end{align}
 which is also valid when $\frac  {\delta \Gamma }{\delta \al(x)} $ is replaced by   $\frac  {\delta \Gamma }{\delta j_B(x)}$.
 The different statistics of the fields and the sources have been used, as well as the assumption that no anomaly occurs in the bulk, namely that  \eqref{bulk_WI_symplectic_bracket} holds true at all order of perturbation theory. 
 
A fundamental difference between the holographic anomalies  possibly appearing in the right hand side of  \eqref{potential_anomaly} and the bulk one in \eqref{wiwwwwano} is that the former can be eliminated  by perturbative redefinitions of the  BRST Noether current.  The physical consequences of such a redefinition are perfectly admissible perturbative corrections to the value of the global corner charges of the gauge theory due to quantum effects.  The  linear  $\Gamma$ dependence  of   the boundary anomaly equation  \eqref{potential_anomaly}  makes it  physically  not destructive if   an  anomalous term  occurs,   in opposition to the consequences of the occurrence of a bulk  anomaly in  \eqref{wiwwwwano} due to its nonlinear dependence in   $\Gamma$.
           
Let us make this reasoning explicit. Suppose that when passing from  the tree level to the one loop level,  \eqref{Noether_ward_identity}  is broken as in \eqref{potential_anomaly} for $n=1$ with a non zero value for~$\mathfrak{a}^{1}_{\rm boundary}$.
     Then,  one  can  put    the higher order equation   \eqref{potential_anomaly}   under the same form as the boundary anomaly free tree level equation 
         \eqref{Noether_ward_identity} by a redefinition of the BRST Noether current at~1-loop
         \begin{equation}
         \label{renormJ}
\J \longrightarrow \J' = \J
- \mathfrak{a}^{1}_{\rm boundary}  \hbar  
  \Delta^1_{d-1}  (\vp) . 
  \end{equation}
    Indeed,  by the same argument  leading to the fact that   \eqref{Boundary_WI_final} implies    \eqref{Noether_ward_identity},  one  finds that 
  \eqref{potential_anomaly} implies
        \begin{equation}
        \label{fakuanom}
\left\langle \left( \int_{\Sigma^+} \J' \right)    \mathcal{O}_1 ... \ \mathcal{O}_n    \right\rangle\bigg\vert_{1-\text{loop}} - \left\langle  \mathcal{O}_1 ... \ \mathcal{O}_n  \left( \int_{\Sigma^-} \J' \right) \right\rangle\bigg\vert_{1-\text{loop}} = 0 . 
\end{equation}
It follows that  the physical interpretation  and consequences   of the boundary Ward identity  \eqref{potential_anomaly} are  the same as those  from  \eqref{Boundary_WI_final} but with a loop corrected Noether current. 

In particular,  the loop effect \eqref{renormJ} can be interpreted as   a loop correction    
to    the  classical global charge $Q=\int_{\pa \Sigma} q_{\rm cl}$ of the gauge theory.     Indeed,  if $\Delta ^1_{d-1}$ is  $d$-exact,  say $\Delta ^1_{d-1} = d \Delta^1_{d-2}$ with $\int_{\pa \Sigma} \Delta^1_{d-2} \neq 0$, its  presence can be absorbed in a further possible redefinition of the  conserved BRST Noether charge $Q$ entering in \eqref{charge_ward_identity}, such that 
\begin{equation}
\label{Charge_loop_correction}
Q \longrightarrow Q' = \int_{\pa \Sigma} \Big( q_{\rm cl} - \mathfrak{a}^{1}_{\rm boundary}  \hbar  \Delta^1_{d-2} \Big)
\end{equation}
 and 
 \begin{equation}
 \label{corrected_WI_charges}
   \bra{\rm out}  [ Q',\mathcal{S} ]  \ket{\rm in}\Big\vert_{1-\text{loop}} = 0 .
 \end{equation}
 Therefore,  \eqref{potential_anomaly} and \eqref{holographic_anomaly} provide a systematic approach to treat the invariance of the quantum $\mathcal{S}$-matrix, namely at all loop, under asymptotic symmetries without ever referring to soft theorems.  If the cohomology $H^1(s_{\rm large} \vert d)$ is empty,  namely if there are no non trivial solutions to \eqref{holographic_anomaly}, one can claim that the large gauge transformations generated by $s_{\rm large}$ are exact symmetries of the $\mathcal{S}$-matrix of the full quantum theory.  On the contrary, if there exists a non trivial $\Delta^1_{d-1}$, this could indicate a potential correction of the boundary Ward identity of the form \eqref{fakuanom}.  In this case,  one has to go through the computation of the potentially anomalous Feynman diagrams to determine the value of the model dependent anomaly coefficient $\mathfrak{a}_{\rm boundary}$.  If this coefficient turns out to be zero,  the non trivial cocycle $\Delta^1_{d-1}$ does not impact the loop validity of the boundary Ward identity. 
This basically reduces to compute the soft factors of the soft theorems associated with \eqref{charge_ward_identity} at loop level and check whether or not they are different from the ones at tree level.  This methodology will be used in Section [\ref{Section_soft_graviton}] to study the invariance of the quantum gravity $\mathcal{S}$-matrix under extended BMS symmetry and its link with the leading and subleading soft graviton theorem.

\section{Yang--Mills}
\label{Section_Yang_Mills}

This work has not yet provided a technical clue as to why the on-shell value of the BRST Noether current should take the form \eqref{Noether_1.5}.  Since this is the cornerstone of all the gauge invariant results of the previous sections, this section tackles the problem by applying the formalism developed in Section [\ref{Section_trigraded}] to Yang--Mills theory with different gauge fixing choices.  One will check the validity of the postulates \eqref{Noether_1.5} and \eqref{claim_Z} in these cases.   
One must however insist on the fact that not having a general proof of \eqref{Noether_1.5} is frustrating since this property is fundamental to generically ensure that \eqref{charge_ward_identity} holds true for all BRST invariant theories with asymptotic symmetries.


\subsection{Trigraded computations in various gauges}

Here,  one applies the concepts introduced in Sections [\ref{Section_adding_ghosts}]-[\ref{subsection_BRST_Noether}] to the gauge fixed Yang--Mills theory case. 
In fact, it was only after  having done such explicit computations that this work was to establish the generic pattern exposed in Section [\ref{subsection_gauge_fixed_fund}]. The technical details presented in this section are to familiarize the reader with most of the manipulations that will be used in the rest of the paper.  In particular, they make quite clear the main differences between the relevant computations  in the ungauge and gauge fixed cases.


Let $A \in \O_{\rm loc}^{0,1,0}(\tilde{\mathcal{F}} \times M)$ and $c \in \O_{\rm loc}^{0,0,1}(\tilde{\mathcal{F}} \times M)$ be a non abelian Yang--Mills gauge field and its associated ghost valued in a $\mathfrak{su}(n)$ Lie algebra.  The nilpotent BRST transformations of these fields are given by 
\begin{align}
\label{delta_BRS}
&L_{V} A =  I_{V} \delta A = s A = - Dc  , 
\nn \\
&L_{V} c = I_{V } \delta c = s c = - \demi [c,c] .
\end{align}
$D$ is the gauge covariant derivative $D(\bullet) \equiv d (\bullet) + [A,\bullet]$ and the commutator $[\bullet,\bullet]$ is taken with the fully antisymmetric structure constants $f_{abc}$ of $\mathfrak{su}(n)$.  The commutator $[c,c]$ in \eqref{delta_BRS} thus reads 
\begin{equation}
sc^a = -\demi f^a_{\ bc} c^b c^c ,
\end{equation}
where $a,b,c,...$ always refer to Lie algebra indices. In what follows,  the Lie algebra indices will consistently be omitted since they will be traced over in every relevant quantity.

The equivalence of $s$ and $L_{V}$ when acting on such simple local forms entirely comes from the $I_{V} \delta $ part. Let us however consider the action of $L_V$ and $s$ separately on $1$-forms in field space to convince the reader that the defining property $[\delta,s]=0$ actually imposes $s=L_V$ and simplify the computation of its action on local forms. Consider for instance $c \delta A \in \O_{\rm loc}^{1,1,1}(\tilde{\mathcal{F}} \times M)$.  Compute then $L_{V} (c \delta A)$ for $V=V_{\rm BRST}$ given by \eqref{V_BRST} and $s (c \delta A)$ using \eqref{nil3}, that is
\begin{align}
\label{ex_claim1}
&L_{V} (c \delta A) = I_{V} ( \delta c \delta A ) + \delta ( c Dc ) = -\demi [c,c] \delta A - \cancel{\delta c Dc} +  \cancel{ \delta c Dc} - c \delta (Dc)  
\end{align}
and
\begin{align}
\label{exemple_s}
&s (c \delta A) = sc \delta A + c \delta (sA) = -\demi [c,c] \delta A - c \delta (Dc).
\end{align}
The two expressions obviously coincide.  This example was just a consistency check on how computations are done in practice using either $L_{V}$ or $s$.  One  understands that it is easier to use $s$ in such simple cases and one  will therefore continue to do so in the rest of the paper, unless stated otherwise.


One can now turn to the computation of the local symplectic potential $\theta$ defined in \eqref{split_theta_gauge} for the gauge fixed Yang--Mills theory. 
In order to fix the gauge, one also needs to introduce antighosts $\bar{c} \in \O_{\rm loc}^{0,0,-1}(\tilde{\mathcal{F}} \times M)$ and their associated Lagrange multipliers $b \in \O_{\rm loc}^{0,0,0}(\tilde{\mathcal{F}} \times M)$, which form a trivial BRST doublet
\begin{align}
\label{trivial_doublet}
s\bar{c} &= b,
\nn \\
sb &= 0.
\end{align}
Recall that a field with ghost number $g$ and spacetime form degree $p$ is denoted $\vp^g_p \in \O_{\rm loc}^{0,p,g}(\tilde{\mathcal{F}} \times M)$.
These are all $0$-forms in field space.  In fact, the field space form degree   always appears through the explicit action of $\delta$ on field space $0$-forms.  The statistics of such local forms is given by \eqref{grading_1}.

\subsubsection{Lorenz gauge with a $(d+s)$-exact gauge fixing term}
\label{subsection_d_s-exact}

Consider first the Yang--Mills case in 4 dimensions when  the   Lorenz gauge  $\pa_\mu A^\mu = 0$  is fixed using the BV setup.
It  introduces both an $s$-exact and a $d$-exact term to the classical Lagrangian.  This means that the boundary term $B$ coming from $sL=dB$ gets a non vanishing contribution $B_{\rm gauge}$.  Everything will be consistent with \eqref{vanishing_q_gauge} by replacing $I_V \theta_{\rm gauge} \to I_V \theta_{\rm gauge} - B_{\rm gauge}$, showing the robustness of the result. 

In fact,  when using 
 the BV formalism to build the BRST invariant gauge fixed Lagrangian for the Lorenz gauge, one must
introduce antifields ${}^* \vp$ for each fields $\vp=\{A,c,\bar{c},b \}$ with grading given by
\begin{equation}
{}^* (\vp^g_p) = ({}^* \vp)^{-g-1}_{4-p} .
\end{equation}
The gauge fixed BV Lagrangian for Yang--Mills theory is
\begin{equation}
L = - \demi F  \star F - {}^*A D c - \demi {}^*c [c,c] + {}^*\bar{c} \ b + s\Big( \bar{\Psi}_{{}^*\vp} \big( {}^*\vp - \frac{\delta Z}{\delta \vp}  \big)  \Big) 
\end{equation} 
where $F = dA + \demi [A,A]$ and $Z = \bar{c}   d (\star A)$ is the gauge fixing functional with $\star$ the Hodge dual operator for spacetime differential forms.  An implicit summation is taken over all fields $\vp$. The trivial BRST doublet $(\bar{\Psi}_{{}^*\vp} , B_{{}^*\vp})$ transforms as \eqref{trivial_doublet} with grading 
\begin{equation}
(\bar{\Psi}_{{}^*\vp})^g_p  \quad \rm{and}   \quad (B_{{}^*\vp})^{g+1}_p
\end{equation}
for $\vp = \vp^g_p$.  After imposing the equations of motion of $B_{{}^*\vp}$,  one gets 
\begin{equation}
\label{L_BV_YM}
L = - \demi F  \star F  + b  d (\star A)  + Dc (\star d) \bar{c} .
\end{equation}
The algebraic equation of motion of the Lagrange multiplier field $b$ imposes
the Lorenz gauge fixing condition $d (\star A) = 0$.  In these equations,   the color indices are not written explicitly but an implicit trace is taken over them.  The exterior products $\w$ and $\curlywedge$ in spacetime and field space are also omitted since they are obvious when multiplying local forms.  In these notations, one has 
\begin{equation}
- \demi F  \star F = - \frac{1}{4}  \text{Tr}(F_{\mu \nu} F^{\mu \nu})d^4x.
\end{equation}
For any forms $\o,\o'$ in $\O_{\rm loc}^{q,1,g}(\tilde{\mathcal{F}} \times M)$, one also has the four dimensional identity 
\begin{equation}
\label{hodge_spacetime}
(\star \o ) \o' = -\o (\star \o').
\end{equation}
Two other identities are to be  used in the following computations.
Given any $\o_1 , \o_2, \o_3$ in $\O_{\rm loc}^{q,p,g}(\tilde{\mathcal{F}} \times M)$  valued in $\mathfrak{su}(n)$, one can perform integration by parts on traces in $\mathfrak{su}(n)$ by using the identity
\begin{equation}
\label{graded_IPP}
\Tr(\o_1 D \o_2) =  (-1)^{\mathfrak{g}(\o_1)+1} \Tr((D \o_1) \o_2) + (-1)^{\mathfrak{g}(\o_1)} d\Tr(\o_1 \o_2)
\end{equation}
and commute traces of commutators by using 
\begin{align}
\label{traces_commutation}
\Tr(\o_1 [\o_2,\o_3]) &= f_{abc} \o_1^a \o_2^b \o_3^c = - f_{abc} \o_1^b \o_2^a \o_3^c = (-1)^{\mathfrak{g}(\o_1) \mathfrak{g}(\o_2) + 1} f_{abc} \o_2^a \o_1^b \o_3^c 
\nn \\
&= (-1)^{\mathfrak{g}(\o_1) \mathfrak{g}(\o_2) + 1} \Tr(\o_2 [\o_1,\o_3]). 
\end{align}

By using \eqref{hodge_spacetime}, \eqref{graded_IPP}, \eqref{traces_commutation} and the graded commutation rules of $(\O_{\rm loc}^{\bullet,\bullet,\bullet}(\tilde{\mathcal{F}} \times~M),\delta,d,s)$, 
one can rewrite the basic relation $\delta L = E + d \theta$ for the Lagrangian \eqref{L_BV_YM} as follows: 
\begin{align}
\label{delta_L_YM}
\delta L &=  - \delta A \Big( D \star F     + (\star d) b   + [ (\star d) \bar{c} ,  c]  \Big)  - \delta b \Big(  (\star d) A   \Big) - \delta \bar{c}  \Big(  d (\star D)c \Big) + \delta c \Big(    D (\star d) \bar{c} \Big)
\nn \\
&\quad +  d \Big(   \delta A \star	F - b \star \delta A + \delta \bar{c} (\star D )c - \delta c (\star d) \bar{c}   \Big).
\end{align}
The local symplectic potential for the Lorenz gauge fixed Yang--Mills theory  in the BV formalism is therefore
\begin{equation}
\label{theta_YM_BV}
\theta = \delta A \star	F - b \star \delta A + \delta \bar{c} (\star D) c - \delta c (\star d) \bar{c} 
\end{equation}
and the symplectic two form is  
\begin{align}
\O &= \int_{\Sigma_3} \delta A  \delta (\star F) - \delta b \star \delta A - \delta \bar{c} \big( (\star D) \delta c - [\star \delta A, c]  \big) + \delta c ( \star d ) \delta \bar{c}    .
\end{align}
Before proceeding to the computation of $I_{V} \O$, one can determine
 the BRST Noether current \eqref{NT1} associated with \eqref{L_BV_YM} and check the validity of \eqref{Noether_1.5}.  
So one must    compute  $B$ for the Lagrangian \eqref{L_BV_YM}  such that $sL = dB$.  
One gets 
\begin{equation}
\label{B_YM_BV}
B = s(\bar{c} (\star D)c) = b (\star D)c .
\end{equation}
The  presence of this term  is due to the fact that  the gauge fixing term of \eqref{L_BV_YM} differs from the usual $s$-exact one by a $d$-exact term, that is 
\begin{equation}
\label{boundary_term_diff}
s\big(\bar{c} d (\star A)\big) = bd (\star A) + Dc (\star d) \bar{c} + d\big(\bar{c} (\star D) c \big).
\end{equation}
The Noether currents associated with these two different gauge fixing functionals differ from one another. One will check however that in both cases, the gauge dependent charges $q_{\rm gauge}$ vanish as advocated in \eqref{vanishing_q_gauge}.

The BRST Noether current for the Lagrangian \eqref{L_BV_YM} can be computed using  \eqref{theta_YM_BV} and \eqref{B_YM_BV}, providing
\begin{align}
\label{Noether_current_YM_BV}
\star J_{\rm BRST} = I_V \theta - B = - (Dc) \star F + b (\star D) c + \demi [c,c] (\star d)\bar{c} = -C - dq_{c} .
\end{align}
To check that  $q_c$ is gauge independent and coincides with the classical Noether charge,   as it should,   one must compute $C$. 
These constraints are defined in \eqref{def_Noether_constraints}, so using the equations of motion $E = E^A_\vp   \delta \vp^A$ in \eqref{delta_L_YM} one gets 
\begin{align}
\label{constraints_YM_BV}
d C \equiv I_V E  &= Dc \Big( D \star F     + (\star d) b   + [ (\star d) \bar{c} ,  c]  \Big)   - b  \Big(  d (\star D)c \Big) - \demi [c , c ] \Big(    D (\star d) \bar{c} \Big)
\nn \\
&= d \Big(  c D \star F  - b (\star D) c + \demi c [(\star d) \bar{c} , c ]  \Big) 
\nn \\
\quad& \Longrightarrow \quad C = c D \star F  - b (\star D) c + \demi c [(\star d) \bar{c} , c ] . 
\end{align} 
Comparing with \eqref{Noether_current_YM_BV} therefore leads to 
\begin{equation}
\label{Q_YM_GF}
q_c = q_{\rm cl} = c \star F ,
\end{equation}
in agreement with \eqref{vanishing_q_gauge}.  These are indeed the corner Noether  charges of ungauge fixed Yang--Mills theory. 
One thus consistently  finds that in the Lorenz gauge, the BRST Noether  charges  are the same as in the classical case,  modulo the interchange of the Faddeev--Popov ghost $c(x)$
 into the  infinitesimal gauge parameter $\lambda(x)$ (both entities are equivalent ways of parameterizing   the infinite number of Noether charges).

%

One may now look for possible modifications of the fundamental canonical relation, that is 
\begin{align}
\label{hamiltonian_vector}
I_{V} \O &= \int_{\Sigma_3} - Dc \star \delta F - \delta A [c,\star F] + \delta b (\star D) c -b \big( (\star D) \delta c - [\star \delta A, c]  \big) 
\nn \\
&\ \ \ \ \ \ \ \ \ - \delta \bar{c} \big( \  \cancel{ -\demi (\star D)[c,c] } + \cancel{ [(\star D) c,c]  }  \ \big) - \demi [c,c] (\star d ) \delta \bar{c} + \delta c (\star d ) b
\nn \\
&\ \hat{=} \int_{\Sigma_3} - \delta \big(   Dc \star F -  b (\star D) c + \demi c [(\star d) \bar{c},c]    \big) - \int_{\pa \Sigma_3} \star F \delta c = \int_{\Sigma_3} \Big[ \delta (\star J_{BRST}) - d q_{\delta c} \Big] 
\nn \\
&\ \hat{=} \int_{\Sigma_3} - \delta \big(  - c (\star d) b - \demi c [(\star d) \bar{c} , c ] - b (\star D) c  \big) - \int_{\pa \Sigma_3} c  \star\delta F
\nn \\
&= \int_{\Sigma_3}  \delta s \big(  c (\star d) \bar{c} - b \star A \big) - \int_{\pa \Sigma_3} c  \star\delta F = \int_{\Sigma_3}  \Big[ \delta s \big(  c (\star d) \bar{c} - b \star A \big) + d \big( \delta q_c - q_{\delta c} \big) \Big]  ,
\end{align}
where $q_{\delta c} = \delta c \star F$.  It is interesting to notice that  only  the equations of motion of the gauge field $A$ have been used to obtain this result. Had one started from the bare Lagrangian $L = -\demi F \star F$ without gauge fixing term, the term supported on $\Sigma_3$ in the last line of \eqref{hamiltonian_vector} would be zero and this relation would constitute the classical relation  \eqref{fund_with_fluxes} for field-dependent gauge parameter $\lambda$.  The only modification of this classical result by a gauge fixing term comes from the $\delta s$-exact term supported on $\Sigma_3$ which, as explained in  \eqref{delta_s-exact},  does not impact the construction of the  brackets on phase space.  The relation \eqref{hamiltonian_vector} is in fact compatible with the off-shell gauge fixed fundamental canonical relation \eqref{off-shell_BRST_fund}, providing an explicit check of the conjecture \eqref{claim_Z}, namely
\begin{align}
\label{CZY_YM}
C \ &\hat{=} \ - s (c (\star d) \bar{c} - b \star A ) ,
\nn \\
Z \ &\hat{=} \ 0 ,
\nn \\
Y \ &\hat{=} \ - q_{\delta c} .
\end{align}
On the way,  \eqref{CZY_YM}  also shows the validity of Noether's 1.5th \eqref{Noether_1.5} for gauge fixed Yang--Mills theory, that is
\begin{equation}
\label{J_BRST}
\star J_{\rm BRST} \  \hat{=} \  s (c (\star d) \bar{c} - b \star A ) - d q_c   .
\end{equation}

However, the computation \eqref{hamiltonian_vector} has a limitation.  In fact,  the various integrations by parts carried out in \eqref{hamiltonian_vector} were guided by the knowledge of the classical fundamental canonical relation for Yang--Mills.  One was able to do so because in this case the non integrable part of \eqref{hamiltonian_vector} is simply given by $q_{\delta c}$, which, as well as $q_c$, does not depend on the gauge fixing term.  This is a specificity of internal local symmetries and will not work for spacetime symmetries.  The more generic approach developed in Section [\ref{subsection_gauge_fixed_fund}] to determine $I_V \O$ off-shell, with a clear definition of the quantities $Z$ and $Y$, will be unavoidable to treat such symmetries.

One can now check that $d \star J_{\rm BRST} \ \hat{=} \ 0$.  As explained in \eqref{on-shell_cocycle}, it amounts to verify that
\begin{equation}
\label{G_1-4}
 d \big(  c (\star d) \bar{c} - b \star A \big) \ \hat{=} \ s G^{-1}_4. 
\end{equation}
This $G^{-1}_4$ should be reminiscent of the gauge fixing term used in the Lagrangian, and it is indeed the case, namely
\begin{equation}
\label{consistency_conserved_J_YM}
d \big(  c (\star d) \bar{c} - b \star A \big) \ \hat{=} \ s \big(   d\bar{c} \star A   \big) = b d(\star A) + Dc (\star d) \bar{c} - d(b \star A) .
\end{equation}
The right hand side of \eqref{consistency_conserved_J_YM} obviously originates from the gauge fixing term of the Lagrangian \eqref{L_BV_YM} modulo an other boundary term $d(b \star A)$. This is in agreement with the generic relation \eqref{sG_s-exact}.
Very interestingly,  only the equations of motion of $c$ and $\bar{c}$ have been used to show \eqref{consistency_conserved_J_YM}.  This could be a very generic feature of any $s$-exact term showing up in the BRST Noether current, as  will be checked for other gauge choices and in the case of gravity.


Before turning to other gauge choices, let us show how the argument \eqref{delta_s-exact} leads to a representation of the asymptotic symmetry algebra in this simple example.  To do so, one   computes the double contraction of the symplectic two-form $\O$ along BRST field space vector fields. Using \eqref{hamiltonian_vector}, one finds 
\begin{align}
\label{I_I_O_YM}
\demi I_{V} I_{V} \O &\ \hat{=} \   \demi \int_{\pa \Sigma_3} (\star F) [c,c]   =  \int_{\pa \Sigma_3} s \big(  c \star F  \big) = \int_{\pa \Sigma_3} s q_c \  .
\end{align}
This relation reads
\begin{equation}
\label{charge_algebra}
\demi I_V I_V \O \ \hat{=} \  s Q_c = \demi Q_{[c,c]}. 
\end{equation}
Recalling that the only non vanishing $Q_c$ are for ghost fields $c$ associated with asymptotic symmetry, the equation \eqref{charge_algebra} shows that  $\demi I_V I_V \O$ expresses the action of asymptotic symmetries on the asymptotic corner charges, namely the bracket between them, and provides a representation of this asymptotic symmetry algebra through $s_{\rm large} Q_{c_{\rm large}} = \demi Q_{[c_{\rm large},c_{\rm large}]}$.


\subsubsection{Lorenz gauge with a $s$-exact gauge fixing term}

One now computes the expression of the  BRST Noether current when  the gauge fixing Lagrangian for the Lorenz gauge $\pa_\mu A^\mu = 0$   is defined  by adding a strictly $s$-exact term to the classical Lagrangian, that is
\begin{equation}
\label{L_gauge_fixed_YM_sexact}
L = - \demi F  \star F   + s\big( \bar{c} d (\star A) \big) =  - \demi F  \star F   + bd (\star A) + Dc (\star d) \bar{c} + d\big(\bar{c} (\star D) c \big) .
\end{equation}
In this case, one has $sL=0$ and therefore $B=0$. The equations of motion of the Lagrangian \eqref{L_gauge_fixed_YM_sexact} are the same as the one of the Lagrangian \eqref{L_BV_YM} since they only differ by a boundary term.  Therefore the BRST Noether constraints $C$ are the same and are given by \eqref{constraints_YM_BV}.  
Nevertheless, the local symplectic potential associated with the gauge fixed Lagrangian \eqref{L_gauge_fixed_YM_sexact} differs from the one of \eqref{L_BV_YM} and is given  by 
\begin{align}
\label{theta_s-exact_YM}
\theta &=  \delta A \star	F -   \delta ( \star A ) \big( b  + [\bar{c} , c ] \big) - \delta c (\star d) \bar{c} - \bar{c} (\star D) \delta c . 
\end{align}
The BRST Noether current is then given by 
\begin{align}
\label{J_BRST-s-exact}
\J = I_V \theta = - Dc \star F + b (\star D) c + \demi (\star d) \bar{c} [c,c] = - C - dq_c . 
\end{align}
Plugging \eqref{constraints_YM_BV} in \eqref{J_BRST-s-exact} leads to $q_c = q_{\rm cl} = c \star F$, thus proving that the difference in $\theta$ exactly compensates the fact that $B=0$, such that 
\begin{equation}
I_V \theta_{\rm gauge} = - C_{\rm gauge} \quad \Longrightarrow \quad q_{\rm gauge} = 0.
\end{equation}
Every other results derived for the BV gauge fixing term are still valid here.  In particular, the fundamental canonical relation \eqref{hamiltonian_vector} is not impacted by boundary shifts of the Lagrangian, as shown in \eqref{boundary_shift_fund}.

\subsubsection{Temporal gauge}

A non trivial modification of the previous results   occurs when one replaces the Lorenz gauge choice by the (although incomplete) temporal gauge $A^0=0$.  By doing so, both the equations of motion and the local symplectic potential are modified.  It is thus  relevant to check the validity of Noether's 1.5th in this case.

The $A^0=0$   gauge fixed Lagrangian is
\begin{equation}
L = - \demi F \star F + s (\bar{c} (\star A ) \delta^0 ) =  - \demi F \star F + \Big( b (\star A )  + \bar{c} (\star D) c \Big) \delta^0
\end{equation}
where $(\star A) \delta^0 = A^\mu \delta^0_\mu {\rm Vol}_M = A^0 {\rm Vol}_M$. One then gets
\begin{align}
\label{EoM_temporal_gauge}
\delta L =& - \delta A \Big( D \star F + (b + [\bar{c},c] )(\star \delta^0)    \Big) - \delta b \Big(  \delta^0 \star A \Big) + \delta \bar{c} \Big( \delta^0 (\star D) c  \Big) + \delta c \Big( \delta^0 (\star D) \bar{c} \Big)
\nn \\
&+ d \Big(  \delta A \star F + \bar{c} \delta c (\star \delta^0)  \Big)
\end{align}
so that $\theta = \delta A \star F + \bar{c} \delta c (\star \delta^0)$ and 
\begin{align}
\O &= \int_{\Sigma_3} \delta A \delta (\star F) + \delta \bar{c} \delta c (\star \delta^0).
\end{align}
Furthermore,  one has $sL=0$ so 
\begin{equation}
\star J_{\rm BRST} = I_V \theta = - Dc \star F - \demi \bar{c} [c,c] (\star \delta^0) = - C - dq_c .
\end{equation}
Finally, the constraints are given by
\begin{align}
I_V E &= Dc \Big( D \star F + (b + [\bar{c},c] )(\star \delta^0)    \Big)  + b \Big( \delta^0 (\star D) c  \Big) - \demi [c,c] \Big( \delta^0 (\star D) \bar{c} \Big)
\nn \\
&= d \Big(   c D \star F + \demi \bar{c} [c,c] (\star \delta^0)  \Big) \quad \Longrightarrow \quad C = c D \star F + \demi \bar{c} [c,c] (\star \delta^0) ,
\end{align}
so that the charge $q_c$ again coincides with the classical one, namely
\begin{equation}
q_c = c \star F .
\end{equation}
Using the equation of motion of $A$ given by \eqref{EoM_temporal_gauge}, one also finds 
\begin{equation}
C \ \hat{=} \ - c \Big( b + \demi [\bar{c}, c ]  \Big) (\star \delta^0) = - s \Big( \bar{c} c  (\star \delta^0)  \Big) , 
\end{equation}
leading to the expected on-shell form for the Noether current 
\begin{equation}
\J \ \hat{=} \ s \Big( \bar{c} c  (\star \delta^0)  \Big) - d q_c .
\end{equation}
The consistency check $d \J \ \hat{=} \ 0$ is trivial here because 
\begin{equation}
\label{trivial_check_temporal_YM}
d \Big( \bar{c} c  (\star \delta^0)  \Big) \ \hat{=} \ 0
\end{equation}
by the equations of motion of the ghosts and antighosts. 
The fundamental canonical relation is then modified as 
\begin{align}
\label{A_0_gauge}
I_{V} \O \ &\hat{=} \  \int_{\Sigma_3} \Big[   \delta s \big( \bar{c} c \star \delta^0 \big) + d \big(  \delta q_c - q_{\delta c} \big) \Big] .
\end{align} 
So far, it appears that not only are the Noether charges the same whatever  the choice of gauge,  but  the representation of the charge algebra is also the same when one replaces the Lorenz gauge by the temporal gauge since  \eqref{A_0_gauge} also leads to \eqref{charge_algebra}.

\subsubsection{Feynman--t'Hooft gauge}

Another  important check is for the Lorentz invariant class of 
 Feynman--t'Hooft gauges with some possible 4 ghosts interactions.  Such gauges are implemented by adding the following $s$-exact term to the  Lagrangian: 
\begin{equation}
\label{Feynman_thooft_gauge}
s \Big( \bar{c} ( \pa_\mu A^\mu +  x b + y [\bar{c} , c ]  )  \Big) ,
\end{equation} 
where $x$ and $y$ are constant gauge parameters.
The gauge function $\mathcal{F}_{x,y}[A,c,\bar{c},b] \equiv \pa_\mu A^\mu +  x b + y [\bar{c} , c ]$ is the most general renormalizable gauge function in $d = 4$ dimensions. Imposing $y = 0$ eliminates the 4 ghosts vertex but when $y \neq 0$, \eqref{Feynman_thooft_gauge} provides a perfectly well defined gauge fixed theory that cannot be described in the genuine bigraded classical CPS, since in this case 
the ghost-antighost dependence of the action is quartic and the path integral cannot  integrate out the ghosts and the antighosts by trivial Berezin formulas as in the simpler Faddeev--Popov case.  It is however quite simple to check that
 \eqref{Q_YM_GF} and \eqref{charge_algebra} remain valid for all possible values of the gauge parameters $x$ and $y$ since the $(x,y)$-dependent part of \eqref{Feynman_thooft_gauge} does not contain any derivative of the fields. 
Indeed, this means that $\theta(x,y)$ equates the symplectic potential \eqref{theta_s-exact_YM} for all values of $x$ and $y$ and thus \eqref{Q_YM_GF} and \eqref{charge_algebra} follow for all gauges of the class $\mathcal{F}_{x,y}$.

This reasoning equally applies when modifying the Lorenz gauge function as 
\begin{equation}
\pa_\mu A^\mu = 0 \quad \longrightarrow \quad \pa_\mu A^\mu = z [v ,  \Phi ] 
\end{equation}
and most conveniently treat the spontaneous symmetry breaking in the Higgs mechanism. 
 Here the scalar field $\Phi$ VEV is $v \neq 0$ and its BRST transformation $s \Phi = - c \Phi$ does not include any derivative of the fields, which means that  changing the value of the t'Hooft parameter $z$ is also not modifying the symplectic potential \eqref{theta_s-exact_YM}.

 The various computations of this subsection  correspond to the larger  
class of Lorentz invariant gauges $\mathcal{F}_{x,y,z}$ indexed by different values  of the real   parameters  $x,y,z$.  The fact that the relations \eqref{Q_YM_GF} and \eqref{charge_algebra} are valid in this whole class of gauges 
 proves  the $x,y,z$  independence 
of the  classical covariant phase space corner charges and their bracket structure. 
 One gets  in these gauges the same results as
    in  the temporal gauge. 
This actually shows the  gauge independence of the results that one obtains with   Noether's second theorem in the genuine ungauge fixed case. 

This gauge independence of $Q$       provides  a more consistent meaning to the  Hamiltonian relation $[Q, \mathcal{S}_{\rm physical} ]=0$.   
Indeed,  the unitary physical  $\mathcal{S}$-matrix  operator
$\mathcal{S}_{\rm physical}$ as well as the physical states  do require a choice of gauge to be possibly expressed in terms of the fields,  so the same is needed  for the global charge $Q$, which completely justifies the  needed efforts to define and compute the     value  of  $Q$ in different gauges.

\subsection{Refined methodology}

The generic guideline presented in Section [\ref{subsection_gauge_fixed_fund}] is now applied to the  gauge fixed Yang--Mills Lagrangian via the BV formalism of  Section [\ref{subsection_d_s-exact}].   This  consistency check is useful in view of   using the method for gauge fixed gravity. 

The goal of this subsection is to show how the splitting between the BRST Noether constraints $C$,  the charges $Q$ and the flux $Z+dY$ appears systematically without having to rely on separate computations for the ungauge fixed case. This provides a direct proof of \eqref{CZY_YM}. 

Let us recall   the gauge fixed Yang--Mills Lagrangian \eqref{L_BV_YM} and its associated equations of motion and symplectic potential, that is
\begin{align}
\label{L,E,theta,YM-BV-2}
L &= - \demi F ( \star F )  + s\big(\bar{c} d (\star A)\big) - d\big(\bar{c} (\star D) c \big)  = - \demi F ( \star F ) + b  d (\star A)  + Dc (\star d) \bar{c}  ,
\nn \\
E &= - \delta A \Big( D \star F     + (\star d) b   + [ (\star d) \bar{c} ,  c]  \Big) - \delta b \Big(  (\star d) A   \Big) - \delta \bar{c}  \Big(  d (\star D)c \Big) + \delta c \Big(    D (\star d) \bar{c} \Big)  ,
\nn \\
\theta &= \delta A \star	F - b \star \delta A + \delta \bar{c} (\star D) c - \delta c (\star d) \bar{c}  .
\end{align}

In order to get the Noether constraints $C$, the Noether charge $q$ and the flux $Z+dY$, one has to compute $B,I_V E,I_V \theta, sE$ and $s\theta$.  Almost all these quantities have already been computed  in the paper (including the Appendix [\ref{Annex_anomaly}]), or at least some combinations of them. The values of $B,C$ and $Z$ are presented in the following equations:
\begin{align}
sL = ds\big( \bar{c} (\star D) c  \big)  \quad &\Longrightarrow \quad  B =  s\big( \bar{c} (\star D) c  \big) = b (\star D) c ,
\nn \\
I_V E = d \Big(  c D \star F  - b (\star D) c + \demi c [(\star d) \bar{c} , c ]  \Big) \quad &\Longrightarrow \quad  C =  c D \star F  - b (\star D) c + \demi c [(\star d) \bar{c} , c ]     ,
\nn \\
sE = d \Big(\delta c  \big(  D \star F     + (\star d) b   + [ (\star d) \bar{c} ,  c]  \big) \Big) \quad &\Longrightarrow \quad  Z = \delta c  \big(  D \star F     + (\star d) b   + [ (\star d) \bar{c} ,  c]  \big) .
\end{align}
It implies
\begin{align}
\label{CZ_YM_BV}
C  \ \hat{=}& \ -  s \big( c (\star d ) \bar{c} - b \star A   \big)   ,
\nn \\
Z =& \  - \delta c  E_A \ \hat{=} \ 0 .
\end{align}
The last equation provides an explicit check of \eqref{claim_Z}.  The only missing quantities in order  to apply the generic relation  \eqref{off-shell_BRST_fund} are $q$ and $Y$.  To obtain them, one has to compute
\begin{align}
\label{I_V_theta+s_theta}
I_V \theta &= -(Dc) \star F + 2b (\star D)c + \demi [c,c] (\star d)\bar{c}   ,
\nn \\
s \theta &=  - D(\delta c) \star F - \delta \big(  b (\star D) c  \big)   + \delta c \big(  (\star d) b  + [(\star d) \bar{c},c]  \big)   .
\end{align}
Knowing $I_V \theta,B$ and $C$,  one gets 
\begin{align}
  I_V \theta - B &= -(Dc) \star F + b (\star D)c + \demi [c,c] (\star d)\bar{c} 
\nn \\
&= - d (c \star F )  - c D \star F + b (\star D)c - \demi  c [(\star d)\bar{c},c]  = - dq - C \nn \\
\quad& \Longrightarrow \quad q = q_c = c \star F .
\end{align}
Notice that the value of $C$  indicates which integrations by parts have to be done to arrive at the BRST Noether charge $q$, which is uniquely defined.  
Having or not  directly computed the Noether charges $q_c\big\vert_{c=\lambda}$ in the classical case does not influence this computation;  it simply turns out afterwards that $q=q_c$.

The value of $Y$ is obtained by computing \eqref{s_theta_BRS2},  so by using the second equation of \eqref{I_V_theta+s_theta}, one has
\begin{align}
s \theta + \delta B - Z =& - D(\delta c) \star F - \delta \big(  b (\star D) c  \big)   + \delta c \big(  (\star d) b  + [(\star d) \bar{c},c]  \big) + \delta \big( b (\star D) c \big) 
\nn \\
&- \delta c  \big(  D \star F     + (\star d) b   + [ (\star d) \bar{c} ,  c]  \big)
\nn \\
=& - d ( \delta c \star F )  \quad \Longrightarrow \quad Y = - \delta c \star F = - q_{\delta c} . 
\end{align}  
This equation,  associated with \eqref{CZ_YM_BV}, 
provides a justification of \eqref{CZY_YM}. In fact, it is a strictly stronger result than \eqref{CZY_YM} since the flux $Y$ already coincides with the classical flux off-shell, in agreement with the discussion of Section [\ref{subsection_gauge_fixed_fund}]. 

One can plug all these results in  \eqref{off-shell_BRST_fund} and one finally gets
\begin{align}
\label{YM_fund_can_rel}
I_V \o =& - \delta	 ( c D \star F  - b (\star D) c + \demi c [(\star d) \bar{c} , c ]  ) + \delta c  \big(  D \star F     + (\star d) b   + [ (\star d) \bar{c} ,  c]  \big) 
\nn \\
&- \delta d ( c \star F ) - d ( \delta c \star F )
\nn \\
\hat{=}& - \delta \Big(   -  s \big( c (\star d ) \bar{c} - b \star A    \big) + d ( c \star F )  \Big) - d ( \delta c \star F ) ,
\end{align}
which is exactly what was obtained in \eqref{hamiltonian_vector}, namely 
\begin{equation}
I_V \O \ \hat{=}  \int_{\Sigma_3}  \Big[ \delta s \big(  c (\star d) \bar{c} - b \star A \big) + d \big( \delta q_c - q_{\delta c} \big) \Big]   . 
\end{equation}
This series of computations constitutes an autonomous derivation of the Noether charge and of the fundamental canonical relation for gauge fixed Yang--Mills theory.  It provides an explicit realization of the  generic construction and conjectures of Section [\ref{subsection_gauge_fixed_fund}].


\def \sgf{( \bar{\O}^I \star M^{I }_a   + \bar{\xi}^i   \delta^i_a \star \delta_+ )}
\def \mgf {(\bar{\O}^I M^{I \mu}_a   + \bar{\xi}^i  \delta^i_a \delta^\mu_+)}
\section{Gravity}
\label{Section_gravity}

Having studied in details the Yang--Mills theory  case,  this section is going to find out which properties are conserved and which  are lost in a theory involving not only internal symmetries, but also spacetime symmetries,  such as   first order gravity with a local Lorentz$\times \Diff$  symmetry.   In this theory,  the classical flux  depends on  $i_\xi \theta$ (see \eqref{fundamental_canon}) and is therefore clearly not gauge invariant since $\theta$ \eqref{split_theta_gauge} receives a contribution  
 issued from the variation of the gauge fixing term. 
One might then expect a difference in the splitting between physical quantities and gauge dependent ones in the fundamental canonical relation \eqref{off-shell_BRST_fund}.   
Therefore, in the case of gauge fixed first order gravity,   a brute force computation of  $I_V \O$ using the knowledge of the ungauge fixed case is not enough to reproduce  \eqref{off-shell_BRST_fund}. 
This is where the generic procedure  developed in Section [\ref{subsection_gauge_fixed_fund}] becomes vital, allows one to understand the gauge dependence of the flux and provides a  non trivial check of the postulates \eqref{Noether_1.5} and \eqref{claim_Z}.  

One can then explicitly derive  the boundary Ward identity \eqref{charge_ward_identity} for the charges associated with extended BMS symmetry by applying the formalism constructed in Section [\ref{Section_WI}] to asymptotically flat gravity in  Bondi gauge.  Loop corrections to the soft graviton theorem are discussed using the holographic anomalies of Section [\ref{Holographic_anomalies}]. 

\subsection{Preliminary}

One starts with the computation of the Noether charges associated with the local Lorentz and diffeomorphism symmetry of the ungauge fixed Lagrangian of first order gravity.  This computation is  in fact needed when going to the gauge fixed case.  It also allows one to compare the classical case with the gauge fixed case and possibly  identify differences between the two fundamental canonical relations that result. 

One works in four spacetime dimensions. The ungauge fixed Lagrangian of first order gravity can be written as a spacetime top-form made from the vierbein (or frame field) $e^a \in  \O_{\rm loc}^{0,1,0}(\tilde{\mathcal{F}} \times M)$, the spin connection $\o^{ab} \in  \O_{\rm loc}^{0,1,0}(\tilde{\mathcal{F}} \times M)$ and the fully anti symmetric Levi--Civita symbol, that is 
\begin{equation}
\label{1st_order_L}
L = \frac{1}{4} \epsilon_{abcd}  R^{ab} e^c e^d 
\end{equation}
where $R^{ab} \equiv d\o^{ab} + \demi [\o,\o]^{ab} =  d\o^{ab} +  \o^{a}_{\ c} \w \o^{cb}$.  The ``gauge" symmetry of first order gravity, under which the Lagrangian \eqref{1st_order_L} is invariant,  is a local Lorentz$\times$diffeomorphism symmetry. 
The action of this symmetry in the BRST CPS is governed by the following nilpotent BRST operator 
\begin{align}
\label{BRST_GR}
s e^a &=         \Lie_\xi e^a - \hat{\O}^{ab} e^b    =  i_\xi T^a - \O^{ac} e^c - D(i_\xi e^a) ,
\nn \\
s \o^{ab} &= \Lie_\xi \o^{ab} - D \hat{\O}^{ab} = i_\xi R^{ab} - D \O^{ab} ,
\nn \\
s \xi^\mu &= \xi^\nu \pa_\nu \xi^\mu  ,
\nn \\
s \hat{\O}^{ab} &= \Lie_\xi  \hat{\O}^{ab} - \demi [\hat{\O}, \hat{\O}]^{ab} ,
\end{align}
where $\xi \in \O_{\rm loc}^{0,0,1}(\tilde{\mathcal{F}} \times M)$ is a spacetime vector ghost for reparametrization symmetry 
and $\hat{\O}^{ab} \equiv \O^{ab} - i_\xi \o^{ab} \in \O_{\rm loc}^{0,0,1}(\tilde{\mathcal{F}} \times M)$ is a ghost for local Lorentz symmetry.  The field space $\tilde{\mathcal{F}}$ of this theory is therefore composed of $\vp^A = ( e^a , \o^{ab} , \xi , \hat{\O}^{ab} )$. When acting on local forms such as $e^a$, the spacetime Lie derivative is defined as $\Lie_\xi e^a = (i_\xi d - d i_\xi)e^a$. Sometimes,  for instance to ensure that $s^2 e^a_\mu = 0$, it is useful to write its action on the components,  which is given by  $\Lie_\xi e^a_\mu = \xi^\nu \pa_\nu e^a_\mu + e^a_\nu \pa_\mu \xi^\nu$.

Acting on the Lagrangian \eqref{1st_order_L} with the exterior derivative in field space $\delta$,   one gets $\delta L = E + d\theta$ as follows
\begin{align}
\label{EoM_classical_gravity}
\delta L &=  \demi \epsilon_{abcd} R^{ab} (\delta e^c ) e^d - \frac{1}{4} \epsilon_{abcd} (d \delta \o^{ab}  + [\o,\delta \o]^{ab} ) e^c e^d
\nn \\
&= \demi \Big( \epsilon_{abcd} R^{ab} e^d   \Big) \delta e^c - \frac{1}{4} \epsilon_{abcd} D (\delta \o^{ab}) e^c e^d
\nn \\ 
&= \demi \Big( \epsilon_{abcd} R^{ab} e^d   \Big) \delta e^c  + \demi \Big(  \epsilon_{abcd} T^c e^d  \Big) \delta \o^{ab} - \frac{1}{4} d \Big(  \epsilon_{abcd} \delta \o^{ab}    e^c e^d      \Big)
\end{align}
where $T^a \equiv De^a$ and $D(\bullet) = d(\bullet) + [\o , \bullet]$. 
The relation $\delta R^{ab} = - D(\delta \o^{ab})$ was used to get this result.
 It will be used repeatedly in the upcoming computations.  One will also need  the Bianchi identities for the curvatures 
\begin{equation}
\label{Bianchi_GR}
DR^{ab} = 0, \quad \quad DT^a = [R,e]^a .
\end{equation}
The variation \eqref{EoM_classical_gravity} implies the following value for the local presymplectic potential
\begin{equation}
\theta = -\frac{1}{4}   \epsilon_{abcd} \delta \o^{ab}    e^c e^d      .
\end{equation}
The corresponding presymplectic two form is therefore
\begin{equation}
\o \equiv \delta \theta = -\frac{1}{2}   \epsilon_{abcd} \delta \o^{ab}    \delta e^c e^d      .
\end{equation}
Using \eqref{BRST_GR}, \eqref{EoM_classical_gravity} and \eqref{Bianchi_GR}, one gets
\begin{align}
\label{steps_cl_gravity}
sL &=  - d i_\xi L    ,
\nn \\
I_V E &= \demi d \Big(  \epsilon_{abcd} (  R^{ab} e^d i_\xi e^c + T^c e^d \O^{ab} )  \Big)  ,
\nn \\
sE &= - d \left(  i_\xi E_e + \demi \epsilon_{abcd} \big[  i_\xi ( T^c e^d ) \delta \o^{ab} + R^{ab} e^d i_{\delta \xi} e^c + T^c e^d \delta \O^{ab}  \big]   \right)      ,
\nn \\
I_V \theta &=  - \frac{1}{4} \epsilon_{abcd}  ( i_\xi R^{ab} - D \O^{ab} ) e^c e^d  =  - \frac{1}{4} \epsilon_{abcd}  ( i_\xi R^{ab}  e^c e^d + 2 T^c e^d \O^{ab} ) + \frac{1}{4} d ( \epsilon_{abcd} \O^{ab} e^c e^d )   ,
\nn \\
s \theta &=  \frac{1}{4} \epsilon_{abcd} \Big(  \delta ( i_\xi R^{ab} - D \O^{ab} ) e^c e^d   -   2 \delta \o^{ab} (  i_\xi T^c - [ \O , e ]^c - D (  i_\xi e^c )   ) e^d \Big)   .
\end{align}
In turn, this provides respectively 
\begin{align}
\label{GR_classical_CQYZ}
B &= - \frac{1}{4}   \epsilon_{abcd}  \left( i_\xi R^{ab} e^c e^d + 2 R^{ab} i_\xi e^c e^d \right) , 
\nn \\
C &= \demi  \epsilon_{abcd} (  R^{ab} e^d i_\xi e^c + T^c e^d \O^{ab} )     , 
\nn \\
Z &= - \left(  i_\xi E_e + \demi \epsilon_{abcd} \big[  i_\xi ( T^c e^d ) \delta \o^{ab} +  R^{ab} e^d i_{\delta \xi} e^c + T^c e^d \delta \O^{ab}  \big]   \right)      , 
\nn \\
\star J_{\rm BRST} &= I_V \theta - B = -C -dq \quad   \Longrightarrow \quad q= - \frac{1}{4} \epsilon_{abcd} \O^{ab} e^c e^d  , 
\nn \\
 dY &= s\theta + \delta B - Z  \quad \quad  \quad \quad \   \Longrightarrow \quad  Y =  \frac{1}{4} \epsilon_{abcd} (  \delta \O^{ab} e^c e^d + 2 \delta \o^{ab} i_\xi e^c e^d )      
\end{align}
where $E_e = \demi ( \epsilon_{abcd} R^{ab} e^d  ) \delta e^c$.  All this provides the known Noether charges and flux of first order gravity. To see it  explicitly,  one writes \eqref{off-shell_BRST_fund}  for \eqref{GR_classical_CQYZ}.
Then, using  $Z \ \hat{=} \ 0$ and $C \ \hat{=} \ 0$,  one gets
\begin{align}
\label{I_V_o_cl_gravity}
I_V \o \ \hat{=} \  d(\delta q + Y) &= d \Big(  - \frac{1}{4} \epsilon_{abcd} \big(  \delta (\O^{ab} e^c e^d) - \delta \O^{ab} e^c e^d - 2 \delta \o^{ab} i_\xi e^c e^d  \big)  \Big)
\nn \\
&=   \demi d \Big( \epsilon_{abcd} \big(   \hat{\O}^{ab} \delta e^c e^d + i_\xi \o^{ab} \delta e^c e^d + \delta \o^{ab} i_\xi e^c e^d \big) \Big) 
\nn \\
&=  - d \Big( \delta q_\xi -  f_\xi + \delta q_{\hat{\O}}  -  f_{\hat{\O}} \Big) 
\end{align}
with 
\begin{align}
q_\xi &\equiv \demi \epsilon_{abcd}  e^c e^d i_\xi \o^{ab},   &  f_\xi &\equiv \demi \epsilon_{abcd} (  i_\xi e^c e^d \delta \o^{ab} - \delta e^c e^d i_\xi \o^{ab} + \delta ( e^c e^d i_\xi \o^{ab} ) ) ,
\nn \\
q_{\hat{\O}} &\equiv \demi \epsilon_{abcd} e^c e^d \hat{\O}^{ab} , & f_{\hat{\O}} &\equiv \demi \epsilon_{abcd} e^c e^d \delta \hat{\O}^{ab} .
\end{align}
This is exactly the usual fundamental canonical relation found in \cite{Freidel1,Freidel} with an explicit split between the charges and the fluxes associated with  diffeomorphism symmetry and those associated with local Lorentz symmetry.

\subsection{Bondi--Beltrami gauge}
\label{Section_Bondi-Beltrami}

Let us see how \eqref{I_V_o_cl_gravity} is impacted by a gauge fixing.  
The motivation is the BRST construction of the extended BMS4 symmetry done in \cite{Baulieu_Tom_BMS}  and the desire of using it and to apply the results of Section [\ref{Section_WI}] to this asymptotic symmetry group.  The gauge is that of \cite{Baulieu_Tom_BMS},   namely the Bondi--Beltrami gauge.

In the context of BMS4 symmetry at future null infinity, it is   convenient to work in light cone coordinates $(\t^+ = t +r ,\t^- = t - r,z,\bar{z})$ or rather in retarded Finkelstein--Eddington coordinates $(u \equiv \t^-,r,z,\bar{z})$. The various coordinates are best represented in the Penrose diagram Fig [\ref{figure_1}].  Spacelike infinity $i^0$ and past and future null infinities $\mathcal{I}^\pm$ sit at infinite radial coordinate $r=+\infty$. Past and future timelike infinities $i^\pm$ are at infinite times $t=\pm \infty$.  Each point in the diagram is a two sphere parametrized by complex coordinates $(z,\bar{z})$ and $u$ goes from $-\infty$ to $+\infty$ in $\mathcal{I}^+$.

To fix the gauge for  local Lorentz symmetry,  one imposes the  covariant Beltrami condition \cite{leafofleaf} on the vierbein.  If $I$ stands for the six indices of local Lorentz symmetry in four dimensions,  this gauge fixing condition can be written as
\begin{equation}
\label{gf1}
M^{I \mu}_a e^a_\mu = 0
\end{equation}
where the $M^{I \mu}_a$'s are field independent matrix elements defined such that the vierbein written in light cone coordinates reduces to 
\begin{equation}
\label{Beltrami_vierbein}
{e^{a}_{\rm Beltrami} }=
\pbM  \exp \frac{\Phi}{2} & 0 & 0 & 0 \\ 0 & \exp \frac{\Phi}{2} & 0 & 0 \\  0 & 0 & \mathcal{M} & 0 \\  0 & 0 & 0 & \mathcal{M}  \peM  \pbM 1 & \m & \mu^z_+ & \mu^z_- \\ \mb & 1 & \mu^\bz_+ & \mu^\bz_- \\ 0 & 0 & 1 & \mu^+_- \\ 0 & 0 & \mu^-_+ & 1  \peM   
\pbM   dz \\ d\bz \\ d\t^+ \\ d\t^-  \peM  .
\end{equation}
This system of coordinates covariantly splits the  ten degrees of freedom of a generic four dimensional metric into eight Weyl invariant ``Beltrami fields" $\mu$'s and two Weyl factors $\Phi$ and $\mathcal{M}$. The gauge fixing \eqref{gf1} is covariant under $\Diff_4$ providing one appropriately expresses the Lorentz ghosts $\hat{\O}^{ab}$ in terms of the vector ghosts $\xi$ only.  

One must therefore still fix the residual diffeomorphism symmetry.  To reach the partial Bondi gauge of \cite{Geiller} starting from \eqref{Beltrami_vierbein}, one simply has to impose a light cone gauge on the vierbein\footnote{See \cite{BMS_Hamiltonian} for an Hamiltonian derivation of the BMS algebra in the light-cone gauge.}
\begin{equation}
\label{gf2}
e^i_+ = \delta^i_a \delta^\mu_+ e^a_\mu = 0  \quad \Longrightarrow \quad \mu^i_+ = 0 ,
\end{equation}
where $i=\{z,\bz,- \}$.  The Bondi metric can then be obtained by computing $g_{\mu\nu} = e^a_\mu \eta_{ab} e^b_\nu$ for the Minkowski metric $\eta_{ab}dx^a dx^b = - d\t^+ d\t^- + 2 r^2 dz d\bar{z}$ in light cone coordinates.  Indeed, imposing \eqref{gf2} in \eqref{Beltrami_vierbein},  this computation leads to
\begin{equation}
\label{lcbb1}
ds^2 = - \mathcal{M}^2 (1 + \mu^+_-) du^2 -2 \mathcal{M}^2 dr du  + 2 r^2   \exp \Phi ||    dz + \m d\bz + \mu^z_- du    ||^2.
\end{equation}
This is the original   Bondi--Sachs metric \cite{bondi,sachs}   by identifying 
\begin{align}
\label{equiv_bel_bondi}
\mu^+_- &= - 1 - \frac{V}{r}  ,
\quad
\mathcal{M}^2 = e^{2 \beta} ,
\quad
\pbM \mu^z_- \\ \mu^\bz_- \peM  = - \Mu  \pbM U^z   \\  U^\bz  \peM  .
\end{align}
Notice that no restriction on $\Phi$, eventually implementing the 4th determinant gauge condition,  has been imposed.  This means that by further adding appropriate falloff conditions  to the fields, the asymptotic symmetry group  associated with \eqref{lcbb1} will be the  enlarged BMS group of \cite{Geiller}.  In \cite{Geiller:2024amx}, the authors have shown that this bigger group leads to new non vanishing corner charges, in addition with the already known charges for supertranslations, superrotations and Weyl transformations \cite{Compere}.  It is therefore important to check the claims \eqref{Noether_1.5} and \eqref{claim_Z} for this more general class of gauge,  hence obtaining the asymptotic Ward identity \eqref{charge_ward_identity} for the most general set of asymptotic charges.

The Bondi--Beltrami gauge fixing conditions \eqref{gf1} and \eqref{gf2} can be imposed consistently by adding the following $s$-exact term to the Lagrangian \eqref{1st_order_L}
\begin{equation}
L_{\rm GF} =  s \Big(   \bar{\O}^I M^{I \mu}_a e^a_\mu + \bar{\xi}^i  \delta^i_a \delta^\mu_+ e^a_\mu  \Big)d^4 x  .
\end{equation}
A few comments are in order.   This gauge fixing requires the introduction of trivial BRST doublets $(\bar{\O}^I, B^I)$ and $(\bar{\xi}^i, b^i)$, where $\bar{\O}^I,  \bar{\xi}^i$ are antighosts for local Lorentz symmetry and  reparametrization symmetry respectively and are both  elements of $\O_{\rm loc}^{0,0,-1}(\tilde{\mathcal{F}} \times M)$.  They transform trivially under the nilpotent BRST symmetry as
\begin{align}
s \bar{\O}^I &= B^I ,  &  s \bar{\xi}^i &= b^i ,
\nn \\
sB^I &= 0 ,   & s b^i &= 0.
\end{align}
This allows one to impose the gauge fixing conditions \eqref{gf1} and \eqref{gf2} through the trivial equations of motion of the Lagrange multipliers $B^I$ and $b^i$.  
Furthermore,  $s$ commutes with $M^{I\mu}_a$ and $ \delta^i_a \delta^\mu_+$ since they are both field independent. The same applies to $d^4x$, which is simply the flat space measure $d^4x = \star 1 = \frac{1}{4!} \epsilon_{\alpha \beta	\gamma \sigma} dx^\alpha \w dx^\beta \w dx^\gamma \w dx^\sigma$ where $\epsilon_{\alpha \beta	\gamma \sigma}$ is  the Levi--Civita symbol without any metric dependence.  Indeed, there is no need to use a covariant measure $\sqrt{-g} d^4x$ for such a gauge fixing term that breaks diffeomorphism invariance anyway.  

The starting point is therefore the following gauge fixed  Lagrangian of first order gravity
\begin{equation}
L = L_{\rm cl} + L_{\rm GF}= \frac{1}{4} \epsilon_{abcd}  R^{ab} e^c e^d + s \Big(  ( \bar{\O}^I M^{I \mu}_a   + \bar{\xi}^i  \delta^i_a \delta^\mu_+ )e^a_\mu  \Big)d^4 x  .
\end{equation}
In order to gather terms coming from $L_{\rm cl}$ and $L_{\rm GF}$ when taking the $\delta$ variation of this Lagrangian,  one must write $L_{\rm GF}$ in terms of spacetime differential forms.  In fact,  it   can be done by viewing $M^{I\mu}_a$ and $ \delta^i_a \delta^\mu_+$ as spacetime vector fields components, just like $\xi^\mu$,  and  defining 
\begin{align}
\star M^I_a &\equiv \frac{1}{4!} \epsilon_{\alpha \beta	\gamma \sigma} dx^\alpha \w dx^\beta \w dx^\gamma M^{I \sigma}_a , 
\nn \\
\star \delta_+ &\equiv \frac{1}{4!} \epsilon_{\alpha \beta	\gamma \sigma} dx^\alpha \w dx^\beta \w dx^\gamma \delta^{\sigma}_+  , 
\nn \\
\star \xi &\equiv \frac{1}{4!} \epsilon_{\alpha \beta	\gamma \sigma} dx^\alpha \w dx^\beta \w dx^\gamma \xi^\sigma  ,
\end{align}
so that 
\begin{equation}
\label{L_bondi_beltrami}
L = L_{\rm cl} + L_{\rm GF}= \frac{1}{4} \epsilon_{abcd}  R^{ab} e^c e^d + s \Big(  ( \bar{\O}^I \star M^{I }_a  + \bar{\xi}^i   \delta^i_a \star \delta_+  )e^a  \Big) .
\end{equation}
One can now compute 
\begin{align}
\label{delta_L_bondi}
\delta L =& \Big(  - \demi \epsilon_{abcd} R^{cb} e^d - (B^I \star M^I_a + b^i \delta^i_a \star \delta_+) + \pa_\nu ( \xi^\nu ( \bar{\O}^I \star M^{I }_a  + \bar{\xi}^i   \delta^i_a \star \delta_+  ) )  
\nn \\
& -  \pa_\mu (\star \xi) ( \bar{\O}^I M^{I \mu}_a   + \bar{\xi}^i  \delta^i_a \delta^\mu_+  )  + \hat{\O}^{ba} ( \bar{\O}^I \star M^{I }_b  + \bar{\xi}^i   \delta^i_b \star \delta_+  )   \Big)  \delta e^a
 + \Big( \demi \epsilon_{abcd} T^c e^d \Big) \delta \o^{ab}
\nn \\
&+ d^4x \bigg[ \Big( M^{I\mu}_a e^a_\mu  \Big) \delta B^I
 + \Big( \delta^i_a \delta^\mu_+ e^a_\mu  \Big) \delta b^i
 - \Big( M^{I\mu}_a s e^a_\mu  \Big) \delta \bar{\O}^I
\nn \\
&- \Big( \delta^i_a \delta^\mu_+ s e^a_\mu  \Big) \delta \bar{\xi}^i -  \Big( ( \bar{\O}^I M^{I \mu}_a   + \bar{\xi}^i  \delta^i_a \delta^\mu_+ ) e^b_\mu \Big) \delta \hat{\O}^{ab}
\nn \\
&+ \Big(  (\pa_\nu e^a_\mu  - \pa_\mu e^a_\nu) (\bar{\O}^I M^{I \mu}_a   + \bar{\xi}^i  \delta^i_a \delta^\mu_+)  - e^a_\nu \pa_\mu (\bar{\O}^I M^{I \mu}_a   + \bar{\xi}^i  \delta^i_a \delta^\mu_+)  \Big) \delta \xi^\nu  \bigg]
\nn \\
&+ d \Big( - \frac{1}{4} \epsilon_{abcd} \delta \o^{ab} e^c e^d - (\bar{\O}^I M^{I \mu}_a   + \bar{\xi}^i  \delta^i_a \delta^\mu_+)(\star \xi) \delta e^a_\mu  +  ( \bar{\O}^I \star M^{I }_a   + \bar{\xi}^i   \delta^i_a \star \delta_+ )e^a_\nu \delta \xi^\nu  \Big).
\end{align}
The local symplectic potential is then given by 
\begin{align}
\theta &= \theta_{\rm cl} + \theta_{\rm gauge}
\nn \\
&= - \frac{1}{4} \epsilon_{abcd} \delta \o^{ab} e^c e^d - \Big[ (\bar{\O}^I M^{I \mu}_a   + \bar{\xi}^i  \delta^i_a \delta^\mu_+)(\star \xi) \delta e^a_\mu - ( \bar{\O}^I \star M^{I }_a   + \bar{\xi}^i   \delta^i_a \star \delta_+ )e^a_\nu \delta \xi^\nu \Big] .
\end{align}
At this stage,  one has to repeat the computations performed in \eqref{steps_cl_gravity} but for the  quantities $L,E,\theta$ coming from \eqref{delta_L_bondi}. The computations are much more involved in this case and   will be explained  step by step.  


Firstly, since  $s^2=0$, the boundary term $B$ is unchanged in the gauge fixed case and still takes the following form 
\begin{equation}
\label{B_gauge-fixed_GR}
B = - \frac{1}{4}   \epsilon_{abcd}  \left( i_\xi R^{ab} e^c e^d + 2 R^{ab} i_\xi e^c e^d \right) .
\end{equation}
The other piece of the BRST Noether current is given by computing $I_V \theta$,  one gets 
\begin{align}
I_V \theta =&  - \frac{1}{4} \epsilon_{abcd}  ( i_\xi R^{ab} + 2 T^c e^d \O^{ab} ) + \frac{1}{4} d ( \epsilon_{abcd} \O^{ab} e^c e^d ) 
\nn \\
&-   (\bar{\O}^I M^{I \mu}_a   + \bar{\xi}^i  \delta^i_a \delta^\mu_+)(\star \xi) s e^a_\mu + ( \bar{\O}^I \star M^{I }_a   + \bar{\xi}^i   \delta^i_a \star \delta_+ )e^a_\nu s \xi^\nu  . 
\end{align}
The only missing piece to get the BRST Noether charges are the constraints $C$,  which are obtained by computing $I_V E = dC$,  namely\footnote{In view of extending the counterexample \eqref{counterexample} to the case of gauge fixed first order gravity,  
a small comment about the cancellations appearing in this computation that make $I_V E$ a boundary term is in order.   If one were to define the action of $I_{V_{\delta \C}} \delta$ on the fields of first order gravity based on \eqref{BRST_GR}, no matter how the coefficients in the definition are fixed, one would find that the linear term in $\delta \hat{\O}^{ab}$ in $I_{V_{\delta \C}} E $ is proportional to 
\begin{equation}
\label{counterexample_GF_gravity}
 (\bar{\O}^I M^{I \mu}_a   + \bar{\xi}^i  \delta^i_a \delta^\mu_+)  \Big(  \pa_\mu \xi^\nu [ \delta \hat{\O} , e_\nu ]^a + \xi^\nu \pa_\nu [ \delta \hat{\O} , e_\mu ]^a \Big)  d^4 x .
\end{equation}
This term clearly neither vanishes nor becomes a pure boundary term, thus constituting an explicit example of \eqref{obstruction_anomaly} in the case of gravity. }
\begin{align}
\label{Constraints_GR_gauge_fixed}
I_V E =& \ \demi d \Big(  \epsilon_{abcd} (  R^{ab} e^d i_\xi e^c + T^c e^d \O^{ab} )  \Big) 
\nn \\
&+  ( \bar{\O}^I \star M^{I }_a   + \bar{\xi}^i   \delta^i_a \star \delta_+ ) \hat{\O}^{ab} \Lie_\xi e^b  - (\bar{\O}^I M^{I \mu}_a   + \bar{\xi}^i  \delta^i_a \delta^\mu_+) e^b_\mu \Lie_\xi \hat{\O}^{ab} d^4 x 
\nn \\
&+ \Big(  (\pa_\nu e^a_\mu  - \pa_\mu e^a_\nu) (\bar{\O}^I M^{I \mu}_a   + \bar{\xi}^i  \delta^i_a \delta^\mu_+)  - e^a_\nu \pa_\mu (\bar{\O}^I M^{I \mu}_a   + \bar{\xi}^i  \delta^i_a \delta^\mu_+)  \Big) s \xi^\nu d^4x
\nn \\
&+ \Big( \pa_\nu ( \xi^\nu ( \bar{\O}^I \star M^{I }_a  + \bar{\xi}^i   \delta^i_a \star \delta_+  ) )   -  \pa_\mu (\star \xi) ( \bar{\O}^I M^{I \mu}_a   + \bar{\xi}^i  \delta^i_a \delta^\mu_+  )  \Big) se^a
\nn \\
=& \ d \bigg[ \demi \Big(  \epsilon_{abcd} (  R^{ab} e^d i_\xi e^c + T^c e^d \O^{ab} )  \Big)    
\nn \\
&+ (\bar{\O}^I M^{I \mu}_a   + \bar{\xi}^i  \delta^i_a \delta^\mu_+)(\star \xi) s e^a_\mu -  ( \bar{\O}^I \star M^{I }_a   + \bar{\xi}^i   \delta^i_a \star \delta_+ )e^a_\nu s \xi^\nu \bigg] .
\end{align}
Then, from  $\star J_{\rm BRST} = I_V \theta - B = -C -dq$,  one obtains the Noether charges
\begin{equation}
\label{q_gf_gravite}
q = - \frac{1}{4} \epsilon_{abcd} \O^{ab} e^c e^d .
\end{equation}
They indeed coincide with the classical charges found in \eqref{GR_classical_CQYZ}.  To obtain the expression of the constraints $C$ on-shell,  one uses the following equations of motion
\begin{equation}
T^a \ \hat{=} \ M^{I \mu}_a s e^a_\mu \ \hat{=} \  \delta^i_a \delta^\mu_+   se^a_\mu \ \hat{=} \ 0,
\end{equation}
along with the on-shell expression of $\epsilon_{abcd} R^{ab} e^d i_\xi e^c$,  leading to 
\begin{align}
C \ \hat{=}& \ s \Big( ( \bar{\O}^I \star M^{I }_a   + \bar{\xi}^i   \delta^i_a \star \delta_+ ) i_\xi e^a \Big) + d i_\xi \Big( ( \bar{\O}^I \star M^{I }_a   + \bar{\xi}^i   \delta^i_a \star \delta_+ ) i_\xi e^a \Big)
\nn \\
&-  \Big( \pa_\mu e^a_\nu (\bar{\O}^I M^{I \mu}_a   + \bar{\xi}^i  \delta^i_a \delta^\mu_+)  + e^a_\nu \pa_\mu (\bar{\O}^I M^{I \mu}_a   + \bar{\xi}^i  \delta^i_a \delta^\mu_+)  \Big) (\star \xi) \xi^\nu 
\nn \\
&- (\bar{\O}^I M^{I \mu}_a   + \bar{\xi}^i  \delta^i_a \delta^\mu_+)  (\star \xi) e^a_\nu \pa_\mu \xi^\nu .
\end{align}
By the equation of motion of $\xi$,  the last two lines vanish and one is left with 
\begin{equation}
C \  \hat{=} \ s \Big( ( \bar{\O}^I \star M^{I }_a   + \bar{\xi}^i   \delta^i_a \star \delta_+ ) i_\xi e^a \Big) + d i_\xi  \kappa_\xi ,
\end{equation}
where $\kappa_\xi \equiv ( \bar{\O}^I \star M^{I }_a   + \bar{\xi}^i   \delta^i_a \star \delta_+ ) i_\xi e^a$. A final simplification occurs by noticing that if one writes the Lagrangian \eqref{L_bondi_beltrami} as $L=L_{\rm cl} + s \Psi$,  one has
\begin{equation}
\label{Psi_kappa_link}
i_\xi i_\xi \Psi \ \hat{=}  - 2 i_\xi \kappa_\xi
\end{equation}
by the equations of motion of $B^I$ and $b^i$. Since $\Psi \ \hat{=} \ 0$ by definition of the gauge fixing functional \eqref{gauge_fixing},  \eqref{Psi_kappa_link} implies $i_\xi \kappa_\xi \ \hat{=} \ 0$ and one finally gets the desired result 
\begin{equation}
\label{final_onshell_constraints_gravite}
C \  \hat{=} \ s \Big( ( \bar{\O}^I \star M^{I }_a   + \bar{\xi}^i   \delta^i_a \star \delta_+ ) i_\xi e^a \Big) .
\end{equation}
The conjectured Noether's 1.5th theorem \eqref{Noether_1.5} is therefore valid for first order gravity in Bondi--Beltrami gauge.

The equations of motion of $\xi , \bar{\xi} , \hat{\O}$ and $\bar{\O}$ provide a consistency check of the conservation of the BRST Noether current on-shell as
\begin{equation}
d \J \ \hat{=} \ s d \Big( ( \bar{\O}^I \star M^{I }_a   + \bar{\xi}^i   \delta^i_a \star \delta_+ ) i_\xi e^a \Big)  \ \hat{=} \ 0.
\end{equation}
So, it does not provide a non trivial $G^{-1}_4$ reminiscent of the gauge fixing term of \eqref{L_bondi_beltrami} as in \eqref{G_1-4} for the covariant gauge $\pa_\mu A^\mu = 0$.  In fact,  it resembles  the Yang--Mills result in the temporal gauge  \eqref{trivial_check_temporal_YM}. This had to be expected since the Bondi--Beltrami gauge is a non covariant gauge of the type $A^0=0$.

To go to the fundamental canonical relation,   one still needs to compute $Z$ and $Y$.  The computation of $Z$ is the most involved. After a long and tedious calculation,  one  arrives at 
\begin{align}
sE = dZ = - d \left( \demi \epsilon_{abcd} \Big( i_\xi (T^c e^d ) \delta \o^{ab} + T^c e^d \delta \O^{ab} \Big) + (\star \xi) ( E_{\hat{\O}} + E_\xi )  + i_\xi E_e + E_{e,a} i_{\delta \xi} e^a \right) ,
\end{align}
where the equations of motion are defined as $E= \sum_\vp E_\vp$ with $E_\vp = E_{\vp, A} \delta \vp^A$ where $A$ stand for whatever indices carried by the field $\vp$.  Note also that $E_{\hat{\O}}$ and $E_\xi$ have to be thought of as spacetime zero forms in this equation, namely as they appear in \eqref{delta_L_bondi} without the $d^4 x$.  The main takeaway of this computation is that in agreement with \eqref{claim_Z},  one has
\begin{equation}
\label{check_Z_vanish}
Z \ \hat{=} \ 0 ,
\end{equation}
meaning that the flux in the fundamental canonical relation is a pure corner term on-shell even in the presence of a gauge fixing term in gravity.  This corner flux is given by $Y$,  which one can now compute using 
\begin{align}
s \theta =& \  s \theta_{\rm cl} + s \theta_{\rm gauge}  ,
\nn \\
=& \    \frac{1}{4} \epsilon_{abcd} \Big(  \delta ( i_\xi R^{ab} - D \O^{ab} ) e^c e^d   -   2 \delta \o^{ab} (  i_\xi T^c - [ \O , e ]^c - D (  i_\xi e^c )   ) e^d \Big)
\nn \\
& - s \mgf (\star \xi) \delta e^a_\mu + \mgf s ((\star \xi) \delta e^a_\mu )
\nn \\
& + s \sgf e^a_\nu \delta \xi^\nu + \sgf s (e^a_\nu \delta \xi^\nu)  .
\end{align}
Then,  after another lengthy calculation,  one gets
\begin{align}
\label{gauge_flux}
dY =& \ s \theta + \delta B - Z 
\nn \\
=&  \ d \bigg(  \frac{1}{4} \epsilon_{abcd} (  \delta \O^{ab} e^c e^d + 2 \delta \o^{ab} i_\xi e^c e^d )  
\nn \\
&  + \mgf i_\xi (\star \xi) \delta e^a_\mu - i_\xi \sgf e^a_\nu \delta \xi^\nu  \bigg)  
\nn \\
=& \ d \big( Y_{\rm cl}  - i_\xi \theta_{\rm gauge} \big).
\end{align}
The classical flux $Y_{\rm cl}$, given by \eqref{GR_classical_CQYZ}, is corrected by a gauge dependent term $i_\xi \theta_{\rm gauge}$.  
This could have been anticipated from the classical expression of the flux given by \eqref{fundamental_canon}, which contains a $i_\xi \theta$ piece. 
Indeed,  one could simply repeat the steps performed in Section [\ref{subsection_gauge_fixed_fund}] but acting with $s - \Lie_\xi$ on $L,\theta,\delta L$ rather than just acting with $s$.  It leads to the following fundamental canonical relation
\begin{equation}
\label{modified_BRST_fund_can}
I_V \o = - \delta (C + dq ) + Z + d (Y' - i_\xi \theta) ,
\end{equation} 
where the flux has been split and is calculated from 
\begin{equation}
d Y' = s \theta - Z + \delta B + d i_\xi \theta .
\end{equation}
However, there is no generic argument supporting that this $Y'$ is the equivalent of $q_{\delta \xi}$ in \eqref{fundamental_canon},  which would in turn imply that the gauge dependence of $Y' - i_\xi \theta$ is entirely contained in $i_\xi \theta_{\rm gauge}$. 
Essentially, one has to do the same calculation as in \eqref{gauge_flux} to show this, but the relation \eqref{modified_BRST_fund_can} at least provides a clear origin for the term $i_\xi \theta$.


By using \eqref{q_gf_gravite}, \eqref{final_onshell_constraints_gravite}, \eqref{check_Z_vanish} and \eqref{gauge_flux},  the fundamental canonical relation \eqref{off-shell_BRST_fund} for the gauge fixed Lagrangian \eqref{L_bondi_beltrami}  takes the form 
\begin{align}
\label{fund_can_gauge_fixed_GR}
I_V \o \ \hat{=}&  - \delta s \kappa_\xi + d \Big(   \delta q_\O   + Y_{\rm cl}   - i_\xi \theta_{\rm gauge} \Big)
\nn \\
\ \hat{=}& - \delta s \kappa_\xi - d \Big( i_\xi \kappa_{\delta \xi} \Big)   + d \Big(   \delta   q_\O   - q_{\delta \O}  - i_\xi \theta_{\rm cl}  \Big)  ,
\end{align}
with 
\begin{align}
q_\O &= - \frac{1}{4} \epsilon_{abcd} \O^{ab} e^c e^d , &  q_{\delta \O} &= -\frac{1}{4} \epsilon_{abcd} \delta \hat{\O}^{ab} e^c e^d - \frac{1}{4} \epsilon_{abcd} i_{\delta \xi} \o^{ab} e^c e^d ,
\nn \\
\kappa_\xi &= \sgf i_\xi e^a ,   &   \kappa_{\delta \xi} &= \sgf i_{\delta \xi }  e^a , 
\nn \\
Y_{\rm cl} &= \frac{1}{4} \epsilon_{abcd} ( \delta \O^{ab} e^c e^d + 2 \delta \o^{ab} i_\xi e^c e^d ) 
\nn \\
&= - q_{\delta \O} - i_\xi \theta_{\rm cl} .
\end{align}
In the last line of \eqref{fund_can_gauge_fixed_GR}, one used the equations of motion of $B^I$ and $b^i$ to show that $i_\xi \theta_{\rm gauge} \ \hat{=} \ i_\xi \kappa_{\delta \xi}$.  Then, one can prove that $i_\xi \kappa_{\delta \xi}$ actually vanishes on-shell by using the following  identities 
\begin{align}
i_\xi i_{\delta \xi} \Psi \ \hat{=}& \ 0 ,
\quad \quad
\delta (i_\xi \kappa_\xi)  \ \hat{=} \ 0,
\quad  \quad
i_\xi i_\xi \delta \Psi  \ \hat{=} \ 0 . 
\end{align}
Finally,  one arrives at 
\begin{equation}
\label{fund_can_gauge_fixed_GR_2}
\boxed{ I_V \o \ \hat{=} \  - \delta s \kappa_\xi     + d \Big(   \delta   q_\O   - q_{\delta \O}  - i_\xi \theta_{\rm cl}  \Big)  } \ .
\end{equation}
One knows from \cite{Freidel} that a bracket providing an exact representation of the asymptotic symmetry algebra, which in this case will be the enlarged BMS algebra \cite{Geiller,Geiller:2024amx},  can be generically constructed from \eqref{fund_can_gauge_fixed_GR_2} by considering\footnote{This bracket is a generalization of the Barnich--Troessaert bracket \cite{Barnich_Charge_algebra}. } 
\begin{equation}
\label{charge_bracket_L}
 \{ Q_\O ,  Q_\O  \}^L \equiv   I_V I_V \tilde{\O} -  \int_{\pa \Sigma} i_\xi i_\xi L ,
\end{equation}
where $\tilde{\O} = \int_\Sigma \o$ and $Q_\O = \int_{\pa \Sigma} q_\O$. This ``gauge fixed" bracket obviously coincides with the ungauge fixed one of \cite{Freidel} since $I_V \delta s \kappa_\xi = s^2 \kappa_\xi =0$ and $i_\xi i_\xi s \Psi \ \hat{=} \ 0$.  A clear advantage of \eqref{charge_bracket_L} is that the symplectic two form $\tilde{\O}$ is non degenerate because of \eqref{fund_can_gauge_fixed_GR_2}. 
This is another  major result of this paper.  It proves the gauge independence of this covariant phase space construction for all known extensions of the BMS group.  With the gauge being properly fixed,  canonical quantization of the phase space can  be performed by promoting the brackets \eqref{charge_bracket_L} into quantum commutators.  

It would be very interesting to check that \eqref{fund_can_gauge_fixed_GR_2} is still valid in the de Donder gauge but with a different $\delta s$-exact term. 
Note that the increase in computational difficulty when going from the light cone Bondi--Beltrami  gauge \eqref{gf1}--\eqref{gf2} to the de Donder gauge where \eqref{gf1} is replaced by $e^a_\mu = e^\mu_a$ and \eqref{gf2} by $\pa_\mu e^a_\mu = 0$ should be similar to that encountered when going from the temporal gauge $A^0=0$ to the Lorenz gauge $\pa_\mu A^\mu=0$ in Yang--Mills.  However, this computation is left aside for now as one is hoping instead for a generic proof of \eqref{fund_can_gauge_fixed_GR_2}.

\subsection{Extended BMS symmetry and soft graviton theorems}
\label{Section_soft_graviton}

Let us apply the formalism of Section [\ref{Section_WI}] to the Noether charges of extended BMS symmetry.  The extended BMS group is the asymptotic symmetry group of asymptotically flat gravity for the following falloff conditions  for the Beltrami fields of \eqref{Beltrami_vierbein} near $\mathcal{I}^+$:
\begin{align}
\label{power_expansion} 
\mu^z_\bz &= \frac{{\mu^z_\bz}_{-1}(u,z,\bz)}{r} + \mathcal{O}(r^{-2})  ,  & \mu^\bz_z &= \frac{{\mu^\bz_z}_{-1}(u,z,\bz)}{r} + \mathcal{O}(r^{-2}) ,
\nn \\
\mu^+_- &= \frac{{\mu^+_-}_{-1}(u,z,\bz)}{r} + \mathcal{O}(r^{-2})   ,    &   \mu^A_u &= \frac{{\mu^A_u}_{-2}(u,z,\bz)}{r^2} + \mathcal{O}(r^{-3})  ,
\nn \\
\mathcal{M} &= 1 +  \frac{\mathcal{M}_{-2}(u,z,\bz)}{r^2} + \mathcal{O}(r^{-3}) , & \exp \Phi &= \o(u,z,\bz) + \frac{\Phi_{-2}(u,z,\bz)}{r^2}  +  \mathcal{O}(r^{-3}) .
\end{align}
The Beltrami differentials ${\mu^z_\bz}_{-1}(u,z,\bz)$ and ${\mu^\bz_z}_{-1}(u,z,\bz)$, which will be denoted $\mu$ and $\bar{\mu}$ from now on, are the two Weyl invariant components of the shear tensor and stand for field representations of the two helicity states of a graviton. 
As shown in \cite{Baulieu_Tom_BMS},  imposing the stability of these falloffs and of the Bondi--Beltrami gauge \eqref{gf1}--\eqref{gf2} under the action of the BRST operator \eqref{BRST_GR} leads to a residual diffeomorphism symmetry at $\mathcal{I}^+$, which is an asymptotic symmetry in the sense of \eqref{asg},  parametrized by\footnote{Strictly speaking, one also needs to impose the $4$th gauge condition on the determinant, which writes $\pa_r \Phi  =  \pa_r (\m \mb)/(\mmb)$, and its stability to get the extended BMS group.   One simply assumes that this additional condition  does not alter the $s$-exactness of \eqref{final_onshell_constraints_gravite}.  This is a reasonable assumption since the charges of \cite{Geiller:2024amx} contain those of supertranslations and superrotations.  The Weyl factor $\o(u,z,\bz)$ is also assumed to be $1$ for simplicity. }
\begin{align}
\label{xi_BMS}
\xi^u_{\rm BMS} &= \alpha (z,\bar{z}) + \frac{u}{2} \pa_A \xi^A ,
\nn \\
\xi^z_{\rm BMS} &= \xi(z),
\nn \\
\xi^\bz_{\rm BMS} &= \bar{\xi}(\bz) .
\end{align}
The ghost $\alpha$ generates angle dependent time translations on $\mathcal{I}^+$, such transformations are called \textit{supertranslations}. The ghosts $\xi^A=(\xi,\bar{\xi})$ are conformal killing vector on the two sphere and generate the so-called \textit{superrotations}.  Both supertranslations and superrotations have a non trivial action on the radiative phase space and are therefore physical transformations.  Their action on the radiative data, that is on the graviton fields $\mu$, $\bar{\mu}$, and the representation of the asymptotic symmetry algebra associated with \eqref{xi_BMS} are given by the action of the ``large" nilpotent BRST operator 
\begin{align}
\label{BRSTBMSoperator}
s \mu &= \Big(  \xi^u_{\rm BMS} \pa_u + \xi^A \pa_A + \frac{3}{2} \bar{\pa} \bar{\xi} - \demi \pa \xi \Big)\mu - \bar{\pa}^2 \xi^u  
\nn \\
s \alpha &=\xi^A \pa_A \alpha + \frac{\alpha}{2} \pa_A \xi^A ,
\nn \\
s\xi  &= \xi  \pa \xi  ,
\nn \\
s\bar{\xi} &= \bar{\xi} \bar{\pa} \bar{\xi}
\end{align}
derived in \cite{Baulieu_Tom_BMS}.  The action of the above BRST operator on the supertranslation and superrotation ghosts reveals the semi-direct structure of the extended BMS group 
\begin{equation}
\text{eBMS   = Superrotations $\ltimes$  Supertranslations}.
\end{equation}
Having a non trivial asymptotic symmetry group means that there is a non empty boundary Ward identity \eqref{Boundary_WI_final}. Indeed,  starting from the generating functional \eqref{exp_W} with $L$ given by \eqref{L_bondi_beltrami} and $B$ given by \eqref{B_gauge-fixed_GR}, one gets the Ward identity 
\begin{equation}
\label{BRST_WI_GR}
 \left\langle \left( \int_{\Sigma^+} \J \right)    \mathcal{O}_1 ... \ \mathcal{O}_n    \right\rangle - \left\langle  \mathcal{O}_1 ... \ \mathcal{O}_n   \left( \int_{\Sigma^-} \J \right) \right\rangle = 0 
\end{equation}
for the BRST Noether current  
\begin{equation}
\label{J_BRST_GR_gf}
\J \ \hat{=} \ - s \Big( ( \bar{\O}^I \star M^{I }_a   + \bar{\xi}^i   \delta^i_a \star \delta_+ ) i_\xi e^a \Big) + d \Big(  \frac{1}{4} \epsilon_{abcd} \O^{ab} e^c e^d \Big) = -s \kappa_\xi - d q_\O . 
\end{equation}
The boundary Ward identity \eqref{BRST_WI_GR} is non empty because some parts of 
\begin{equation}
\label{charges_BMS}
Q^+[\xi_{\rm BMS}] \equiv \int_{\Sigma^+} dq_{\O} = \int_{-\infty}^{+\infty} du \  \pa_u  \left(  \int_{S^2} \ \frac{1}{4}  \epsilon_{abcd} \O^{ab}(\xi_{\rm BMS}) e^c e^d  \right)
\end{equation}
 are non vanishing when the $S^2$ integration is at $r=\infty$.  Since the gauge has been fixed by \eqref{gf1}--\eqref{gf2},  $e^a$ is the Beltrami vierbein \eqref{Beltrami_vierbein} with falloffs given by \eqref{power_expansion} and the explicit $\xi$ dependence of $\O^{ab}$ is to ensure the consistency of \eqref{gf1}. The expression of $\O^{ab}(\xi)$ is given in Eq.(B.30) of \cite{Baulieu_Tom_BMS}.  
 
 The goal of this paper is not to dissect the $r$-dependence of $q_\O$ in the Bondi--Beltrami gauge with falloffs \eqref{power_expansion} and check which parts of  $Q^+[\xi_{\rm BMS}]$ are non zero.\footnote{In fact, this dissection is very close to that of \cite{Freidel1} since their gauge fixing Eq.(5.14) for local Lorentz symmetry shares four conditions with the Beltrami parametrization \eqref{gf1}.}  Rather, one understands that an identity of the type \eqref{J_BRST_GR_gf} should be derivable in any formulation of gravity,  meaning that one can work in its favorite framework to derive the charges for supertranslations and superrotations.  To do this, in any case,   one has to assume additional falloff conditions for the fields when $u \to \pm \infty$ because of the integration over $u$ in \eqref{charges_BMS}.  In fact, the shear $\mu$ and the news $N \equiv \pa_u \mu$ (encoding gravitational radiation) can be split into soft and hard parts as\footnote{Without loss of generality, one works in a superrotation frame in which the vacuum news tensor vanishes.} 
 \begin{align}
 \label{hard-soft_split}
 \mu &=   \mu^0(z,\bz) + \tilde{\mu}(u,z,\bz) , 
 \nn \\
 N &=  \tilde{N} (u,z,\bz),
 \end{align}
 where $\tilde{N} (u,z,\bz)\equiv \pa_u \tilde{\mu}(u,z,\bz)$.  Analogous relations hold for the complex conjugated quantities.  The denomination ``soft" simply means that soft quantities do not parametrize gravitational radiation, namely they are $u$-independent,  as opposed to the ``hard" quantities.  The BRST transformation laws of each piece of \eqref{hard-soft_split} under extended BMS symmetry is readily available by plugging the decomposition \eqref{hard-soft_split} into \eqref{BRSTBMSoperator}.  

 One must then impose $u$-falloff conditions  on the elements of \eqref{hard-soft_split} that are sufficiently weak to allow for gravitational tails \cite{Blanchet:1987wq}. They are given by
  \begin{align}
 \label{u_falloffs_new}
 \tilde{\mu}(u,z,\bz) &= \frac{\mu^{L,\pm}(z,\bz)}{u} + o(u^{-1}) ,  &   \tilde{N} (u,z,\bz)  &= - \frac{\mu^{L,\pm}(z,\bz)}{u^2} + o(u^{-2})  , 
 \nn \\
 \tilde{\mu}\Big\vert_{u=\pm \infty} &= \mp 2 \bar{\pa}^2 N^0,  &  m_B \Big\vert_{u=+\infty} &= 0 .
 \end{align}
The presence of $\mu^{L,\pm}$ indicates that the peeling property \cite{Sachs:1961zz} is lost and it is actually necessary to incorporate gravitational tails \cite{Sahoo:2018lxl, Saha:2019tub}. 
One has $N^0 = \demi (C_+ - C_-)$ where $C_\pm$ are the supertranslation modes at $\mathcal{I}^+_\pm$ (see the last term in the r.h.s. of $s\mu$ in \eqref{BRSTBMSoperator}) and the Bondi mass $m_B$ is given by $m_B(u,z,\bz) = - \demi {\mpm}_{-1}$ in terms of Beltrami fields.  It will   also be useful to  define the so-called  leading soft news $\mathcal{N}^0$ and the subleading soft news $\mathcal{N}^1$ as 
 \begin{equation}
 \label{def_N0_N1}
 \mathcal{N}^0 \equiv \int_{-\infty}^{+\infty} du \ \tilde{N} = -4 \bar{\pa}^2 N^0, \quad \quad   \mathcal{N}^1 \equiv \int_{-\infty}^{+\infty} du \ u  \tilde{N}. 
 \end{equation}
 
 With the definitions \eqref{def_N0_N1},  the $u$-falloffs \eqref{u_falloffs_new} and with the asymptotic symmetry group being the extended BMS group \eqref{BRSTBMSoperator},   the charges \eqref{charges_BMS} must at least contain 
 \begin{align}
 \label{supertranslation_superrotations_charges}
 Q^+_\alpha &=   Q^{\text{soft}}_{\alpha}  +  Q^{\text{hard}}_\alpha =  \frac{2}{\kappa^2} \left[ \int_{\mathcal{I}^+_-}  d^2z \  \alpha \Big( \pa^2   \mathcal{N}^0 + \bar{\pa}^2  \bar{\mathcal{N}}^0 \Big) -  \int_{\Sigma^+} du d^2z \  \alpha \Big( \tilde{N} \tilde{\bar{N}}  \Big) \right] ,
 \nn \\
 Q^+_{\xi} &= Q^{\text{soft}}_{\xi}  +  Q^{\text{hard}}_\xi = - \frac{2}{\kappa^2} \left[ \int_{\mathcal{I}^+_-} d^2z \ \xi \pa^3  \mathcal{N}^1 + \int_{\Sigma^+} du d^2z \ \xi \Big(   \frac{3}{2}  \tilde{\bar{N}} \pa \tilde{\mu} + \demi \tilde{\mu} \pa \tilde{\bar{N}} - \frac{u}{2} \pa ( \tilde{N} \tilde{\bar{N}}  )   \Big) \right]   
 \end{align}
where $\kappa^2 = 32 \pi G$.  
Even though these charges are non conserved and non integrable because of the non vanishing flux in \eqref{fund_can_gauge_fixed_GR_2}, 
they have to be building blocks of the more generic finite and integrable ones that have been singled out in the literature \cite{Donnay:2021wrk,Compere} since they generate the leading and subleading soft graviton theorem at tree level.  As one is interested in studying possible loop corrections to these tree level theorems through \eqref{renormJ}, the charges \eqref{supertranslation_superrotations_charges} will be enough for the following discussion.  
Indeed,  \eqref{BRST_WI_GR} and \eqref{J_BRST_GR_gf} lead to the Ward identity for the charges \eqref{charge_ward_identity},  which splits in two parts that are 
\begin{align}
\label{leading}
\bra{\rm out} \big( Q^+_\alpha   \mathcal{S} - \mathcal{S}  Q^-_\alpha     \big) \ket{\rm in} = 0 \quad  &\iff  \quad   \text{tree\ level\ leading\ soft\ graviton\ theorem}   ,   
 \\
 \label{subleading}
 \bra{\rm out} \big(   Q^+_{\xi} \mathcal{S} - \mathcal{S}   Q^-_{\xi}   \big) \ket{\rm in} = 0  \quad &\iff \quad  \text{tree\ level\ subleading\ soft\ graviton\ theorem}  .
\end{align}
For the original proof of the equivalence \eqref{leading}, see \cite{BMS_soft_graviton}; and for the equivalence \eqref{subleading}, see~\cite{Kapec,Campiglia}.

 \subsubsection{Supertranslation Ward identity and leading soft graviton theorem}

 Let us make the statement \eqref{leading} more precise.  The leading soft graviton theorem, also known as the Weinberg soft graviton theorem \cite{PhysRev.140.B516}, can be formulated as 
\begin{equation}
\label{leading_soft_theorem}
\lim_{\o \to 0} \o \left. \bra{\rm out}  a^{\rm out}_\pm (\o \hat{q}) \mathcal{S}  \ket{\rm in}  \right\vert_{\rm tree} =  S^{(0)}_\pm    \left. \bra{\rm out}    \mathcal{S} \ket{\rm in} \right\vert_{\rm tree}
\end{equation}
at tree level.  In this equation, $\left. \bra{\rm out}    \mathcal{S} \ket{\rm in} \right\vert_{\rm tree} = \mathcal{M}_n^{\rm tree}(p_1, ..., p_n)$ refers to any scattering process at tree level with $n$ outgoing hard massless particles with momenta $p_k$ (all of them can be taken to be outgoing without loss of generality), and $a^{\rm out}_\pm(\o \hat{q})$ is the creation operator for an outgoing graviton with momentum $q=\o \hat{q}$ and helicity $\epsilon^\pm_{\mu \nu}(\hat{q})$.  This theorem is a factorization theorem. In fact, factorization occurs by taking the soft limit $\o \to 0$ for the outgoing graviton and the universal leading soft factor $S^{(0)}_\pm$ is given by 
\begin{equation}
\label{leading_factor}
S^{(0)}_\pm = \frac{\kappa}{2} \sum_{k=1}^n \frac{p_k^\mu p_k^\nu \epsilon_{\mu \nu}^\pm(\hat{q})}{p_k \cdot\hat{q}}  . 
\end{equation}
By \eqref{leading}, the loop validity of \eqref{leading_soft_theorem} is equivalent to that of the Ward identity for the supertranslation charges.  So far, the only way to investigate the    validity of this Ward identity at the quantum level was to look for possible loop corrections to \eqref{leading_soft_theorem}. If, by Feynman diagram arguments, one can show that \eqref{leading_soft_theorem} is valid at all loops, then one could claim that the same is true for the associated Ward identity \eqref{leading}.  With the framework developed in Section [\ref{Holographic_anomalies}], one can now go the other way around.  One can look for an holographic cocycle $\Delta^1_3$ for the Ward identity \eqref{leading}, which is a $3$-form in spacetime, linear in the supertranslation ghost $\alpha$ and satisfying the consistency condition \eqref{holographic_anomaly}
\begin{equation}
\label{supertranslation_anomaly?}
s \int_{\Sigma^+} \Delta^1_3 = 0 , 
\end{equation}
where $s$ is given by \eqref{BRSTBMSoperator}. If there exists no non trivial solution to \eqref{supertranslation_anomaly?}, then the Ward identity \eqref{leading} is valid at all orders of perturbation theory, and the same is true for the leading soft graviton theorem.  

By inspection,  no solution to \eqref{supertranslation_anomaly?} was found in \cite{Baulieu_Tom_BMS}.  This shows that supertranslation symmetry is an exact symmetry of any gravitational scattering 
and provides an alternative way of proving the exactness of the leading soft graviton theorem \eqref{leading_soft_theorem} at all orders of perturbation theory without ever referring to Feynman diagrams.  Indeed,  Weinberg showed in \cite{PhysRev.140.B516} that the leading universal soft factor \eqref{leading_factor} receives no loop correction.

\subsubsection{Superrotation Ward identity and subleading soft graviton theorem}

The tree level subleading soft graviton theorem \cite{Cachazo} can be written as
\begin{equation}
\label{subleading_soft_theorem}
\lim_{\o \to 0} (1 + \o \pa_\o) \left. \bra{\rm out}  a^{\rm out}_\pm (\o \hat{q}) \mathcal{S}  \ket{\rm in}  \right\vert_{\rm tree} =  S^{(1)}_\pm    \left. \bra{\rm out}    \mathcal{S} \ket{\rm in} \right\vert_{\rm tree} .
\end{equation}
The operation $\lim_{\o \to 0} (1 + \o \pa_\o)$ projects out the leading contribution in $\o$, which is given by the leading soft factor \eqref{leading_factor}. 
The subleading soft factor $S^{(1)}_\pm$ is given by
\begin{equation}
\label{subleading_factor}
S^{(1)}_\pm = - \frac{i \kappa}{2} \sum_{k=1}^n \frac{p_k^\mu   \epsilon_{\mu \nu}^\pm(\hat{q}) q_\lambda }{p_k \cdot\hat{q}}  J_k^{\lambda \nu} ,
\end{equation}
where  the $J_k$'s are the angular momenta of the hard particles. 

As opposed to \eqref{leading_soft_theorem},  \eqref{subleading_soft_theorem} is not valid at all orders of perturbation theory because the subleading soft factor $S^{(1)}_\pm$ receives loop corrections \cite{Bern:2014oka,He_2014}.  The authors of these papers showed that at $1$-loop, the subleading soft factor is corrected as 
\begin{equation}
\label{shift_soft}
S^{(1)}_\pm \longrightarrow S^{(1)}_\pm + \frac{\kappa^2}{\epsilon}  \Big( \hat{\sigma}'_{n+1} S^{(0)}_\pm - S^{(1)}_\pm \sigma_n   \Big) ,
\end{equation}
with 
\begin{align}
\hat{\sigma}'_{n+1} &\equiv \frac{\hbar}{2(4 \pi)^2} \sum_{k=1}^n  (p_k \cdot \hat{q}) \ln \left( \frac{p_k \cdot \hat{q}}{\mu}  \right) , 
\nn \\
\sigma_n &\equiv \frac{\hbar}{(8 \pi)^2} \sum_{k,l=1}^n (p_k \cdot p_l) \ln \left(\frac{p_k \cdot p_l}{\mu} \right),
\end{align} 
and where $\epsilon$ is the infrared regulator in the loop.  The subleading soft graviton theorem is then $1$-loop exact. 
 Since \eqref{subleading_soft_theorem} is equivalent to the Ward identity for the superrotation charges, 
 one can  wonder whether such a correction might not come from  an holographic anomaly for superrotations $\Delta^1_3 = d \Delta^1_2$ at 1-loop, which would be linear in the superrotation ghosts and with a non trivial $\Delta^1_2$.  For this to be true, the  anomaly must  respect the consistency condition \eqref{supertranslation_anomaly?}. It must also be such that the induced 1-loop correction on the superrotation charge given by \eqref{Charge_loop_correction}, which writes
\begin{equation}
\label{correction_superrotation_charge}
 Q^+_{\xi}  \longrightarrow  {Q^+_{\xi}}' =  Q^+_{\xi}  - \mathfrak{a}^{1}_{\rm boundary}  \hbar \int_{\mathcal{I}^+_-}   \Delta^1_{2} ,
\end{equation}
reproduces the shift \eqref{shift_soft} by considering the corrected Ward identity  \eqref{corrected_WI_charges}
\begin{equation}
\bra{\rm out} \big(   {Q^+_{\xi}}' \mathcal{S} - \mathcal{S}   {Q^-_{\xi}}'   \big) \ket{\rm in} \Big\vert_{1-\text{loop}} = 0.
\end{equation}
A correction of the type \eqref{correction_superrotation_charge} that correctly reproduces the shift  \eqref{shift_soft} was found in \cite{He:2017fsb}, it is given by
\begin{align}
\label{andy_correction}
\int_{\mathcal{I}^+_-}   \Delta^1_{2} &= \int_{\mathcal{I}^+_-} d^2 z \  \xi \Big(  2 \bar{ \mathcal{N} }^0 \pa  \mathcal{N}^0 + \pa ( \bar{\mathcal{N}}^0 \mathcal{N}^0)  \Big) = \int_{\Sigma^+} du \w dz \w d\bz \   \pa_u \left[ \xi   \Big(  2 \tilde{\bar{\mu }} \pa  \mathcal{N}^0 + \pa ( \tilde{\bar{\mu }} \mathcal{N}^0)  \Big) \right]     
\end{align}
and $\mathfrak{a}^{1}_{\rm boundary} = \frac{i}{\epsilon \pi \kappa^2}$.
However, the authors of this paper did not verify the crucial condition  \eqref{supertranslation_anomaly?}  for such a correction to take place.  It amounts to compute the $s$ variation of \eqref{andy_correction} for the BRST operator \eqref{BRSTBMSoperator} evaluated at $\alpha = 0$.  After a tedious calculation, one gets 
\begin{align}
\label{invalidating_loop_correction}
s \int_{\Sigma^+} du \wedge dz \wedge d\bz \   \pa_u \left[ \xi   \Big(  2 \tilde{\bar{\mu }} \pa  \mathcal{N}^0 + \pa ( \tilde{\bar{\mu }} \mathcal{N}^0)  \Big) \right]   &= -3 \int_{\Sigma^+} du \wedge dz \wedge d\bz \   \pa_u \Big( \xi \pa \xi \tilde{\bar{\mu}} \pa \mathcal{N}^0  \Big) 
\nn \\
&= -3 \int_{\mathcal{I}^+_-} d^2 z  \     \xi \pa \xi \Big( \bar{ \mathcal{N} }^0 \pa \mathcal{N}^0  \Big)   \neq 0 . 
\end{align}
This result indicates that the 1-loop charge renormalization of \cite{He:2017fsb} cannot occur. 

The correct way of getting the shift \eqref{shift_soft} was exposed in \cite{Donnay:2022hkf}.  It consists in starting with the correct expression for the integrable superrotation soft charges \eqref{supertranslation_superrotations_charges} at tree level, namely 
\begin{equation}
\label{new_superrotation_charges}
Q^{\text{soft, new}}_{\xi} = - \frac{2}{\kappa^2}  \int_{\mathcal{I}^+_-} d^2z \ \xi \Big(   \pa^3  \mathcal{N}^1 + \pa^3 ( C^0 \mathcal{N}^0 )  - 3 \bar{\pa}^2 \bar{\mathcal{N}}^0 \pa C^0 - C^0 \pa \bar{\pa}^2 \bar{\mathcal{N}}^0 \Big) ,
\end{equation}
where $C^0 (z,\bz)= \demi \big( C_+ (z,\bz) + C_- (z,\bz) \big)$ is the supertranslation Goldstone mode of spontaneously broken supertranslation symmetry.  One can insert such modes into the $\mathcal{S}$-matrix by using 
\begin{align}
\label{celestial_correlator_1-loop}
\left. \bra{\text{out}}  C^0 (z,\bz) \mathcal{S}  \ket{\text{in}}  \right\vert_{\text{tree}} &= 0 \  ,
\nn \\
\left. \bra{\text{out}}  C^0 (z,\bz) \mathcal{S}  \ket{\text{in}}  \right\vert_{k-\text{loop}} &= - \frac{i \kappa^2}{\epsilon} \hat{\sigma}'_{n+1} \left. \bra{\text{out}}  \mathcal{S}  \ket{\text{in}}  \right\vert_{k-\text{loop}} \   
\end{align}
for $k \geq 1$. 
These identities are derived by studying the celestial correlator $\left< C^0 (z_1,\bz_1) C^0 (z_2,\bz_2) \right>$ and the IR factorization property of the $\mathcal{S}$-matrix \cite{Donnay:2022hkf,Himwich:2020rro,Nguyen:2021ydb}. The subleading soft graviton theorem \eqref{subleading_soft_theorem}  with the correct shift \eqref{shift_soft} at 1-loop is  then equivalent to the Ward identity for the superrotation charges given by \eqref{new_superrotation_charges}. 

Since the computation \eqref{invalidating_loop_correction} seems to invalidate the loop correction to the superrotation charges advocated in \cite{He:2017fsb}, it would be interesting to check if something goes wrong in \cite{Pasterski:2022djr}, in which the author tries to link the result of \cite{He:2017fsb} to that of \cite{Donnay:2022hkf}. 
Notice that the result \eqref{invalidating_loop_correction} should also be welcomed in the context of \cite{Agrawal:2023zea}. Indeed,  the authors of this paper showed that when massive particles are incorporated,  the superrotation Ward identity for   $Q^{\text{soft, new}}_{\xi}$ plus additional charges $Q^{i^\pm}_\xi$ at timelike infinities
 is also responsible for the all-loop exact logarithmic soft graviton theorem \cite{Sahoo:2018lxl}, so superrotation invariance of the gravitational $\mathcal{S}$-matrix better not be anomalous.




\def\ss{{\tilde s}}

\subsubsection{Non trivial holographic cocycle for extended BMS symmetry}

Although supertranslation and superrotation symmetries seem to be exact symmetries of the quantum gravity $\mathcal{S}$-matrix by the previous points, it doesn't mean that there are no non trivial holographic cocycles of the form \eqref{holographic_anomaly} for extended BMS symmetry.  In fact,   we showed in \cite{Baulieu_Tom_BMS} that such a cocycle does appear.  Indeed, for the BRST operator associated with extended BMS symmetry \eqref{BRSTBMSoperator},  there is a non trivial solution to the local cocycle equation 
\begin{equation}
\label{WZC}
s\Delta ^g_{ 4-g    } +d \Delta ^{g+1}_{3-g    }  =0
\end{equation}
for $g=1,2,3,4$. 
The full cocycle  $\tilde{\Delta}_4$ in $\Sigma^+$ is given by
 \begin{align}\label{solsolsol} 
 \Delta^1_{3}   =& \  du\w dz\w d\bz    \ (\mu \pa^3   \xi     +  \bar{\mu} \bar{\pa}^3 \bar{\xi}) ,
 \nn \\
{\Delta^2_2 }  =& \ dz\w  d\bz \  \xi^u ( \mu \pa^3 \xi + \bar{\mu} \bar{\pa}^3 \bar{\xi}  )    - du \w d\bz \  \big( \xi ( \mu \pa^3 \xi + \bar{\mu} \bar{\pa}^3 \bar{\xi}  ) + \pa \xi^u \bar{\pa}^3 \bar{\xi} \big) + du\w  dz \ \big( c.c. \big),
\nn \\
{\Delta^3_1 } =& - du \  \big(\xi \bar{\xi} ( \mu \pa^3 \xi + \bar{\mu} \bar{\pa}^3 \bar{\xi} ) + \xi (\pa^3 \xi) (\bar{\pa} \xi^u) - \bar{\xi} (\bar{\pa}^3  \bar{\xi}) (\pa \xi^u )  \big)
\nn \\
& - d\bz \ \xi^u \big( \xi (\mu \pa^3 \xi + \bar{\mu} \bar{\pa}^3 \bar{\xi} ) + \pa \xi^u \bar{\pa}^3 \bar{\xi} \big)  +dz 
\ \xi^u (c.c.)   )  , 
\nn \\
{\Delta^4_0}  =&\   \xi \pa^3 \xi (\bar{\xi} \xi^u \mu  + \xi^u \bar{\pa} \xi^u ) - {\bar{\xi}} \  {\bar{\pa}^3} \bar{\xi}  (  \xi \xi^u \bar{\mu}  + \xi^u \pa \xi^u ) ,
\end{align}
where $\xi^u \equiv \xi^u_{\rm BMS}$.   In order for $\Delta^1_3$ to be an holographic cocycle one must verify that it fulfils the global consistency condition \eqref{supertranslation_anomaly?} when the falloffs of the fields satisfy   \eqref{u_falloffs_new} near $u=\pm \infty$.
In fact,  \eqref{supertranslation_anomaly?}  is only satisfied by   the hard part of $\Delta^1_3$ in \eqref{solsolsol}, namely its $\tilde{\mu}$ dependent part and
when $s$ only involves the superrotations. This   operation $\ss$  is  defined  by \eqref{BRSTBMSoperator} evaluated at $\alpha=0$.  With these restrictions, one gets\begin{align}
\label{consistent_multiplicative_renorm}
\ss \int_{\Sigma^+} \tilde{\Delta}^1_3 &= \int_{\Sigma^+} du \w dz \w d\bz \ \pa_u \Big(   \frac{u}{2} \pa_A \xi^A ( \tilde{\mu} \pa^3 \xi + \tilde{\bar{\mu}} \bar{\pa}^3 \bar{\xi}  ) \Big) 
\nn \\
&= \Big[   \frac{u}{2} \pa_A \xi^A ( \tilde{\mu} \pa^3 \xi + \tilde{\bar{\mu}} \bar{\pa}^3 \bar{\xi}  ) \Big]^{u=+\infty}_{u=-\infty} =  0.
\end{align}
There is of course  a conjugate cocycle on $\Sigma^-$.

One may observe that the  consistent and non trivial  superrotations cocycle  $ \tilde{\Delta}^1_3 $     does not break the charge Ward identity \eqref{subleading}.
In fact, it is  nothing but the original soft charge for superrotations \eqref{supertranslation_superrotations_charges} modulo a factor $-\kappa^2/2$.  One has indeed    
\begin{align}
\label{superrotation_charges_tree_level}
 Q^{\text{soft}}_\xi + Q^{\text{soft}}_{\bar{\xi}}  \propto   \int_{\mathcal{I}^+_-} d^2z  \Big( \xi \pa^3  \mathcal{N}^1 + \bar{\xi} \bar{\pa}^3 \bar{\mathcal{N}}^1  \Big)  &=    \int_{\Sigma^+} du \w dz \w d\bz  \Big(  \xi \pa^3 ( u \pa_u \tilde{\mu} ) + \bar{\xi} \bar{\pa}^3 ( u \pa_u \tilde{\bar{\mu}} )   \Big)
\nn \\
&= - \int_{\Sigma^+} du \w dz \w d\bz  \Big(  \xi \pa^3 \tilde{\mu} + \bar{\xi} \bar{\pa}^3 \tilde{\bar{\mu}} \Big) =   \int_{\Sigma^+} \tilde{\Delta}^1_3 \ .
\end{align}
Having a non vanishing coefficient $\mathfrak{a}^{1}_{\rm boundary}$ in \eqref{correction_superrotation_charge} for this cocycle would therefore mean that the superrotation charges generating the tree level subleading soft graviton theorem get multiplicatively renormalized. It clearly cannot  be the case since the renormalization of the subleading soft factor \eqref{shift_soft} at 1-loop is not of that kind, so $\mathfrak{a}^{1}_{\rm boundary}=0$.

  Another superrotation  cocycle  exists that  was not present in \cite{Baulieu_Tom_BMS} and requires the weakened $u$-falloffs  \eqref{u_falloffs_new}, namely the loss of peeling.
If one defines $\Delta \mu^{L} \equiv \mu^{L,+} - \mu^{L,-}$, this quantity transforms as  
\begin{equation}
s \Delta \mu^{L} = (  \xi^A \pa_A + \bar{\pa} \bar{\xi} - \pa \xi ) \Delta \mu^{L}
\end{equation}
under the extended BMS symmetry \eqref{BRSTBMSoperator}. This new quantity  leads to the following   non trivial solution of \eqref{supertranslation_anomaly?} 
\begin{equation}
\label{soft_log_charge}
Q_{\xi,\text{soft}}^{\log} = - \frac{2}{\kappa^2} \int_{\Sigma^+} du \w dz \w d \bar{z} \  (\pa^3 \xi)  \pa_u (u^2 \pa_u \tilde{\mu}) =  \frac{2}{\kappa^2} \int_{S^2} dz \w d \bar{z} \  (\pa^3 \xi) \Delta \mu^{L} 
\end{equation}
with 
\begin{align}
\label{consistent_log_renormalization}
s Q_{\xi,\text{soft}}^{\log} &= \frac{2}{\kappa^2} \int_{S^2} dz \w d \bar{z} \  \Big(  (2 \pa \xi \pa^3 \xi + \xi \pa^4 \xi) \Delta \mu^{L}  - \pa^3 \xi ( \xi^A \pa_A + \bar{\pa} \bar{\xi} - \pa \xi ) \Delta \mu^{L}    \Big) = 0.
\end{align}
Here   part   integrations on $\xi \pa^4 \xi \Delta \mu^{L}$ and  on $\pa^3 \xi  \bar{\xi} \bar{\pa} \Delta \mu^{L} $ have been used.   The quantity $Q_{\xi,\text{soft}}^{\log}$ was introduced  in 
\cite{Choi:2024ygx,Geiller:2024ryw}  as   the 1-loop correction to the superrotation soft charge  linked to the 1-loop logarithmic corrections to the soft graviton theorem.    The cocycle equation  \eqref{consistent_log_renormalization} shows   that such a renormalization procedure is perfectly admissible.
  If one compares~the    soft charge 1-loop correction of    \cite{Choi:2024ygx} with \eqref{correction_superrotation_charge} for the consistent and non trivial cocycle $Q_{\xi,\text{soft}}^{\log}$, one gets $\mathfrak{a}^{1,\log}_{\rm boundary} = \frac{1}{\epsilon}$.\footnote{This comparison  requires the use of $\epsilon^{-1} \longleftrightarrow \ln(\lambda^{-1})$ where $\epsilon$ comes from dimensional regularization while $\lambda$ is an IR cutoff, see the discussion around Eq.(2.12) of \cite{Agrawal:2023zea}.}
This observation   makes the results of \cite{Choi:2024ygx,Geiller:2024ryw} consistent with the rules of perturbative quantum field theory.  It is interesting to underline the   formal ressemblance between the cocycle \eqref{soft_log_charge} and a 2d Virasoro type anomaly $  \pa_z c^z\pa_z^2\mu^z_\bz$  that was looked for in \cite{Baulieu_Tom_BMS}.

The link between the two ways \cite{Agrawal:2023zea, Choi:2024ygx} of getting the 1-loop logarithmic corrections of the soft graviton theorem remains unclear although it must be closely related to the 1-loop correction of the celestial correlator \eqref{celestial_correlator_1-loop}.

Let us conclude this section by noting that   the bulk computation of the consistent and non trivial holographic cocycle  $\tilde{\Delta}^1_3$   for    superrotation symmetry  may provide  a solid starting point in the quest of finding an exact holographic matching between bulk and boundary quantities in flat space.  
Indeed,  if one considers a Carrollian field theory in $\mathcal{I}^+$ dual to the bulk theory, then the cocycle $\tilde{\Delta}^1_3$ must be computable from this intrinsic boundary theory.  One could for example construct the purely three dimensional horizontality conditions  defining the BRST operator associated to the conformal Carroll symmetry and look for the cocycle from there.   This  quest  is to be pursued in a  near future.

\section{Conclusion}

This work studies the gauge dependence    
 of the  covariant phase space (CPS)   construction of corner  charges generating asymptotic symmetries on phase space.  
 It does so by extending the genuine bigraded CPS of a theory with  a given gauge symmetry parametrized by $\lambda(x)$  into a trigraded BRST CPS involving ghosts, antighosts and Lagrange multiplier fields. 
The~BRST CPS geometrically encodes the nilpotent BRST symmetry operator $s$ associated with the gauge symmetry as a field space Lie derivative $L_{\V}$ along a specific ghost number one   field space vector field $\V$   acting on the entire BRST CPS.

This extended framework makes it possible to construct the   gauge dependent BRST Noether current  $\J$ with ghost number one that is  associated with  the global BRST~invariance  of any given   gauge fixed action.  
 
 Based on explicit computations of this current in Yang--Mills theory and first order gravity in various gauges, 
the paper    puts forward  the   conjecture that the   generic field dependence  of   $\J$ can be decomposed on-shell 
as a sum of BRST and  $d$ exact terms.  As a matter of fact, one finds in the studied cases that
\begin{equation}
\label{Noether_2_conclusion}
\J \ \hat{=} \ s G_{\rm gauge} (  \vp_{\rm cl}, c,\bar{c}, b ) - d q_{\rm cl} (\vp_{\rm cl} , c)  .
\end{equation}
The gauge dependent part  $G_{\rm gauge} (  \vp_{\rm cl}, c,\bar{c}, b )$ can be    computed   unambiguously as  a local  functional of  the  gauge  function.   Its  dependence 
of the BRST corner Noether charge~$ q_{\rm cl}$ on the ghost~$c$ is  linear.  The relationship between
 $ q_{\rm cl}$ and the standard  corner Noether charges~$q_\lambda$ (that originate from Noether's second theorem and that probe asymptotic symmetries)~is
\begin{equation}
\star J_{\lambda} \ \hat{=} \ d q_\lambda \quad  \text{with} \quad  q_\lambda  ( \vp_{\rm cl}, \lambda)  =  \left.q_{\rm cl} (\vp_{\rm cl},c)\right\vert_{c = \lambda} .
\end{equation}

The article  puts forward a second conjecture  concerning 
 the on-shell value of the non integrable part of $I_{V_{\rm BRST}} \O$ for the non degenerate BRST symplectic two form $\O$,  
which is in the same spirit as   \eqref{Noether_2_conclusion} and which is  verified in the same particular cases. 
The gauge independence of   asymptotic symmetries  and of their action on phase space 
  is then proved by the $s$-exactness of the gauge dependent part of $\J$  in \eqref{Noether_2_conclusion} and by this second conjecture.  
  These results imply the gauge independence of the phase space canonical quantization method in Yang--Mills theory and gravity if  one  were to    promote   the brackets $I_{V_{\rm BRST}} I_{V_{\rm BRST}} \O$ into quantum commutators.  



Another important part of the paper is the determination 
of a unified BRST Ward~identity for   small and large gauge transformations.  It splits into a bulk part for small gauge transformations that ensures the ultraviolet renormalization of the theory via the BRST~BV quantum master equation and into a boundary part for large gauge transformations that 
ensures the conservation of $\J$ at the quantum level.  In addition with \eqref{Noether_2_conclusion},  the boundary  i.e. holographic  Ward identity   provides   a Lagrangian justification of the Hamiltonian argument of the invariance of the physical $\mathcal{S}$-matrix under asymptotic symmetries, $[Q,\mathcal{S}]=0$, and allows one to verify its   validity  at any given   finite order of  perturbation theory.

In fact, the paper   shows that  a consistent breaking of the holographic Ward identity may occur at higher loop levels. It is  related to  the existence of   non trivial  solutions of a codimension one Wess--Zumino consistency condition.  Such solutions are  called      holographic cocycles.  
   They imply   perturbative  renormalizations of  the  large gauge transformation charges, namely  loop corrections to    soft theorems.

%

In asymptotically flat gravity,   these 
   holographic cocycles are for  the extended BMS~symmetry. 
Their study makes it possible to establish a link between this work and  the existing literature on  the 1-loop correction of the   soft graviton theorem.
 It   
  puts
 the approaches   of   \cite{Donnay:2022hkf,Agrawal:2023zea,Geiller:2024ryw,Choi:2024ygx} on a firmer footing.

\begin{center}
\textbf{Acknowledgments}
 \end{center}
\indent It is a pleasure to thank  Luca Ciambelli,  Marc Henneaux, Andrea Puhm,   Romain Ruzziconi and especially Marc Bellon for insightful discussions.

\appendix

\section{Noether's Second Theorem}
\label{Annexe_A}

The derivation of Noether's second theorem presented in this appendix closely follows the one of \cite{Avery_Schwab}. This theorem applies to a field theory $(\mathcal{F},S)$ with a field space composed of classical fields $\vp=\vp_{\rm cl}$ only,  and with an action $S = \int_M d^dx \ L$ which is invariant under infinitesimal symmetry transformations of the form\footnote{More generic transformations with higher order derivatives can also be considered, as Noether originally did \cite{Noether_1971}.}
\begin{equation}
\label{local_transformations}
 \delta_\lambda \vp = f(\vp) \lambda + f^\mu(\vp) \pa_\mu \lambda
\end{equation}
for an arbitrary local parameter $\lambda(x)$. It thus concerns local symmetries.  

One may first recall the local property $\delta L = E_\vp \delta \vp + \pa_\mu \theta^\mu$. Using \eqref{semi_sym} and \eqref{local_transformations} leads to 
\begin{align}
\label{local_identity}
\pa_\mu B^\mu_\lambda = \delta_\lambda L &= E_\vp \delta_\lambda \vp + \pa_\mu (I_{V_\lambda} \theta^\mu)
= \lambda \Big(  E_\vp f - \pa_\mu ( E_\vp f^\mu ) \Big)  + \pa_\mu \Big( I_{V_\lambda} \theta^\mu + \lambda E_\vp f^\mu  \Big) .
\end{align}
This local identity turns into a global one when varying the action, namely 
\begin{equation}
\label{global_Noether}
0 = \int_M d^d x \ \lambda \Big(  E_\vp f - \pa_\mu ( E_\vp f^\mu ) \Big) + \int_M d^d x \ \pa_\mu \Big( I_{V_\lambda} \theta^\mu - B^\mu_\lambda + \lambda E_\vp f^\mu  \Big) .
\end{equation}
One can now take advantage of this identity by noting that it must be valid for any arbitrary function $\lambda(x)$. In particular, one can choose $\lambda$ to have compact support, in which case the second term in the r.h.s.  of \eqref{global_Noether} vanishes and one gets 
the local identity 
\begin{equation}
\label{Delta_constraints}
E_\vp f - \pa_\mu ( E_\vp f^\mu )  = 0  .
\end{equation}
Using this equation, the identity \eqref{local_identity}
 and the definition of the Noether current $J_\lambda$ \eqref{classical_Noether_current}, one gets 
 \begin{equation}
 \pa_\mu ( J^\mu_\lambda + \lambda E_\vp f^\mu  ) = 0 .
 \end{equation}
 Finally, one can turn this equation into an identity on spacetime differential forms and use the algebraic Poincar\'e lemma to obtain 
\begin{equation}
\label{demo_Noether_2}
\star J_\lambda = - \lambda E_\vp (\star f) + d q_\lambda \ \hat{=} \ d q_\lambda.
\end{equation}
This is Noether's second theorem, where the Noether constraints that are vanishing on-shell are given by $C_\lambda = \lambda E_\vp (\star f)$.  


The reason why Noether's second theorem is fundamentally not valid for a residual global BRST symmetry of a gauge fixed Lagrangian of the form \eqref{gauge_fixing} is readily understood by looking at \eqref{local_transformations}. Indeed, this gauge fixing involves the trivial BRST fields $(\bar{c},b)$ that transform as $s \bar{c} = b$ and $sb=0$ under BRST symmetry. So if one proceeds to a ghostification of the local parameter $\lambda(x)$ into the ghost fields $c(x)$, it is straightforward to see that the transformation of the antighost $\bar{c}$ cannot be put into the form \eqref{local_transformations}.  Consequently, the reasoning leading to \eqref{demo_Noether_2} cannot be applied. 

\section{Comments on the Anomaly Operator}
\label{Annex_anomaly}


%
%

The generic separation of $I_V \O$ into integrable and non integrable parts is very well known in the classical covariant phase space. To show it in full generality, it requires using the so-called anomaly operator \cite{Freidel,Hopfmuller:2018fni,Chandrasekaran:2020wwn}.   The goal of this appendix is to remind how this operator is defined in the classical covariant phase space and what it is used for.  Then, using the classical case as a guideline,  one  tries to extend its definition to the BRST CPS of Section [\ref{Section_trigraded}].  Ambiguities are present in this extension.  In fact, using a counter example, one can show that the generic property that the anomaly operator must obey is ruled out in the gauge fixed BRST CPS, no matter how its extended action on the non physical fields is defined.  

\subsection{Generic case for reparametrization symmetry}

The relevance of introducing the anomaly operator appears when dealing with theories that possess a spacetime symmetry. Suppose that this symmetry is parametrized by the vector ghost $\xi$. In such cases, it is known that the action of $s=L_{V_\xi}$ can be compared to the action of $\Lie_\xi - I_{V_{\delta \xi}}$. Most of the time, these two operators coincide (see e.g. the derivation above Eq.(4.16) of \cite{Gomes:2016mwl} or Eq.(A.12) of \cite{Speranza:2017gxd}). This is very useful when computing quantities such as $sL = \Lie_\xi L=-d(i_\xi L)$, or $s\theta = \Lie_\xi \theta - I_{V_{\delta \xi}} \theta$.  In particular, one can use this fact to get a quick derivation of the fundamental canonical relation by writing 
$
I_{V_\xi} \o = s \theta + \delta I_{V_\xi} \theta = \Lie_\xi \theta - I_{V_{\delta \xi}} \theta + \delta I_{V_\xi} \theta 
$. \\
\indent However, when one fixes the gauge and/or adds non trivial boundary conditions, this correspondence is not necessarily true anymore.  This paper is precisely concerned by the former.  Indeed,  suppose reparametrization symmetry is fixed by adding an $s$-exact term to the Lagrangian, then $sL \neq \Lie_\xi L$ because $s$ does not act as a spacetime Lie derivative on the ghosts, antighosts and Lagrange multipliers (see \eqref{s_RG}).  Hence,  if one wants to take advantage of using $\Lie_\xi$ when varying an object, it is mandatory to keep track of the difference with the real variation $s$ by defining the   \textit{anomaly operator}
\begin{equation}
\label{Def_anomaly_operator}
\Delta_{\xi} \equiv  s - \Lie_{\xi} + I_{V_{\delta \xi}} = \hat{s} + I_{V_{\delta \xi}} .
\end{equation}
Classically, this operator generalizes the derivation of \eqref{fund_with_fluxes} to every theory with specific boundary conditions \cite{Freidel}.  
Notice that strictly speaking, in the BRST case,  one should have written $s \equiv L_{V_{\mathsf{C}}}$ where $\C = \{ \xi,\bar{\xi},b \}$ and this is actually what $s$ means in the previous equation.  One sticks with the $\xi$ notation to be closer to the existing literature on the subject and also because the Lie derivative in spacetime is only defined along $\xi$, not along $\bar{\xi}$ or $b$.  Let us then reproduce the derivation of the fundamental canonical relation using this operator, by taking into account the anticommuting nature of $\xi$ and by pointing out which assumptions fail to hold in the gauge fixed trigraded BRST CPS.

The following graded commutation rules hold: 
\begin{align} 
\delta i_\xi &= i_\xi \delta +i_{\delta \xi}
\nn \\
\delta \Lie_\xi &= - \Lie_\xi \delta + \Lie_{\delta \xi}
\nn \\
\delta \Delta_{\xi} &= - \Delta_{\xi} \delta + \Delta_{\delta \xi}
\end{align}
with $\Delta_{\delta \xi} \equiv L_{V_{\delta \xi}} - \Lie_{\delta \xi}$. 

Let us use the example of conventional GR with field content $(g_{\mu \nu}, \xi^\mu, \bar{\xi}^\mu, b^\mu)$ to clarify the notations. The BRST operator $s \equiv L_{V_{\xi}}$ acts on the fields as 
\begin{align}
\label{GR_example_delta}
s g_{\mu \nu} = \Lie_\xi g_{\mu \nu} \quad &\Longrightarrow \quad  \Delta_{{\xi}} g_{\mu \nu} = 0 ,
\nn \\
s \xi^\mu = \demi \Lie_\xi \xi^\mu \quad &\Longrightarrow\quad   \Delta_{{\xi}} \xi^\mu  = - \demi \Lie_\xi \xi^\mu ,
\nn \\
s \bar{\xi}^\mu = b^\mu  \quad &\Longrightarrow  \quad \Delta_{\xi} \bar{\xi}^\mu = b^\mu - \Lie_\xi  \bar{\xi}^\mu ,
\nn \\
s b^\mu = 0 \quad &\Longrightarrow \quad  \Delta_{\xi} b^\mu = -\Lie_\xi b^\mu .
\end{align}
When acting on $0$-forms in field space, $\Delta_\xi$ is simply $\hat{s}$, which has the important property $s^2 = 0 \iff \hat{s}^2 = 0$.
As mentioned above, notice that although $V_\xi$ only contains $\xi$ in its subscript, it generates through the Lie derivative in field space all the BRST transformation of the fields, even for the antighosts. 
One still has to define the action of $I_{V_{\delta \xi}} \delta$ on each fields. 
Classically,  this action is understood as being a diffeomorphism symmetry transformation with  parameter  $\delta \xi$, which is a spacetime vector field valued in the field space 1-forms. 
In the BRST CPS case, whereas the odd operator $I_{V_\xi} \delta$  is defined on fields such that $s^2=L_{V_\xi}^2=0$,  $I_{V_{\delta \xi}} \delta$ is an even operator, so there is no reason to define it such that $L_{V_{\delta \xi}}^2=0$.  In fact,  it is not even possible to do so.\footnote{Usually, the nilpotency of $s$ on the ghosts comes from the graded Jacobi identity $\{ \xi, \{ \xi, \xi \} \} = 0$. Here, one would find $L^2_{V_{\delta \xi}} \xi \propto [\delta \xi, [\xi, \delta \xi]] \neq 0$. Putting $\mathfrak{a}_2 = 0$ in \eqref{I_V_delta_g} does not save the day since it becomes impossible to ensure $L^2_{V_{\delta \xi}} g_{\mu \nu} = 0$ without also requiring $\mathfrak{a}_1 = 0$, in which case $I_{V_{\delta \xi}}$ is completely useless.} 
 So sticking with the classical interpretation and looking at \eqref{GR_example_delta},  its most general and coherent action on fields is   given~by
\begin{align}
\label{I_V_delta_g}
I_{V_{\delta \xi}} \delta g_{\mu \nu} &=\mathfrak{a}_1 \Lie_{\delta \xi} g_{\mu \nu}   ,
\nn \\
I_{V_{\delta \xi}} \delta \xi^\mu &= \mathfrak{a}_2 \Lie_{\delta \xi} \xi^\mu \equiv \mathfrak{a}_2 [\delta \xi,\xi]^\mu,
\nn \\
I_{V_{\delta \xi}} \delta \bar{\xi}^\mu &= \mathfrak{a}_3 \delta b^\mu    ,
\nn \\
I_{V_{\delta \xi}} \delta b^\mu &= 0   .
\end{align}
At this stage,  the definition of this operator in the BRST CPS is arbitrary because of the coefficients $(\mathfrak{a}_1,\mathfrak{a}_2,\mathfrak{a}_3)$. However, if one really tries to stick with the classical derivation of the fundamental canonical relation and extend it to the BRST CPS, these coefficients have to be fixed such that the crucial property $\Delta_\xi E = 0$ holds true,  where $E= \delta L - d \theta$.  Here is why. 

One can exploit the fact that a Lagrangian can be ''\textit{anomalous}"\footnote{Be aware that the term anomaly here has nothing to do with gauge anomalies in the quantum field theory sense.}  in the sense that 
\begin{align}
sL = dl_\xi  \quad \quad \text{and} \quad \quad \Lie_\xi L = - d i_\xi L \neq s L . 
\end{align}
The anomaly of a Lagrangian is then defined as 
\begin{align}
\label{anomaly_L}
\Delta_\xi L &= d a_\xi \quad \text{with} \quad a_\xi = l_\xi + i_\xi L.
\end{align}
Classically, this definition also leads to the relation
\begin{equation}
\label{delta_anomaly}
\Delta_{\delta \xi} L = d a_{\delta \xi}.
\end{equation}
The failure of the generalization of the classical procedure to the gauge fixed BRST CPS originates from the non validity of this equation.  More precisely, the only piece that can possibly not be a boundary term is 
\begin{equation}
\label{obstruction_anomaly}
\Delta_{\delta \xi} L = ( L_{V_{\delta \xi}} - \Lie_{\delta \xi}) L \supset I_{V_{\delta \xi}} E   \neq dX.
\end{equation}
An explicit example of this fact will be provided in Appendix [\ref{subsection_concrete}].  It will occur no matter how the coefficients $(\mathfrak{a}_1,\mathfrak{a}_2,\mathfrak{a}_3)$ are fixed.  
Notice that even without this counterexample, if one were to believe \eqref{delta_anomaly} in the BRST CPS,  one wouldn't be able to generically define the quantity $a_{\delta \xi}$ if $a_\xi$ wasn't linear in the ghosts, antighosts and Lagrange multipliers.

To see what are the consequences of this failure in the rest of the reasoning, let us assume that \eqref{delta_anomaly} holds true and check where does the contradiction reappear.  One then deduces some relations from the fundamental identity $\delta L = E + d \theta$ and from the definition of the constraints $I_V E = dC$. One has
\begin{align}
d a_\xi = \Delta_\xi L &= I_{V_\xi} \delta L + d i_\xi L
\nn \\
&= I_{V_\xi} E + d I_{V_\xi} \theta + d i_\xi L
\nn \\
&= d ( C_\xi + I_{V_\xi} \theta + i_\xi L  ) ,
\end{align}
so that the Noether current is given by
\begin{equation}
\star J_{\rm BRST} = I_{V_\xi} \theta - l_\xi = I_{V_\xi} \theta - a_\xi + i_\xi L \equiv - C_\xi - dq_\xi .
\end{equation}
In order to get $I_{V_\xi} \o$,  one can then evaluate\footnote{The graded identity $I_{V_{\delta  \xi}} \theta + i_{\delta \xi} L = + ( d q_{\delta \xi} + C_{\delta \xi} - a_{\delta \xi})$ was used, where the unexpected plus sign comes from the change of sign in the definition of the Lie derivatives along $\delta \xi$ and $V_{\delta \xi}$.} 
\begin{align}
\label{Delta_theta1}
\Delta_\xi \theta &= I_{V_\xi} \delta \theta - \delta ( I_{V_\xi} \theta + i_\xi L  ) + ( I_{V_{\delta  \xi}} \theta + i_{\delta \xi} L  )  + di_\xi \theta + i_\xi E
\nn \\
&= I_{V_\xi} \o + \delta (  dq_\xi + C_\xi - a_\xi  ) + (  d q_{\delta \xi} + C_{\delta \xi} - a_{\delta \xi} ) + di_\xi \theta + i_\xi E .
\end{align}
An other way of writing  $\Delta_\xi \theta$, with which the previous equation must be compared,  is obtained by considering the following quantity
\begin{align}
\label{delta_Delta_L}
- d \delta a_\xi = \delta \Delta_\xi L &= - \Delta_\xi \delta L + \Delta_{\delta \xi} L = d ( \Delta_\xi \theta + a_{\delta \xi} ) - \Delta_\xi E .
\end{align}
Here comes the contradiction again. In the classical case,  one uses the fact that the equations of motion are non anomalous, namely that $\Delta_\xi E = 0$ off-shell. This means that the anomaly for the presymplectic potential can be defined as
\begin{equation}
\label{Delta_theta2}
\Delta_\xi \theta \equiv d A_\xi - \delta a_\xi - a_{\delta \xi}.
\end{equation}
However, the same counterexample that provides $I_{V_{\delta \xi}} E   \neq dX$ now gives $\Delta_\xi E \neq 0$ and, even worse,  $\Delta_\xi E \neq dY$.  This is because of the property \eqref{def_Z}
\begin{equation}
\label{def_Z1}
d \delta l_\xi = s \delta L = sE - ds\theta \quad \Longrightarrow \quad sE \equiv dZ ,
\end{equation}
leading to
\begin{equation}
\Delta_\xi E = s E - \Lie_\xi E + I_{V_{\delta \xi}} E = d( Z + i_\xi E ) + I_{V_{\delta \xi}} E .
\end{equation}
This means that one cannot use \eqref{Delta_theta2} and plug it in \eqref{Delta_theta1} to obtain the usual fundamental canonical relation 
\begin{equation}
\label{fundamental_canon}
 - I_{V_\xi} \o =   \delta \Big(  C_\xi + dq_\xi \Big)  + C_{\delta \xi} + d \Big( q_{\delta \xi} + i_\xi \theta - A_\xi \Big) .
\end{equation}
Classically, integrating this equation over $\Sigma$ and putting it on-shell leads to \eqref{fund_with_fluxes} for a spacetime symmetry. 
One can now expose the counterexample invalidating \eqref{fundamental_canon}.


\subsection{A concrete example}
\label{subsection_concrete}

An explicit example of \eqref{obstruction_anomaly} appears in gauge fixed gravity and is highlighted in \eqref{counterexample_GF_gravity}.  This example is certainly the most relevant for the previous construction since it appears in a theory with a spacetime symmetry.  Here,  however, one chooses   to present the exact same type of example for the gauge fixed Yang--Mills theory in four spacetime dimensions because the computations are easier.  Indeed, even though
the previous construction  was mainly designed for spacetime symmetries, nothing goes wrong when applying it to a theory with only internal symmetries.  
To do so, one simply considers  the anomaly operator 
\begin{equation}
\Delta_\C \equiv s + I_{V_{\delta \C}},
\end{equation}
where $\C$ now stands for the ghosts, antighosts and Lagrange multipliers $\{c,\bar{c},b \}$.  The same reasoning as above applies,   forgetting all the terms that come from $\Lie_\xi$.  To illustrate how this work,  one starts with Yang--Mills theory without gauge fixing term.  The analogue of \eqref{I_V_delta_g} in this case would be 
\begin{align}
I_{V_{\delta \C}} \delta A &= \mathfrak{a}_1 D(\delta c) ,
\nn \\
I_{V_{\delta \C}} \delta c &=  \mathfrak{a}_2  [\delta c , c] ,
\nn \\
I_{V_{\delta \C}} \delta \bar{c} &= \mathfrak{a}_3 \delta b ,
\nn \\
I_{V_{\delta \C}} \delta b &= 0.
\end{align}
Studying the ungauge fixed case will fix the coefficient $\mathfrak{a}_1$.  One has $\delta L = E + d\theta$ with 
\begin{align}
L &= - \demi F \star	F ,
\nn \\
E &= - \delta A (D \star F) ,
\nn \\
\theta &= \delta A \star F. 
\end{align}
The quantities that one needs to compute are $l_\C= a_\C, a_{\delta \C}, C_\C,C_{\delta \C}, A_\C, q_\C$ and $q_{\delta \C}$. One~gets
\begin{align}
\label{YM_classical_important_quantities}
\Delta_\C L = s L = 0 \quad &\Longrightarrow \quad a_\C = 0 ,
\nn \\
\Delta_{\delta \C} L = I_{V_{\delta \C}} \delta L \propto \delta c [F,\star F] = 0 \quad &\Longrightarrow \quad a_{\delta \C}  = 0 ,  
\nn \\
I_{V_\C} E = Dc (D \star F) = d   (c D \star F) \quad &\Longrightarrow \quad 
\left\{ \begin{aligned}
&C_\C =c D \star F ,  
\nn \\
&C_{\delta \C} = - \mathfrak{a}_1 \delta c D \star F ,  
\end{aligned} \right. 
\nn \\
\Delta_\C \theta = (\mathfrak{a}_1-1) D(\delta c) \star F \quad &\Longrightarrow \quad A_\C = 0 \quad \text{if\ } \mathfrak{a}_1 = 1 ,
\nn \\
\star J_{\rm BRST}   = - Dc \star F = - c D \star F - d (c\star F) = - C_\C - d q_{\C} \quad &\Longrightarrow \quad q_\C =  c \star F  ,
\nn \\
  d q_{\delta \C}  = I_{V_{\delta  \C}} \theta  - C_{\delta \C} + a_{\delta \C} =   \mathfrak{a}_1 d ( \delta c \star F ) \quad &\Longrightarrow \quad q_{\delta \C} =  \mathfrak{a}_1 \delta c \star F  ,   
\end{align}
where the identity $D (D \star F) \propto [F,\star F]=0$ was used. Notice that $\mathfrak{a}_1$ can't be different from $1$, otherwise the crucial property \eqref{Delta_theta2} is not fulfill. With these results, every steps performed previously for a spacetime symmetry  are legitimate and one can conclude 
\begin{align}
\label{I_V_O_YM_ungauge}
I_{V_\C} \O =&\  - \int_{\Sigma_3} \Big(  \delta (c D \star F) - \delta c D \star F \Big) + \int_{\pa \Sigma_3} \Big(    -  \delta c \star F + \delta ( c \star F )    \Big)
\nn \\
\hat{=}& \  \int_{\pa \Sigma_3} \delta (c \star F) - \delta c \star F .
\end{align}
Noether's second theorem $C_\C \ \hat{=} \ 0$ is of course satisfied  for the BRST symmetry associated with the ungauge fixed Yang--Mills Lagrangian.  Eq.\eqref{I_V_O_YM_ungauge}  shows that the generic procedure that uses the anomaly operator described above gives the correct fundamental canonical relation in the ungauge fixed case. To reach this conclusion,  one only had to fix $\mathfrak{a}_1$.  The hope would be that this holds true in the gauge fixed case by imposing further constraints on $\mathfrak{a}_2$ and $\mathfrak{a}_3$.  As already announced, this will not be the case and will constitute the counterexample \eqref{obstruction_anomaly}.

One therefore starts again with the gauge fixed Lagrangian  \eqref{L_BV_YM}, leading to $\delta L = E + d \theta$~with 
\begin{align}
\label{L,E,theta,YM-BV}
L &= - \demi F ( \star F )  + s\big(\bar{c} d (\star A)\big) - d\big(\bar{c} (\star D) c \big)  = - \demi F ( \star F ) + b  d (\star A)  + Dc (\star d) \bar{c}  ,
\nn \\
E &= - \delta A \Big( D \star F     + (\star d) b   + [ (\star d) \bar{c} ,  c]  \Big) - \delta b \Big(  (\star d) A   \Big) - \delta \bar{c}  \Big(  d (\star D)c \Big) + \delta c \Big(    D (\star d) \bar{c} \Big)  ,
\nn \\
\theta &= \delta A \star	F - b \star \delta A + \delta \bar{c} (\star D) c - \delta c (\star d) \bar{c}  .
\end{align}
In this case,  there is no need to compute all the quantities of \eqref{YM_classical_important_quantities} to show the incoherence.  In fact, as explained around \eqref{obstruction_anomaly},  one just has to compute
\begin{align}
\label{counterexample}
I_{V_{\delta \C}} E &= - D(\delta c)  \Big( D \star F     + (\star d) b   + [ (\star d) \bar{c} ,  c]  \Big) - \mathfrak{a}_3 \delta b  \Big(  d (\star D)c \Big) + \mathfrak{a}_2  [\delta c , c] \Big(    D (\star d) \bar{c} \Big) 
\end{align}
and check if there is a choice for $\mathfrak{a}_2, \mathfrak{a}_3$ that makes it a boundary term.  Clearly, no choice will work because the term $- D(\delta c)  (\star d) b$ cannot cancel with any other and is not a boundary term since $D(\star d) b \neq 0$.
The equation \eqref{counterexample} therefore constitutes   the desired counterexample, satisfying 
\begin{equation}
I_{V_{\delta \C}} E   \neq dX.
\end{equation}
This concludes the justification of why the most precise classical fundamental canonical relation \eqref{fundamental_canon} cannot be generically derived in the gauge fixed BRST CPS.

\newpage

\bibliography{biblio}
\bibliographystyle{kp}

\end{document}